\documentclass[aps,prd,superscriptaddress,showpacs,twocolumn,nofootinbib,preprintnumbers]{revtex4}
\usepackage{graphicx}
\usepackage{dcolumn}
\usepackage{bm}
\usepackage{natbib}
\usepackage{amsmath}
\usepackage{amssymb}
\usepackage{epstopdf}
\usepackage{hhline}

\newcommand{\be}{\begin{equation}}
\newcommand{\ee}{\end{equation}}
\newcommand{\bea}{\begin{eqnarray}}
\newcommand{\eea}{\end{eqnarray}}

\begin{document}


\title{Indirect Dark Matter Signatures in the Cosmic Dark Ages II.

Ionization, Heating and Photon Production from Arbitrary Energy Injections}
\preprint{
MIT-CTP/4683}

\author{Tracy R. Slatyer}
\email{tslatyer@mit.edu}
\affiliation{Center for Theoretical Physics, Massachusetts Institute of Technology, Cambridge, MA 02139, USA}


\begin{abstract} Any injection of electromagnetically interacting particles during the cosmic dark ages will lead to increased ionization, heating, production of Lyman-$\alpha$ photons and distortions to the energy spectrum of the cosmic microwave background, with potentially observable consequences. In this note we describe numerical results for the low-energy electrons and photons produced by the cooling of particles injected at energies from keV to multi-TeV scales, at arbitrary injection redshifts (but focusing on the post-recombination epoch). We use these data, combined with existing calculations modeling the cooling of these low-energy particles, to estimate the resulting contributions to ionization, excitation and heating of the gas, and production of low-energy photons below the threshold for excitation and ionization. We compute corrected deposition-efficiency curves for annihilating dark matter, and demonstrate how to compute equivalent curves for arbitrary energy-injection histories. These calculations provide the necessary inputs for the limits on dark matter annihilation presented in the accompanying Paper I, but also have potential applications in the context of dark matter decay or de-excitation, decay of other metastable species, or similar energy injections from new physics. We make our full results publicly available at \texttt{http://nebel.rc.fas.harvard.edu/epsilon}, to facilitate further independent studies. In particular, we provide the full low-energy electron and photon spectra, to allow matching onto more detailed codes that describe the cooling of such particles at low energies. \end{abstract}


\maketitle

\section{Introduction}

Between recombination and reionization, the universe experienced an epoch of extremely low ionization, known as the ``cosmic dark ages''. If new physics were to inject electromagnetically interacting particles into the universe during this period -- with the classic examples being dark matter (DM) annihilation or decay -- and consequently induce increased ionization, it could broaden the last scattering surface and have striking effects on the anisotropies of the cosmic microwave background (CMB) \cite{Adams:1998nr, Chen:2003gz, Padmanabhan:2005es}. Furthermore, heating of the gas induced by such energy injections could have observable effects on the 21cm line from neutral hydrogen \cite{Furlanetto:2006wp, Valdes:2007cu, Natarajan:2009bm, Valdes:2012zv}, and the production of additional low-energy photons could distort the blackbody spectrum of the CMB \cite{Zavala:2009mi, Hannestad:2010zt, 2012MNRAS.419.1294C, Chluba:2013wsa}.

A critical question in studies of the observational consequences of energy injection is what fraction of the injected power proceeds into the various observable channels, and over what period of time. As discussed in \cite{Slatyer:2009yq}, photons and $e^+ e^-$ pairs injected around the electroweak scale (a typical scenario in annihilating DM models) promptly convert the bulk of their energy into photons with energies lying within a redshift-dependent semi-transparent ``window'' \cite{Chen:2003gz}, where the dominant cooling mechanisms have timescales comparable to a Hubble time. Some fraction of these photons never scatter again, slowly redshifting and contributing to X-ray and gamma-ray background radiation in the present day; others eventually partition their energy into lower-energy photons and electrons, which are either efficiently absorbed by the gas or contribute to distortion of the CMB spectrum. Accordingly, energy may be deposited and contribute to observable signatures at times long after its original injection.

In this work we employ the code initially described in \cite{Slatyer:2009yq} and refined in \cite{2013PhRvD..87l3513S, Galli:2013dna} to describe the energy deposition histories corresponding to particle injection at arbitrary energies and redshifts. In \cite{2013PhRvD..87l3513S} we computed the partition between ``deposited'' energy and free-streaming high-energy photons, for injections of photons, electrons and positrons at arbitrary energy and redshift. In that work ``deposited'' energy was taken (as in \cite{Slatyer:2009yq}) to encompass low-energy particles in general, including distortions to the CMB spectrum and ionization, excitation and heating of the gas. 

In order to convert from this overall ``deposited energy'' to the individual deposition channels, articles in the literature have generally followed \cite{Chen:2003gz} in employing a simple prescription for the fraction of deposited power proceeding into ionization, excitation and heating, based on studies of this fraction for 3 keV electrons \cite{1985ApJ...298..268S}. More careful modeling of the cooling of electrons and photons had supported this estimate \cite{2010MNRAS.404.1869F, Valdes:2008cr, Valdes:2009cq, MNR:MNR20624, Galli:2013dna}, in particular for the fraction of deposited power proceeding to ionization (which is the most important channel for determining the impact on the CMB anisotropy spectrum, e.g. \cite{Padmanabhan:2005es}). However, \cite{Galli:2013dna} demonstrated that this prescription can be quite inaccurate in general, as somewhat higher-energy electrons (between a few keV and a few MeV in kinetic energy) deposit the bulk of their energy into distortions of the CMB spectrum rather than through interactions with the gas. By employing the code developed for \cite{Slatyer:2009yq} to model the cooling of high-energy particles (and the secondary particles produced by their cooling) down to 3 keV energies, and then matching onto a separate code handling the cooling of electrons below 3 keV, \cite{Galli:2013dna} and \cite{Madhavacheril:2013cna} presented updated estimates of the power proceeding into ionization -- and hence the constraints from the CMB -- for a selection of DM models. 

In this note, we present a similar update for \emph{all} injections of photons and $e^+ e^-$ pairs at redshifts during the cosmic dark ages, with injection energies in the $\mathcal{O}$(keV-TeV) range.\footnote{We do not in this work provide a detailed study of the energy losses of protons and antiprotons; an approximate method for including these contributions to energy deposition can be found in \cite{Weniger:2013hja}.} Furthermore, we provide estimates for the power proceeding into Lyman-$\alpha$ photons, heating, and CMB spectral distortion (by continuum photons below 10.2 eV), as well as ionization. We  provide the full spectra of photons and electrons below 3 keV produced by the high-energy code at all timesteps, for all injection energies and redshifts, to facilitate the interfacing of these results with more detailed and precise models for the cooling of the low-energy particles.

In Section \ref{sec:review} we review the issues that mandate an improved treatment of the energy deposition, with a focus on setting constraints on energy injection via the anisotropies of the CMB. In Section \ref{sec:code} we review the key elements of the code employed, and describe the resulting publicly available dataset. In Section \ref{sec:deposition} we review two procedures for converting the low-energy spectra into estimates for the deposition to various channels, and show updated results for the total energy deposited into the various channels under these prescriptions. In Section \ref{sec:fcurves} we review how to determine so-called $f(z)$ curves -- the power deposited at any given redshift, normalized to the power injected at that same redshift -- for any energy injection history; as an example, we present $f(z)$ curves corrected for the systematic effects identified in \cite{Galli:2013dna}, for general DM annihilation models, suitable for use with studies that employed earlier injection-energy-independent prescriptions for the fraction of deposited power proceeding into ionization. Finally, we present our conclusions. Appendix \ref{app:files} provides detailed descriptions for the files containing our results, available online at \texttt{http://nebel.rc.fas.harvard.edu/epsilon}.

\section{Continuum photon losses and energy deposited to the gas}
\label{sec:review}

To determine the constraints on any model of new physics that injects electromagnetically interacting particles into the universe during the cosmic dark ages, the key figure of merit is the power deposited into the relevant channel(s) at any given redshift. For example, for constraints based on the anisotropies of the CMB, the most important channel is ionization,\footnote{There is a subdominant effect from the production of additional Lyman-$\alpha$ photons, since atoms in an excited state can be more easily ionized by the ambient CMB photons; however, neglecting this effect entirely has been shown to change the constraints at only the $\sim 5 \%$ level \cite{Hutsi:2011vx, Galli:2013dna}, justifying a simplified approximate treatment of this contribution.} and so the constraints are determined by the power deposited into ionization of the gas as a function of redshift. The distortions to the CMB energy spectrum and the gas temperature are non-zero, but the constraints arising from those channels are much weaker than from the impact of extra ionization on the CMB anisotropies (e.g. \cite{Zavala:2009mi}). Once computed, the power deposited as extra ionization can be incorporated into public codes describing recombination -- such as RECFAST \cite{Seager:1999}, CosmoRec \cite{Chluba:2010ca} or HyRec \cite{AliHaimoud:2010dx} -- as described in \cite{Chen:2003gz, Padmanabhan:2005es}, and the resulting ionization history used to compute the effects on the CMB. 

Prior studies (e.g. \cite{Chen:2003gz, Padmanabhan:2005es,  Zhang:2007zzh, Galli:2009zc, Slatyer:2009yq, Kanzaki:2009hf, Hisano:2011dc,Galli:2011rz, Lopez-Honorez:2013cua, Diamanti:2013bia, Madhavacheril:2013cna, Planck:2015xua}) have divided the calculation of the power deposited to each channel into two steps: (1) computing the total deposited power as a function of redshift, and (2) computing the fraction of deposited power that proceeds into each channel. The result of step 2 has been presumed to be a function of the gas ionization fraction only, independent of the details of the energy injection; under this assumption, all the dependence on the energy injection model is partitioned into step 1. The results of step 1 have frequently been approximated as a constant efficiency factor $f$, so that the deposited power at any redshift is simply $f \times$ the power injected at that redshift, and $f$ captures all model-dependence in how the energy is deposited (e.g. \cite{Padmanabhan:2005es, Galli:2009zc}).

This approach assumes that the fraction of deposited power proceeding into each channel is independent of the spectrum of particles marked ``deposited''; that once the energy contained in free-streaming high-energy photons has been removed, only the total energy of the remaining particles matters. As shown in \cite{Galli:2013dna}, this approach can lead to constraints that are incorrect at the factor-of-two level (even within the limited parameter space explored there). 

\begin{figure}[ht]
\includegraphics[width=0.45\textwidth]{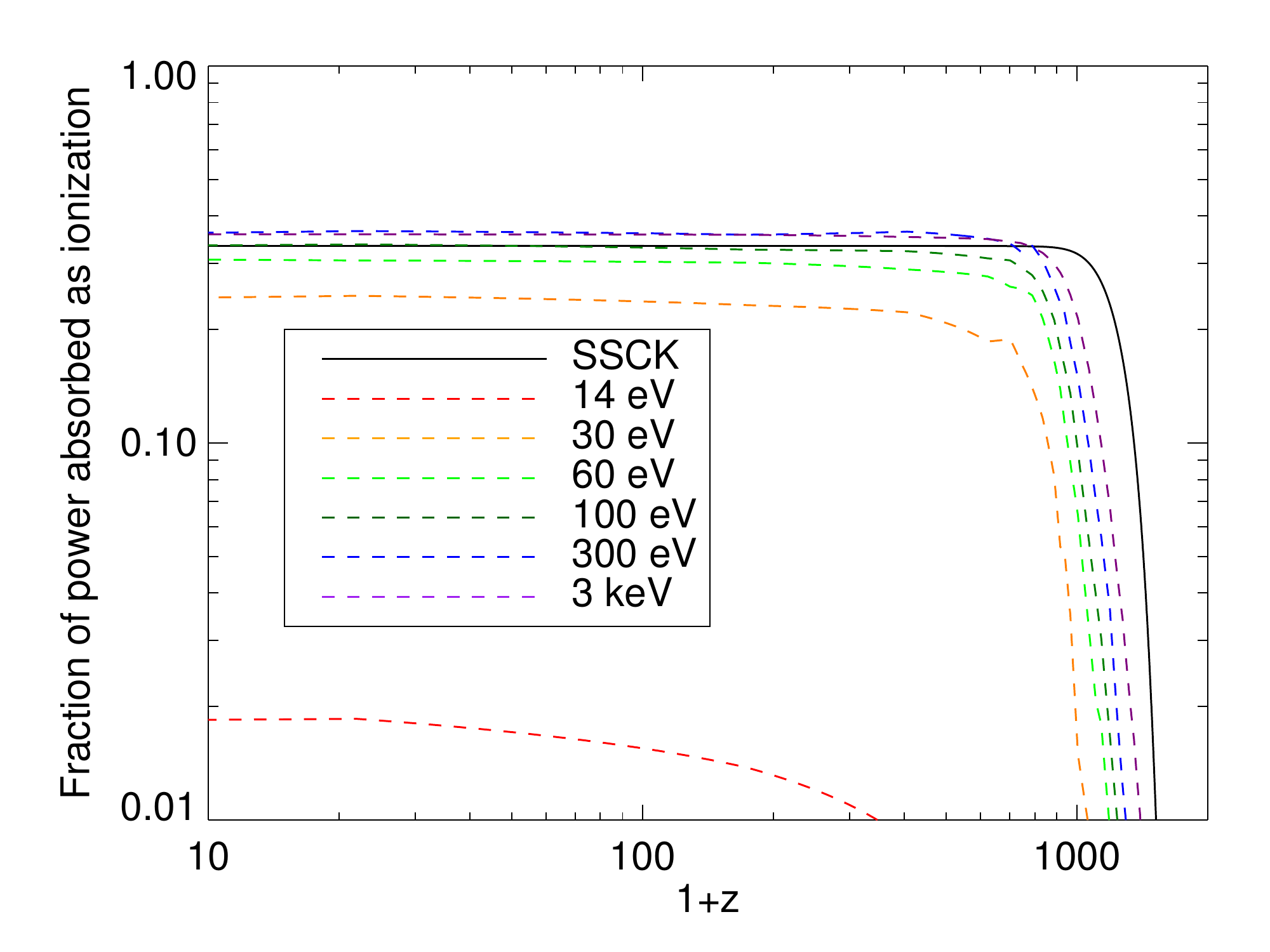}
\caption{\label{fig:ionfractions}
Fraction of deposited power that contributes to (hydrogen) ionization as a function of redshift, employing the ``SSCK'' prescription (solid black line), and the results presented in \cite{Galli:2013dna} for electrons at a range of energies. The ``3 keV'' prescription corresponds to the purple dashed line.
}
\end{figure}

The power into ionization has frequently been estimated in the literature using a simple ansatz first developed by Chen \& Kamionkowski \cite{Chen:2003gz}, based on an earlier numerical result for ionization by 3 keV electrons \cite{1985ApJ...298..268S}: $(1 - x_e)/3$ of the absorbed power goes into ionization, where $x_e$ is the ambient hydrogen ionization fraction. We denote this ansatz as ``SSCK'' (Shull, van Steenberg, Chen \& Kamionkowski). This estimate has been supported by more recent and detailed calculations \cite{2010MNRAS.404.1869F, Valdes:2008cr, Valdes:2009cq, MNR:MNR20624, Galli:2013dna}. \cite{Galli:2013dna} studied the effect of taking into account the spectrum of electrons below 3 keV, as opposed to simply using the results for 3 keV electrons, in determining the fate of the deposited energy; the effect was found to be small for the CMB constraints on the DM models studied in that work. The reason is that the fraction of power deposited as ionization is relatively stable for $\sim 100$ eV - $3$ keV electrons; we show this fraction as a function of redshift (employing a background ionization history from RECFAST as in \cite{Wong:2007ym}, with no new energy injection) in Figure \ref{fig:ionfractions}, for a range of electron energies. We also show the estimate from the ``SSCK'' energy-independent prescription. We will refer to the approach of taking the results presented in \cite{Galli:2013dna} for 3 keV electrons, and applying them to estimate the fraction of deposited power proceeding into the different channels, as the ``3 keV'' prescription; this is the approach employed by the \emph{Planck} Collaboration in setting constraints on annihilating DM \cite{Planck:2015xua}.

However, as noted in \cite{Galli:2013dna}, higher-energy electrons (at kinetic energies of a few keV to a few MeV) lose a very large fraction of their initial kinetic energy to inverse Compton scattering on the CMB; for mildly relativistic electrons, the resulting upscattered photons have too little energy to further interact with the gas after recombination (being well below the excitation and ionization thresholds). Accordingly, the amount of energy going into ionization, excitation or heating of the gas is suppressed, relative to the case for 3 keV electrons. This effect was underestimated in some earlier studies \cite{Valdes:2009cq, Evoli:2012qh} due to a mistake in the expression for the cooling time due to inverse Compton scattering, for non-relativistic electrons (the correct expression is given in e.g. \cite{2010MNRAS.404.1869F}). Consequently, for particles injected at high energy, the details of the low-energy electron and photon spectra produced by their cooling can significantly influence the fraction of deposited power that proceeds into ionization.

Equivalently, describing power degraded to low energy scales as ``deposited'' can be somewhat misleading, since photons at energies comparable to the CMB are not absorbed by the gas. The two-step approach described above could be improved by redefining ``deposited'' energy to exclude photons below 10.2 eV as well as free-streaming high-energy photons. However, such photons are also produced by the cooling of $\sim$ keV and lower-energy electrons; thus with this definition there is no range of energies within which particles can be treated as contributing solely to ``deposited energy''.

\begin{figure}
\includegraphics[width=0.5\textwidth]{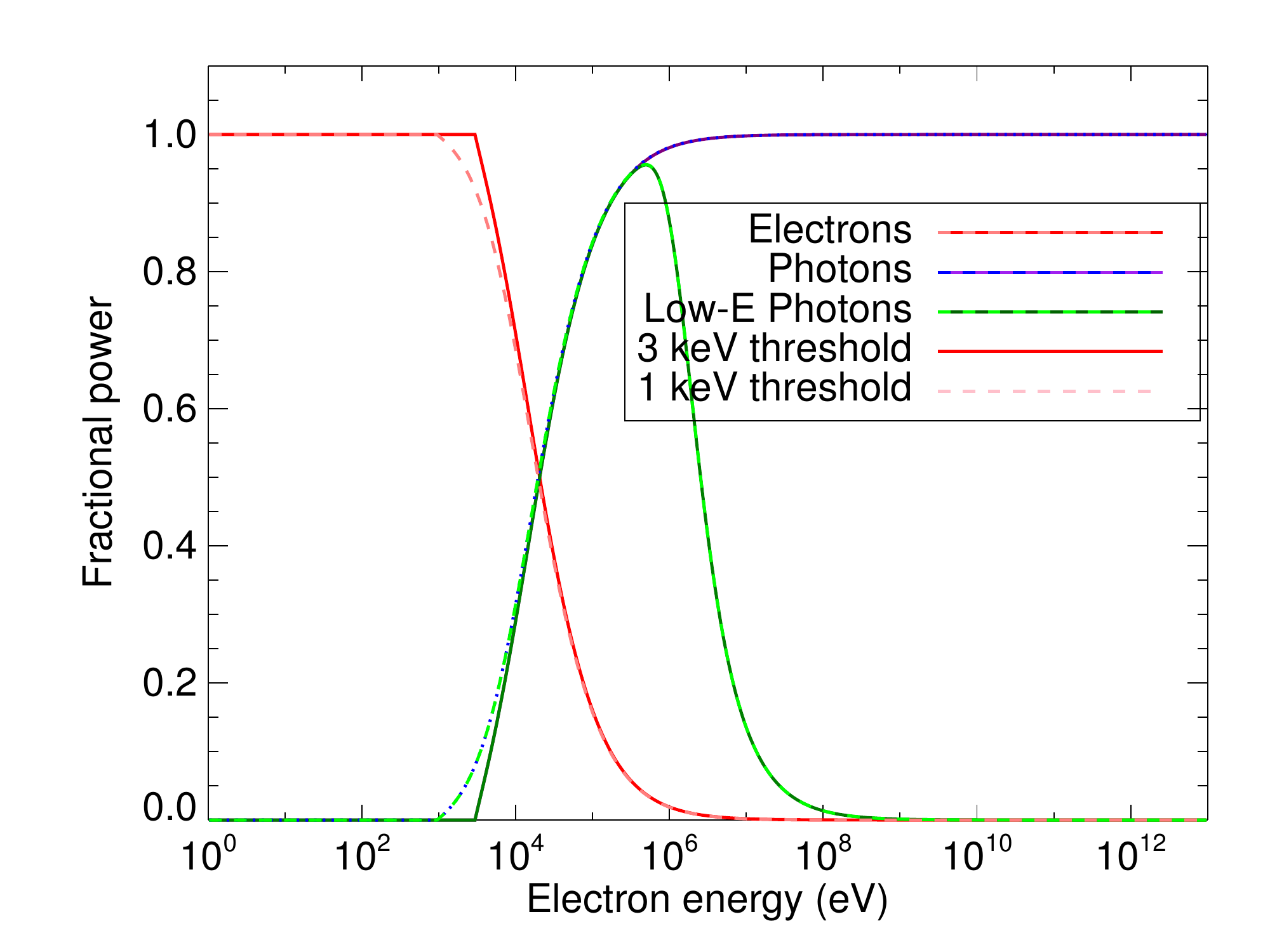}
\caption{\label{fig:contfrac}
Fraction of electron kinetic energy eventually partitioned into (red/pink) excitation/ionization/heating of the gas, plus electrons below the energy cutoff where atomic cooling processes dominate, and (blue/violet) photons produced by inverse Compton scattering of the electron on the CMB. Photons below 10.2 eV, which no longer interact significantly with the gas, constitute a subset of the latter contribution (light/dark green). All curves are for electrons depositing their energy at redshift 600. Dashed and dotted lines (pink,  blue, light green) correspond to a 1 keV threshold, whereas solid lines (red, violet, dark green) correspond to a 3 keV threshold; we see that the behavior is fairly independent of the chosen threshold except very close to it.
}
\end{figure}

Figure \ref{fig:contfrac} demonstrates the magnitude of the fractional energy loss to photons too low-energy to excite hydrogen, as a function of the initial electron energy, at a redshift of $z=600$ (where DM annihilation typically has its greatest effect on the CMB anisotropies \cite{Finkbeiner:2011dx}). Electrons that cool to some threshold (set to 1 keV or 3 keV) are presumed to lose all their energy to atomic processes, which dominate inverse Compton scattering in that energy range and are well-described by existing low-energy codes (for example, the fraction of power proceeding into ionization is well-characterized by the curves shown in Figure \ref{fig:ionfractions}). An electron above this threshold loses some fraction of its energy to atomic processes and some to inverse Compton scattering, in the process of cooling down to the threshold, producing photons on the way. As shown in the figure, when electrons are injected with (kinetic) energies between a few keV and $\sim 10$ MeV, the bulk of its energy is lost into photons below 10.2 eV in energy, for which the universe is approximately transparent.

It is difficult to properly model these inverse Compton losses in the context of Monte Carlo simulations for the low-energy atomic processes, due to the huge number of nearly-elastic collisions that are involved. As discussed above, this difficulty was avoided in \cite{Galli:2013dna} by using two separate codes to treat the low-energy (below 3 keV) and high-energy (above 3 keV) electrons. The high-energy code degrades the initial particles from their injected energy down to the 3 keV threshold, fully taking into account the effects of both inverse Compton scattering and redshifting. The results of the high-energy code -- spectra of electrons and photons below 3 keV -- can then be fed as inputs into a low-energy Monte Carlo code that treats the complex atomic cooling processes in detail. We take the same approach here, and make the results public.

\section{Description of numerical results}
\label{sec:code}

\subsection{Review of the numerical method}

The code developed for \cite{Slatyer:2009yq} takes as an input some injection of photons and electrons, with a specified redshift and energy dependence. Backreaction on the CMB photons and gas is \emph{not} included, as large modifications to the ionization history or CMB spectrum are ruled out by observational constraints, and consequently the problem is (to a good approximation) linear. We thus populate individual energy bins with electrons/positrons or photons at a specific redshift and track the spectral evolution with redshift. Our 40 energy bins are log-spaced between 1 keV and 10 TeV, in photon energy and electron kinetic energy. We employ 63 log-spaced redshift bins spanning the range from $1+z =$ 10 to 3000 (for the redshift at which the energy is injected -- for the redshift at which the energy is deposited, more finely-binned results are available upon request).

At each timestep, the photon spectrum is updated with the results of the various scattering and pair production processes described in \cite{Slatyer:2009yq}, and redshifted. Photons at sufficiently low energies are tagged as ``deposited'', stored, and removed from the part of the code that describes redshifting. The threshold for this ``deposition'' occurs when the photon would on average photoionize an atom once per timestep, and as in \cite{Slatyer:2009yq}, we choose a timestep of $d \ln (1+ z) = 10^{-3}$ (it was confirmed in that work that the results were converged at such a timestep).

As described in \cite{Slatyer:2009yq}, the free electrons produced at each timestep (by direct injection, pair production cascades, Compton scattering, etc) lose their energy on timescales much shorter than a Hubble time, and are handled by a separate module that includes inverse Compton scattering and atomic cooling processes, and resolves their energy deposition entirely at each timestep.

We choose a threshold of 3 keV as the separation scale between ``low'' and ``high'' energies. At ``low'' energies, as described above, inverse Compton scattering of electrons and redshifting of photons can be neglected and the results of a detailed Monte Carlo code will be employed to model electron cooling through atomic processes. ``Deposited'' photons above 3 keV (that is, efficient photoionizers that will produce electrons above the threshold) are assumed to immediately photoionize hydrogen, converting their energy into a spectrum of secondary electrons (the sub-percent energy loss to the ionization itself is neglected for these photons). ``Deposited'' photons below 3 keV are not processed further by the high-energy code, but are stored as an output at every timestep, to facilitate their use as an input for more detailed codes describing the low-energy cooling. Photons below 3 keV are only entirely tagged as ``deposited'' if the timescale for photoionization is less than $10^{-3} \times$ a Hubble time (for that photon energy and redshift); otherwise, only the fraction of the photons that would photoionize the gas in that timestep are tagged as ``deposited'' and added to the output (since their photoionization will create secondary electrons below 3 keV), while the remainder are tracked in the main code.

Electrons above the 3 keV threshold lose energy dominantly through inverse Compton scattering, as discussed above, but simple estimates for energy losses to excitation, ionization, and heating by electrons above the threshold are stored as outputs to the code (and constitute a small contribution to the overall excitation, ionization and heating due to the choice of threshold, typically $\sim 5-15$ \% or less). These estimates are based on the implementation of the excitation, ionization and heating cross sections described in \cite{Slatyer:2009yq}, which have been confirmed to give results fairly consistent with the detailed low-energy code \cite{Galli:2013dna}, with differences at the level of a few percent. Calculations which treat the low-energy cooling in detail can separate Lyman-$\alpha$ photons from sub-10.2 eV photons produced by collisional excitation, but for the subdominant contribution from the high-energy code, we simply assign all the power deposited via excitation to the Lyman-$\alpha$ channel.

Sub-threshold photons produced by inverse Compton scattering of these above-threshold electrons are also stored as outputs; since for mildly relativistic electrons the boost from inverse Compton scattering is small, we must also take into account the \emph{removal} of the original photons from the CMB. (That is, if a 1 eV photon is upscattered to 2 eV, it is described as the removal of 1 eV of energy from the CMB and the addition of a 2 eV photon to the output photon spectrum.) For the electron energies where the impact on the CMB is non-negligible, the inverse Compton scattering cross section is independent of the initial photon energy, so the removal of photons does not change the spectral shape and it is sufficient to characterize it in terms of energy loss. Once electrons cool to the 3 keV threshold or below (electrons produced by photoionization, Compton scattering or pair production may be produced with less than 3 keV of kinetic energy), they are added to the output low-energy electron spectrum for that timestep.

As in \cite{2013PhRvD..87l3513S}, we are primarily interested in charge-neutral sources of energy injection, so we will generally consider injecting $e^+ e^-$ pairs rather than electrons alone. This is largely irrelevant for the energy-loss mechanisms important for high-energy particles, but when the positrons have cooled far enough, they annihilate with ambient electrons, producing gamma rays. For a relativistic positron, the vast majority of its energy will be deposited via inverse-Compton-scattered photons and their subsequent cooling, and these processes do not depend on the charge of the original particle; for a non-relativistic positron, the bulk of its energy will be contained in its mass energy, and be deposited via the photons from annihilation. Accordingly, aside from tracking the annihilation photons we do not distinguish between positrons and electrons; since the low-energy electrons produced dominantly arise from photoionization and Compton scattering, they are indeed electrons (not positrons) and can be treated as such for detailed low-energy codes. Injection of electrons without accompanying positrons can therefore be modeled using the results for $e^+ e^-$ pairs, together with the results for photon injection to remove the impact of the photons from annihilation (see \cite{2013PhRvD..87l3513S} for an example of the procedure).

We truncate the calculation at $z=9$, and advise caution in using our results for $z \lesssim 30$, due to two simplifying assumptions in the code\footnote{We thank Aaron Vincent for raising this question.}: first, no model for reionization is included in our baseline ionization history, and second, interstellar radiation fields other than the CMB are not included, and may be relevant after the onset of star formation. To properly include these effects would involve considering a range of models for ionization and star formation, and testing the sensitivity of the results to these choices; we thus defer a detailed study of low redshifts to future work. The effect of reionization on the cooling of high-energy electrons and photons is likely to be rather small, as inverse Compton scattering and the main energy-loss processes for high-energy photons are largely insensitive to the ionization fraction \cite{Slatyer:2009yq}, so one might expect the output spectra of low-energy photons and electrons described in this section to be fairly independent of the choice of reionization model. However, the rate at which low-energy photons are absorbed by photoionization, and the partition of the low-energy photons' and electrons' energy into the various deposition channels (described in Section \ref{sec:deposition}), will both depend strongly on the background ionization fraction.

\subsection{Structure of the outputs}

The results of the high-energy code are expressed as entries in a three-dimensional grid for each of the particle types, describing -- for a particular redshift-of-injection, initial energy (for $e^+ e^-$ pairs, this corresponds to the initial \emph{kinetic} energy of one member of the pair), and redshift-of-deposition -- the following outputs:

\begin{itemize}
\item An estimate of the energy deposited by electrons above 3 keV in this timestep, as a fraction of the injected energy (for $e^+e^-$ pairs, this includes the mass energy), into the following channels:
\begin{enumerate}
\item Ionization of the H gas,
\item Ionization of the He gas (set to zero for this high-energy deposition, but listed for completeness as this channel will be populated later),
\item Excitation of the gas / production of Lyman-$\alpha$ photons which can excite neutral hydrogen, 
\item Heating of the gas, 
\item Production of photons with insufficient energy to either excite or ionize the gas (i.e. distortion of the CMB spectrum). 
\end{enumerate}
\item The total energy of photons removed from the CMB by upscattering, in this timestep, as a fraction of the injected energy.
\item The spectrum of low-energy electrons (below 3 keV) produced in this timestep (and the corresponding array of energy bins), expressed as the spectrum $dN/dE$ of electrons per pair of injected particles (i.e. in the case where DM annihilation is the source of energy injection, this is equivalent to the spectrum per annihilation).
\item The spectrum of low-energy photons (below 3 keV) produced in this timestep (and the corresponding array of energy bins), expressed as the spectrum $dN/dE$ of photons per pair of injected particles.
\end{itemize}
As a reminder, all timesteps have $d\ln(1+z) = 10^{-3}$; each of these results can be divided by this quantity to obtain (approximately) timestep-independent functions for the energy deposition and low-energy spectra. These results are available online at \texttt{http://nebel.rc.fas.harvard.edu/epsilon}, in the form of \texttt{FITS} and \texttt{.dat} files, and are described in detail in Appendix \ref{app:files}.

As outlined in Table \ref{tab:definitions}, we label the resulting spectra and energy injections by $S^\mathrm{species}_{c,ijk}$ or $S^\mathrm{species}_{\mathrm{sec},ijkl}$. The label ``species'' is either ``$\gamma$'' (for injected photons) or ``$e^+e^-$'' (for injected pairs). The label $c$ runs from 1 to 5 and indexes the absorption channels: ionization on hydrogen, ionization on helium, excitation, heating, and production of photons too low-energy to interact with the gas. Since the direct energy absorption from high-energy electrons is generally subdominant ($\lesssim 15 \%$), we do not distinguish between ionization on hydrogen and helium from these electrons, assigning the ionization contribution entirely to channel $c=1$. Due to numerical issues (truncation of the energy binning for both photons and electrons), the power lost from electrons due to inverse Compton scattering is not identical to the power gained by (tracked) photons (the former quantity is generally slightly larger); the difference is assigned to channel $c=5$, so it will later be added to the power stored in low-energy continuum photons, and verified to be small. The depletion of photons from the CMB spectrum is tracked in the $S^\mathrm{species}_{\mathrm{loss},ijk}$ array; when we compute the partition of deposited energy into each channel, this array will be subtracted from the fifth channel $S^\mathrm{species}_{5,ijk}$.

The $i, j, k$ labels index the redshift of deposition, energy of injection and redshift of injection respectively. The ``sec'' label indicates the species of low-energy (below 3 keV) secondary particles being described, and $l$ indexes the energy of these secondaries (or kinetic energy, in the case of electrons).

These data hold all the key information from the high-energy code, and we recommend their use as inputs for detailed studies of the electron and photon cooling at low energies. However, for ease of use we also provide estimates of these quantities based on the coarsely-binned results for the partition of electron energy given in \cite{Galli:2013dna}, following the methods outlined in that work.

\begin{table*}
\center{\begin{tabular}{|c|l|}
\hline
 & Definition \\
\hline
Channel $c$  & Hydrogen ionization: $c=1$ or ``ionH''\\
& Helium ionization: $c=2$ or ``ionHe'' \\
& Excitation: $c=3$ or ``exc'' \\
& Heating: $c=4$ or ``heat'' \\
& CMB spectral distortion or ``continuum'' photons: $c=5$ or ``cont''. \\
& ``ion'' indicates the sum of the ``ionH'' and ``ionHe'' channels. \\
& ``corr'' indicates the contribution to channel $c=5$ from the low-energy photons produced at each \\
& time-step by the cooling of high-energy particles (as opposed to the cooling of sub-3-keV electrons); this  \\ 
& term was not properly taken into account in  earlier prescriptions. \\
& ``loss'' indicates the depletion of the CMB by scattering (see text). \\
\hline
$S^\mathrm{species}_c(z, E, z^\prime) $ & describes the dimensionless rate (normalized to injected energy, and differential with respect to  \\ & $d\ln(1+z)$) at 
which energy is deposited into  channel $c$, via the interactions of (primary and secondary)    \\ 
& particles as they cool down to 3 keV (kinetic) energy, for a particle of the indicated species injected at  \\ & redshift $z^\prime$ and energy $E$ (see text for details). 
\\
\hline
$S^\mathrm{species}_{c,ijk}$ & $S^\mathrm{species}_c(z^i_\mathrm{dep},  E^j, z^k_\mathrm{inj}) d\ln(1 + z^i_\mathrm{dep}) = $ discretized form of the function above, for a specified grid of \\
& injection/absorption redshifts and injection energies $\{z^i_\mathrm{dep}, E^j, z^k_\mathrm{inj}\}$.
\\
\hline
$S^\mathrm{species}_\mathrm{sec}(z, E, z^\prime, E_\mathrm{sec}) $ & describes the rate $\frac{dN_\mathrm{sec}}{dE_\mathrm{sec} d\ln(1+z)}$ at which sub-3-keV secondary photons (electrons) of (kinetic) energy $E_\mathrm{sec}$  \\
& are produced (``sec'' labels the species of the secondaries), by the interactions of (primary and secondary) \\ & particles as they cool down to 3 keV (kinetic) energy,  for a particle of the indicated species injected at \\ & redshift $z^\prime$ and energy $E$. \\
\hline
$S^\mathrm{species}_{\mathrm{sec},ijkl}$ & $S^\mathrm{species}_\mathrm{sec}(z^i_\mathrm{dep}, E^j, z^k_\mathrm{inj}, E^l_\mathrm{sec}) d\ln(1 + z^i_\mathrm{dep}) =$ discretized version of the function immediately above,  for a  \\
& specified grid of injection/absorption redshifts, injection energies and secondary-particle energies \\ & $\{z^i_\mathrm{dep}, E^j, z^k_\mathrm{inj}, E^l_\mathrm{sec}\}$. \\
\hline
$T^\mathrm{species}_c(z, E, z^\prime) $ & describes the dimensionless rate (differential with respect to $d\ln(1+z)$) at which energy is absorbed into    \\ & channel $c$, for a particle of the indicated species injected at redshift $z^\prime$ and energy $E$ (see text for details). \\
& Derived from the $S$ functions described above.
\\
\hline
$T^\mathrm{species}_{c,ijk}$ & $T^\mathrm{species}_c(z^i_\mathrm{dep}, E^j, z^j_\mathrm{inj}) d\ln(1 + z^i_\mathrm{dep}) = $ discretized form of the function immediately above, for a specified  \\
&grid of injection/absorption redshifts and injection energies $\{z^i_\mathrm{dep}, E^j, z^k_\mathrm{dep}\}$.
\\
\hhline{|=|=|}
$f_c(z)$ &  (energy deposited to channel $c$ in a redshift interval $dz$) / (energy injected in the same interval $dz$) \\
\hline
$f_{\mathrm{high},c}(z)$ & as $f_c(z)$, but only including energy deposited by electrons as they cool down to 3 keV. \\
\hline
$f(z)$ &  $ \sum_c f_c(z) =$ (total energy absorbed to all channels in a redshift interval $dz$) / (energy injected in the \\
& same interval $dz$) \\
 \hline
$\chi_c(z)$ & $f_c(z)/f(z) = $ fraction of total absorbed energy at redshift $z$ proceeding into channel $c$.  \\
\hline
$\chi_c^\mathrm{base}(z)$ & fraction of  total absorbed energy at redshift $z$  proceeding into channel $c$, under an earlier simplified    \\
& prescription labeled by ``base'', corresponding to either the ``SSCK'' or ``3 keV'' prescriptions (see text).\\
\hline
$f^{c,\mathrm{base}}(z)$ & $f_c(z) / \chi_c^\mathrm{base}(z) = $ should be used to replace $f(z)$ in analyses where ``base'' prescription was assumed and the  \\
&  signal  is determined by channel $c$. Would be equal to $f(z)$ (\emph{not} $f_c(z)$) if ``base'' prescription were correct. \\
\hline
$f^{\mathrm{sim}}(z)$ & $f(z) (1 -\chi_\mathrm{corr}(z) ) = $ simplified estimate for $f^{c,\mathrm{base}}(z)$, for $c=1-4$, that is independent of channel $c$ and  \\ & prescription ``base''. Can be used to estimate the corrected constraints/detectability for signals that    \\ & depend on a  combination of ionization, excitation and heating. \\
\hline
$F^\mathrm{sec}(z,E_\mathrm{sec})$ & (spectrum $dN/dE$ of secondary particles, with species labeled by ``sec'', produced in a redshift interval $dz$) \\
& / (pairs of primary particles injected in the same interval)  \\
\hline
\end{tabular}}
\caption{Definitions of the various functions and index labels used in this article. Those entries above the double line describe completely general energy injections, whereas those below the double line require specification of an energy injection history.} \label{tab:definitions}
\end{table*}

\section{Computing the deposited energy by channel}
\label{sec:deposition}

\subsection{The low-energy photons and electrons}

Our next goal is to estimate how low-energy electrons and photons lose their energy, and thus convert the derived low-energy electron and photon spectra (those included in the $S^\mathrm{species}_{\mathrm{sec},ijkl}$ arrays) into contributions to the five channels described above. These contributions can then be added to those obtained directly from the high-energy code (the latter are generally subdominant). Our final result will be a three-dimensional grid $T^\mathrm{species}_{c,ijk}$ for each channel $c$ and injected species, where as previously $i, j$ and $k$ respectively index the redshift of deposition, energy of injection and redshift of injection. The elements of this array will correspond to the energy deposited to channel $c$ in the relevant timestep for deposition, as a fraction of the total injected energy. In general we have:
\begin{align} T^\mathrm{species}_{1-4,ijk} & = S^\mathrm{species}_{1-4,ijk} \nonumber \\
&  + \text{contributions from } S^\mathrm{species}_{\mathrm{sec},ijkl} \nonumber \\
T^\mathrm{species}_{5,ijk} & = S^\mathrm{species}_{5,ijk} - S^\mathrm{species}_{\mathrm{loss},ijk} \nonumber \\
& + \text{contributions from } S^\mathrm{species}_{\mathrm{sec},ijkl}. \end{align}

For the purposes of this work, we will use simplified results from existing Monte Carlo codes that model the atomic cooling processes, as presented in \cite{Galli:2013dna}. However, these results are only directly applicable to low-energy electrons. For low-energy photons (below 3 keV) above 13.6 eV, we follow the ``best'' method outlined in \cite{Galli:2013dna}:
\begin{itemize}
\item For ``deposited'' photons with $13.6$ eV $< E < 3$ keV, we assume prompt photoionization leading to a secondary electron and a contribution (of $13.6$ eV per ionization) to channel $c=1$. It was shown in \cite{Galli:2013dna} that separating out ionization on helium has a negligible impact on the results. This contribution is given by:
\begin{equation} \Delta T^\mathrm{species}_{1,ijk} = \sum_{l>13.6 \mathrm{eV}} 13.6 \mathrm{ eV} \frac{ S^\mathrm{species}_{\gamma,ijkl} E^l_\mathrm{sec} d\ln E_\mathrm{sec}}{ 2 (E^j + m^\mathrm{species})}, \end{equation} 
where the denominator is the total injected energy (i.e. twice the injection energy for photons, or twice the (injection kinetic energy + electron mass) for $e^+ e^-$ pairs).
\item For photons with $10.2$ eV $< E < 13.6$ eV, we assign the associated energy to channel 3, since such photons cannot ionize neutral hydrogen, but can excite it to a state from which it can be more readily ionized. This contribution is given by:
\begin{equation} \Delta T^\mathrm{species}_{3,ijk} = \sum_{10.2\mathrm{eV}<l<13.6 \mathrm{eV}} \frac{ S^\mathrm{species}_{\gamma,ijkl} (E^l_\mathrm{sec})^2 d\ln E_\mathrm{sec}}{ 2 (E^j + m^\mathrm{species})}. \end{equation} 
\item For photons with $E < 10.2$ eV, we assign the associated energy to channel 5:
\begin{align} \Delta T^\mathrm{species}_{5,ijk} & = \sum_{l<10.2\mathrm{eV}} \frac{ S^\mathrm{species}_{\gamma,ijkl} (E^l_\mathrm{sec})^2 d\ln E_\mathrm{sec}}{ 2 (E^j + m^\mathrm{species})}.\end{align} 
\end{itemize}
This approach may be inaccurate at redshifts significantly prior to recombination, when the photoionization rate is highly sensitive to the small neutral fraction, and prompt photoionization may not be possible (or may only be possible on helium). However, small changes to the gas ionization fraction at these same redshifts do not affect the CMB anisotropies (since the universe is opaque to CMB photons); accordingly, this inaccuracy does not impact CMB constraints (see e.g. \cite{Madhavacheril:2013cna} where this was explicitly tested). After recombination, the photoionization rate is always fast relative to the Hubble time for some range of (low) photon energies \cite{Slatyer:2009yq}.

This procedure converts the low-energy photon spectrum to a low-energy electron spectrum, plus contributions to channels 1, 3 and 5. We simultaneously define the related quantity,
\begin{align} T^\mathrm{species}_{\mathrm{corr}, ijk}  & \equiv S^\mathrm{species}_{5,ijk} - S^\mathrm{species}_{\mathrm{loss},ijk} \nonumber \\
& +  \sum_{l<10.2\mathrm{eV}} \frac{ S^\mathrm{species}_{\gamma,ijkl} (E^l_\mathrm{sec})^2 d\ln E_\mathrm{sec}}{ 2 (E^j + m^\mathrm{species})} ,  \label{eq:contlosses}  \end{align}
which describes the \emph{net} energy proceeding into channel 5 \emph{omitting} the direct contribution from low-energy electrons (so this quantity can be used to correct the results of previous studies, which accounted only for the latter).

To characterize the fate of the low-energy electrons, we employ the results presented in \cite{Galli:2013dna}, based on \cite{Valdes:2008cr} and with refinements as described in \cite{Valdes:2009cq, MNR:MNR20624}. These data describe the energy deposition fractions by channel, for electrons with energies from 14 eV to 10 keV, for background ionization fractions ranging from $10^{-4}$ to 1. These results are partially reproduced in Figure \ref{fig:ionfractions}, showing the fraction of power proceeding into ionization as a function of redshift and electron energy.

We interpolate these results logarithmically in both energy and background ionization fraction. The energy deposition is partitioned into ionization on hydrogen, ionization on helium, Lyman-$\alpha$ photons, heating, and continuum photons. Integrating our low-energy electron spectrum over this transfer function at each triple of input redshift, injection energy and output redshift, we obtain the remaining contributions to all five channels $c=1-5$. To evaluate the transfer function, we assume the baseline ionization history given by \texttt{RECFAST} (as shown in e.g. \cite{Wong:2007ym}), since the energy partition is not sensitive to changes in the ionization fraction \cite{Galli:2013dna} at the $\mathcal{O}(10^{-4})$ level that is currently allowed by observations (e.g. \cite{Madhavacheril:2013cna, Planck:2015xua}). Specifically, if the background ionization fraction is $x_e$ and the (interpolated) fraction of energy deposited to channel $c$ by an electron of kinetic energy $E$ is $F_c(x_e, E)$, the contribution to a channel $c$ from the cooling of these low-energy electrons is given by:
\begin{equation} \Delta T^\mathrm{species}_{c,ijk} = \frac{\sum_j d\ln E_\mathrm{sec} \left( E_\mathrm{sec}^l\right)^2 S_{e, ijkl} F_c(x_e(z^i), E^l_\mathrm{sec})}{2 \left(E^j + m^\mathrm{species} \right)} . \end{equation}

This completes our main calculation of the $T^\mathrm{species}_{c, ijk}$ grid.

\subsubsection{An alternate simplified method}
\label{subsubsec:approx}

Alternatively, and as a cross-check on our method, we can employ the ``approx'' procedure described in \cite{Galli:2013dna}. In this approach, rather than compute the energy proceeding into each deposition channel separately, we note that the large corrections relative to the ``3 keV'' and ``SSCK'' prescriptions are driven by energy losses into continuum photons produced by the cooling of high-energy electrons, which are not accounted for by the modeling of 3 keV electrons. This contribution to channel 5 is stored in the array $T^\mathrm{species}_{\mathrm{corr},ijk}$, as defined in Equation \ref{eq:contlosses}. If this contribution is subtracted from the overall deposited energy, then the fractions of the \emph{remaining} deposited energy proceeding into channels 1-4 can be approximated by the ``SSCK'' or ``3 keV'' prescriptions. If the fraction of deposited energy proceeding into a particular channel $c$ is $\chi_c^\mathrm{base}(z)$ in these prescriptions, where ``base'' can be ``SSCK'' or ``3 keV'', this approach leads to an alternate $T^\mathrm{species}_{c, ijk}$ grid given by:
\begin{equation} T^\mathrm{species; base}_{c, ijk} = \left[ \left(\sum_{c^\prime} T^\mathrm{species}_{c^\prime, ijk} \right) - T^\mathrm{species}_{\mathrm{corr},ijk} \right] \chi_c^\mathrm{base}(z^i). \label{eq:tbase} \end{equation}
This provides a reasonable approximation for channels 1-4 but should not be used for channel 5 (one approach is to re-add the $T^\mathrm{species}_{\mathrm{corr},ijk}$ contribution to channel 5).

\subsection{Results for the deposited energy by channel} 

\begin{figure*}
\includegraphics[width=0.3\textwidth]{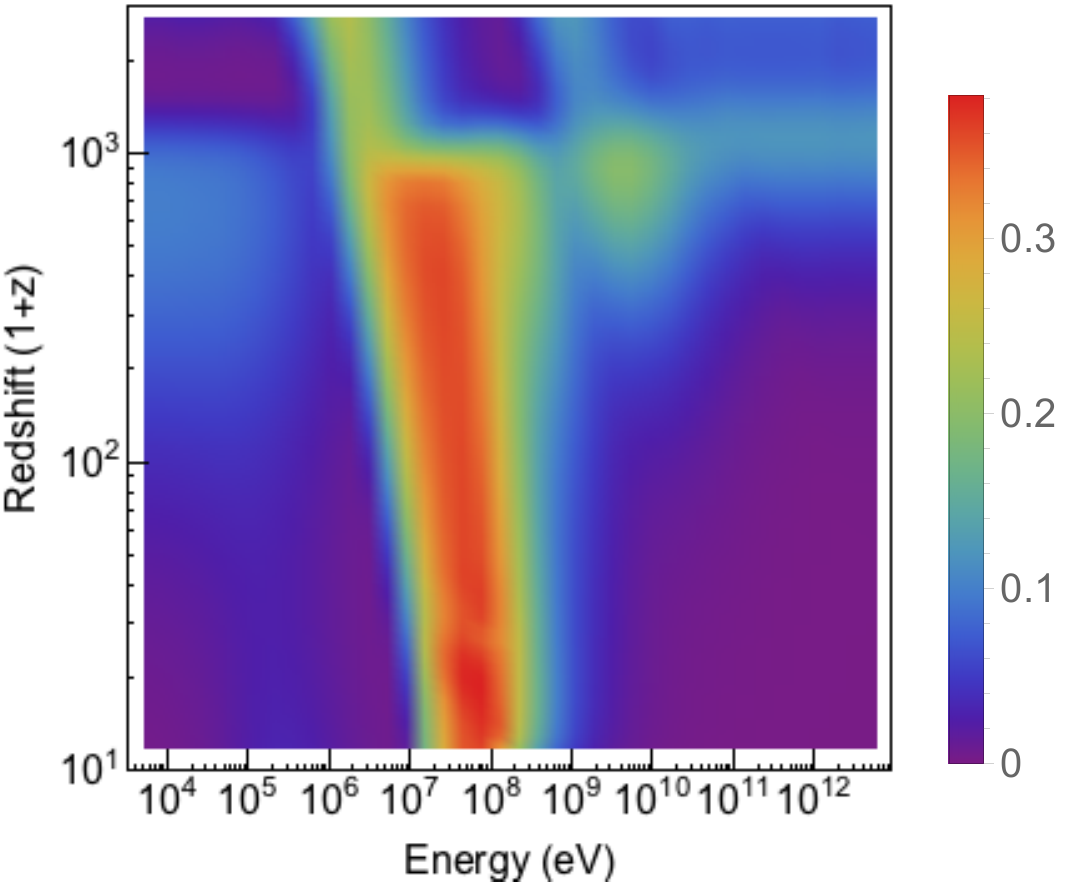}
\includegraphics[width=0.3\textwidth]{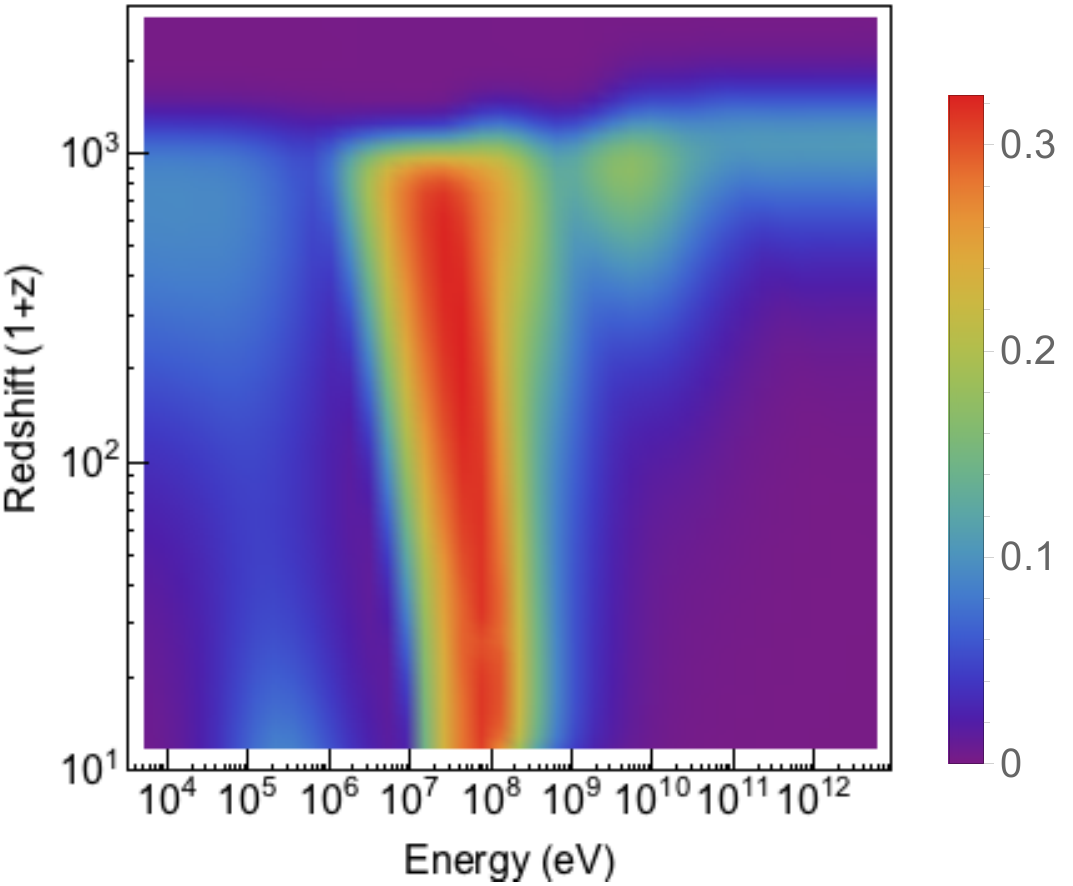}
\includegraphics[width=0.3\textwidth]{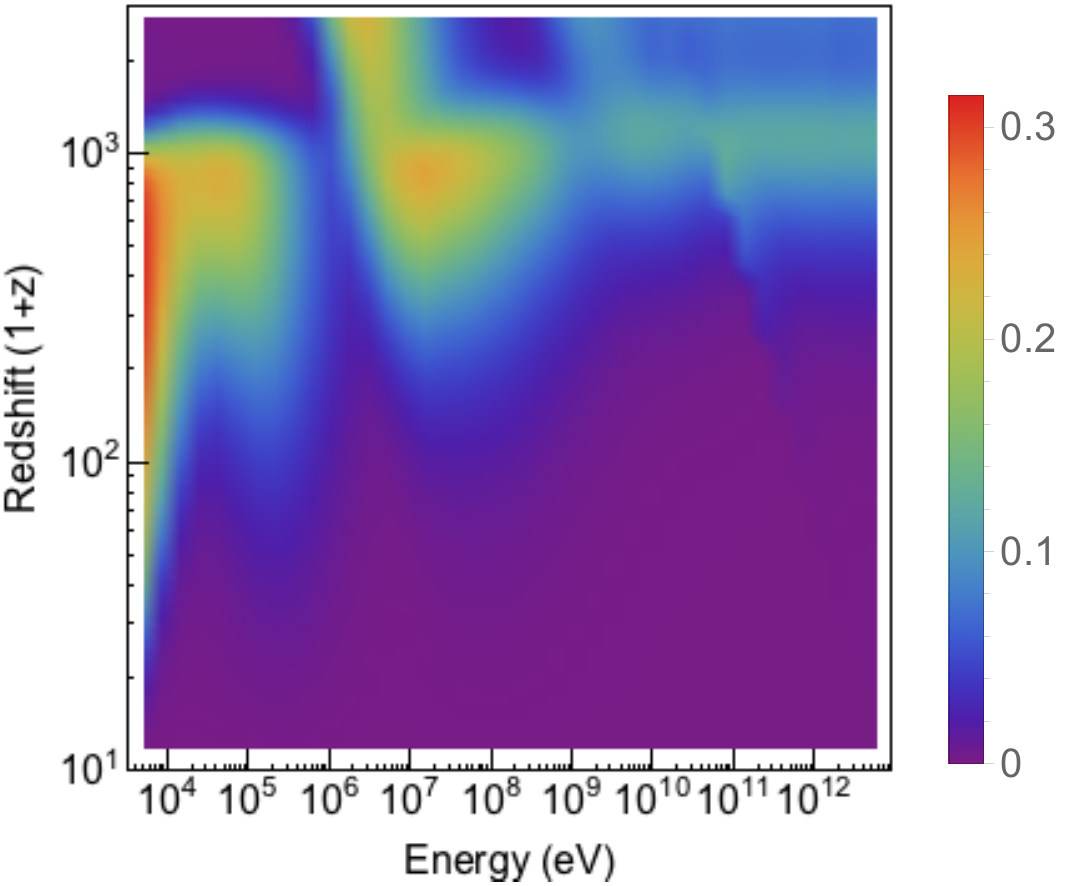} \\
\includegraphics[width=0.3\textwidth]{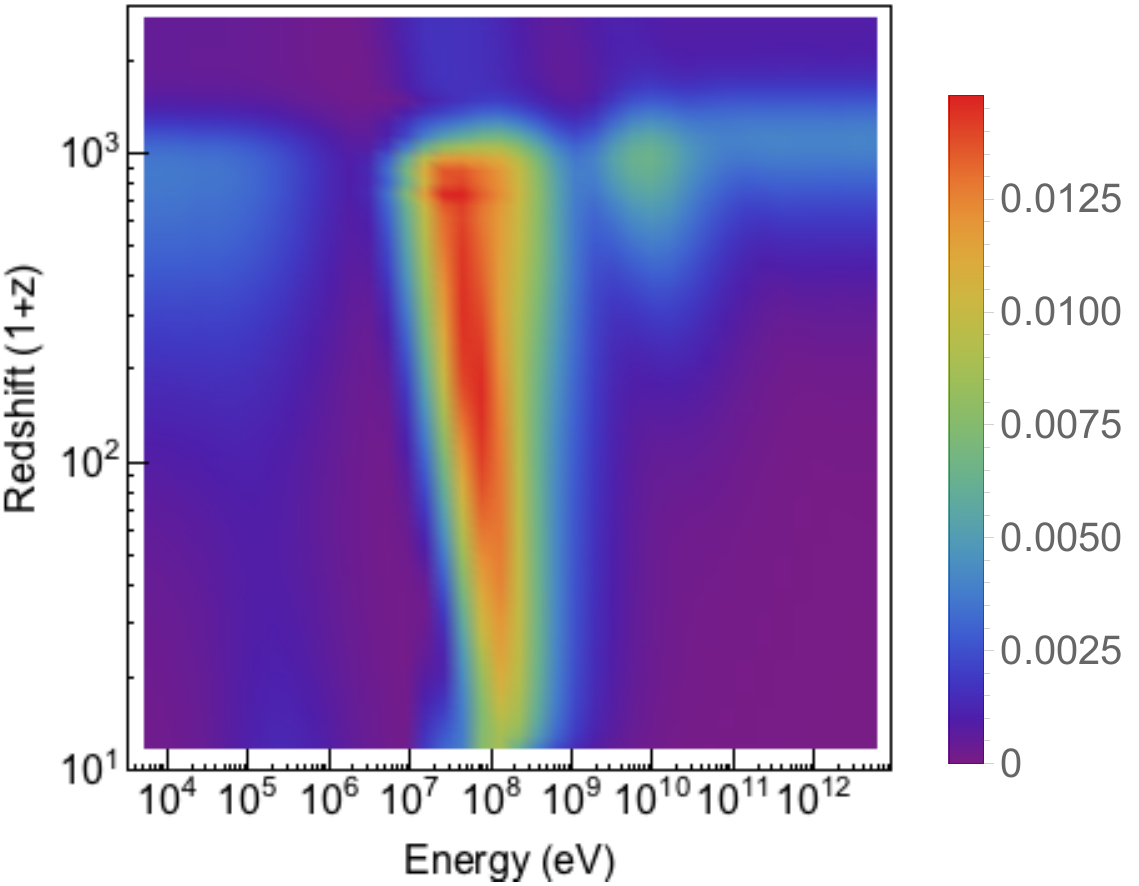}
\includegraphics[width=0.3\textwidth]{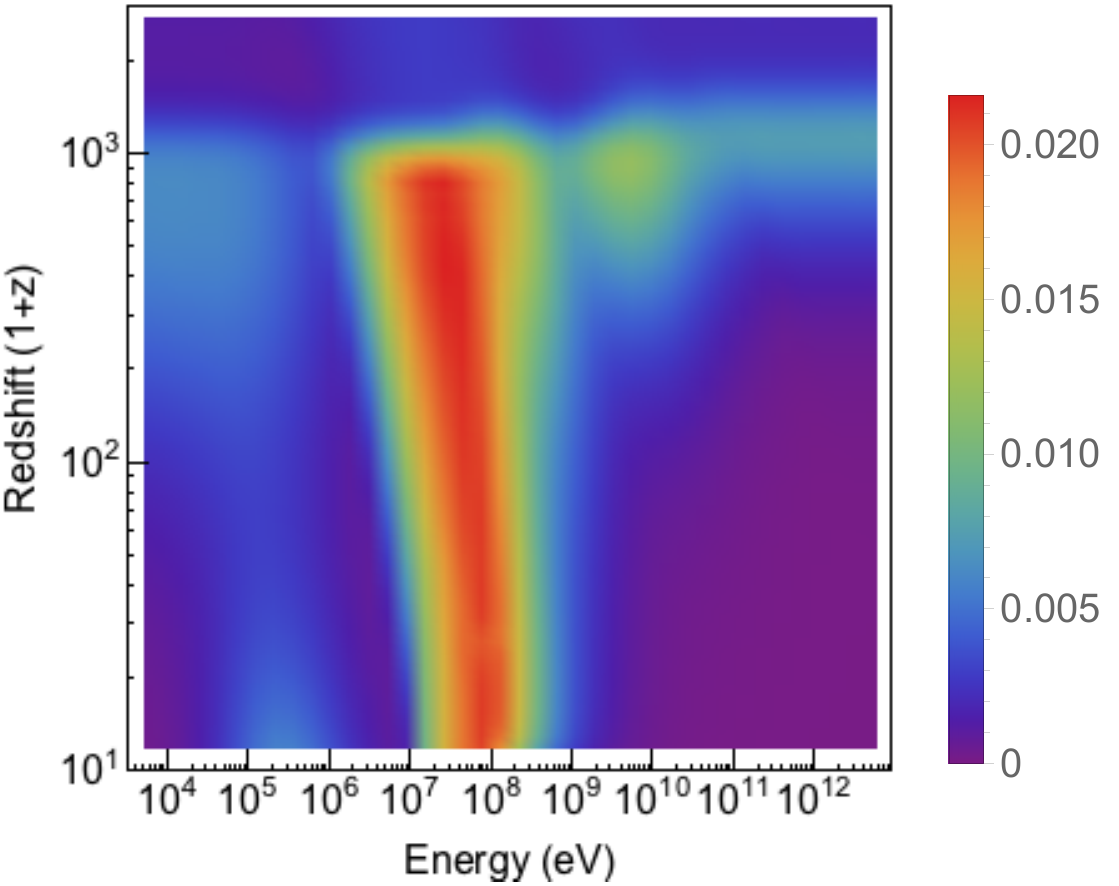}
\includegraphics[width=0.3\textwidth]{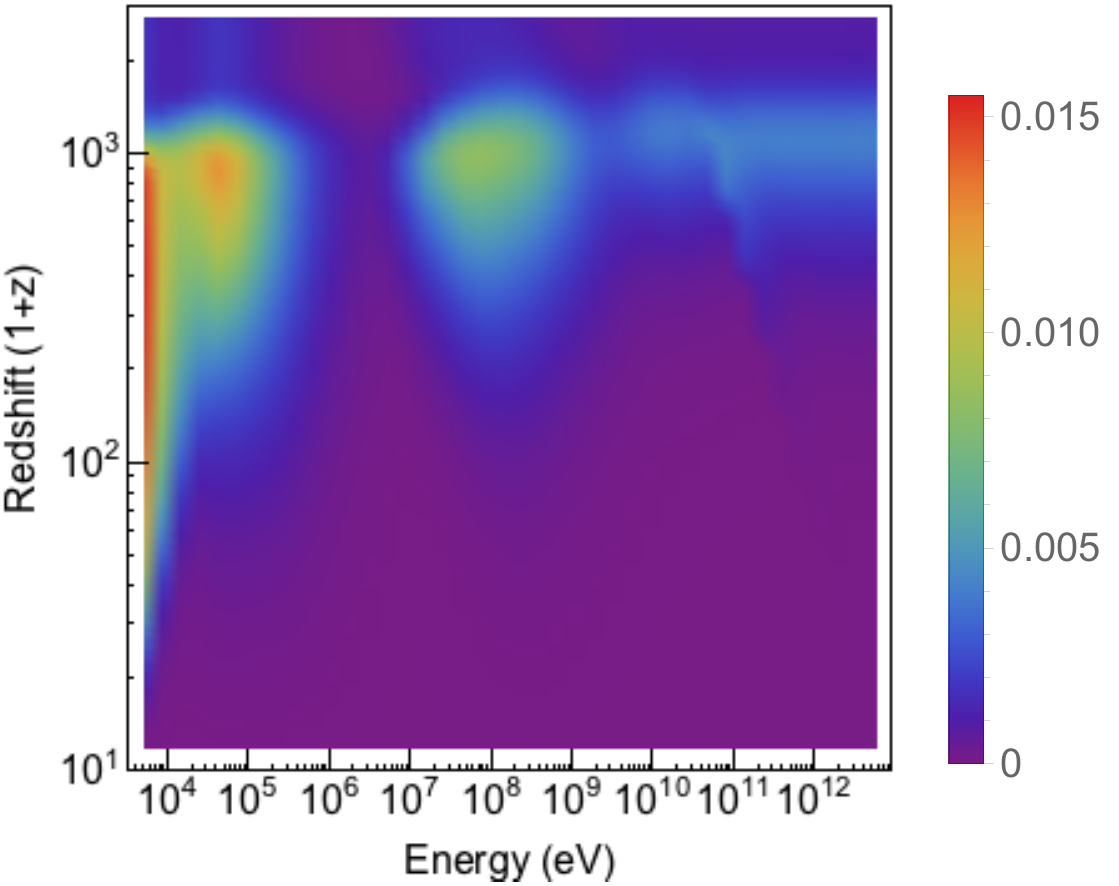} \\
\includegraphics[width=0.3\textwidth]{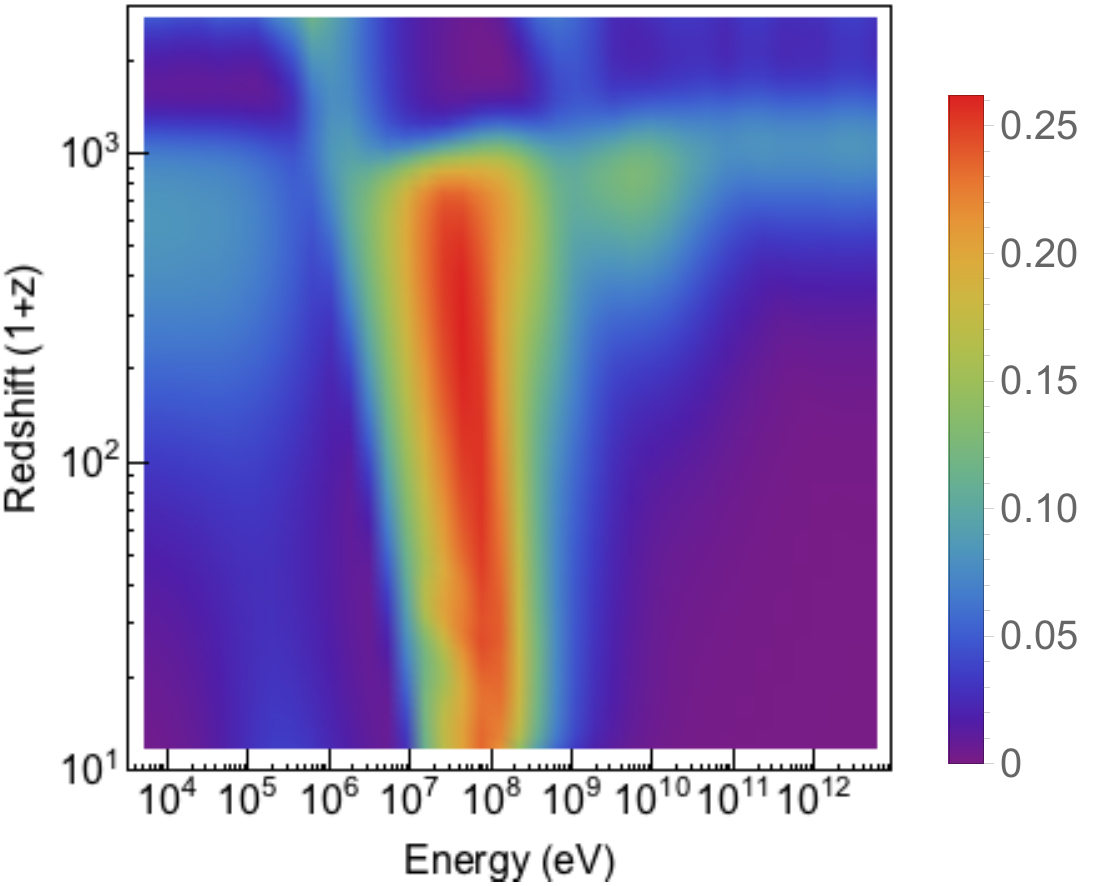}
\includegraphics[width=0.3\textwidth]{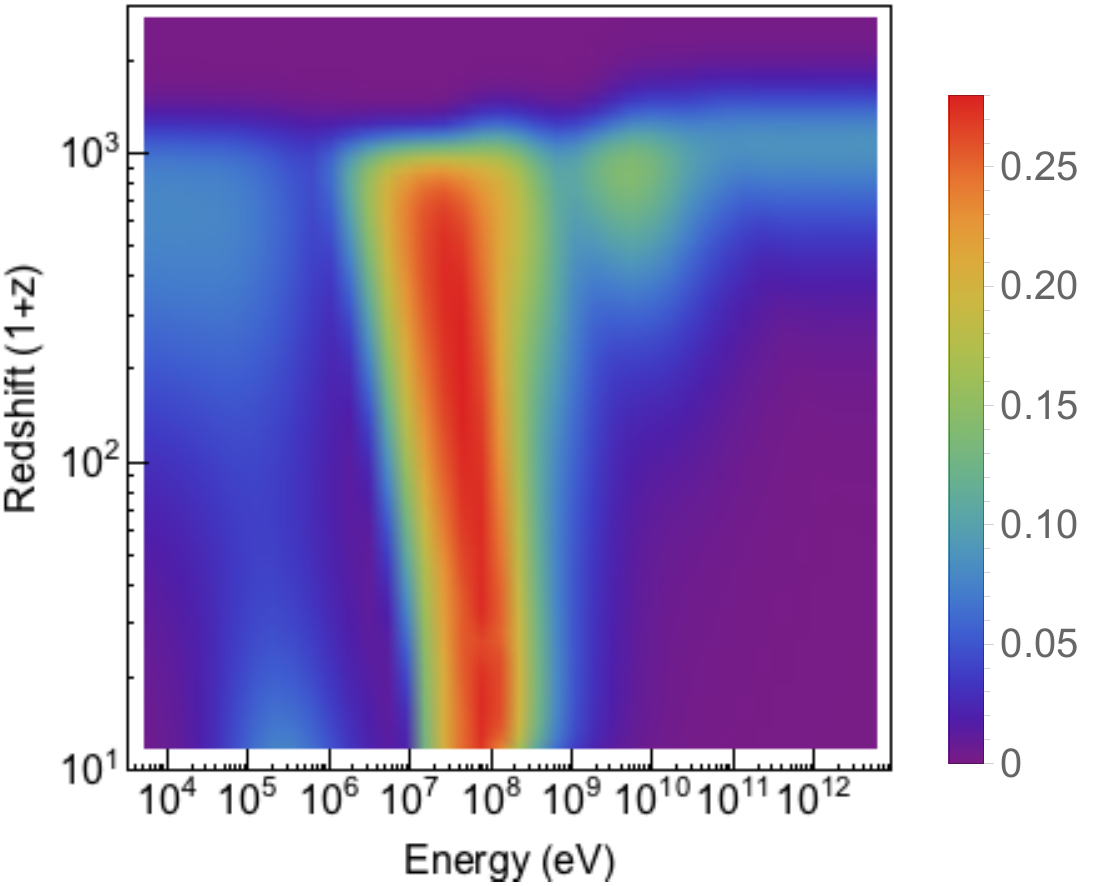}
\includegraphics[width=0.3\textwidth]{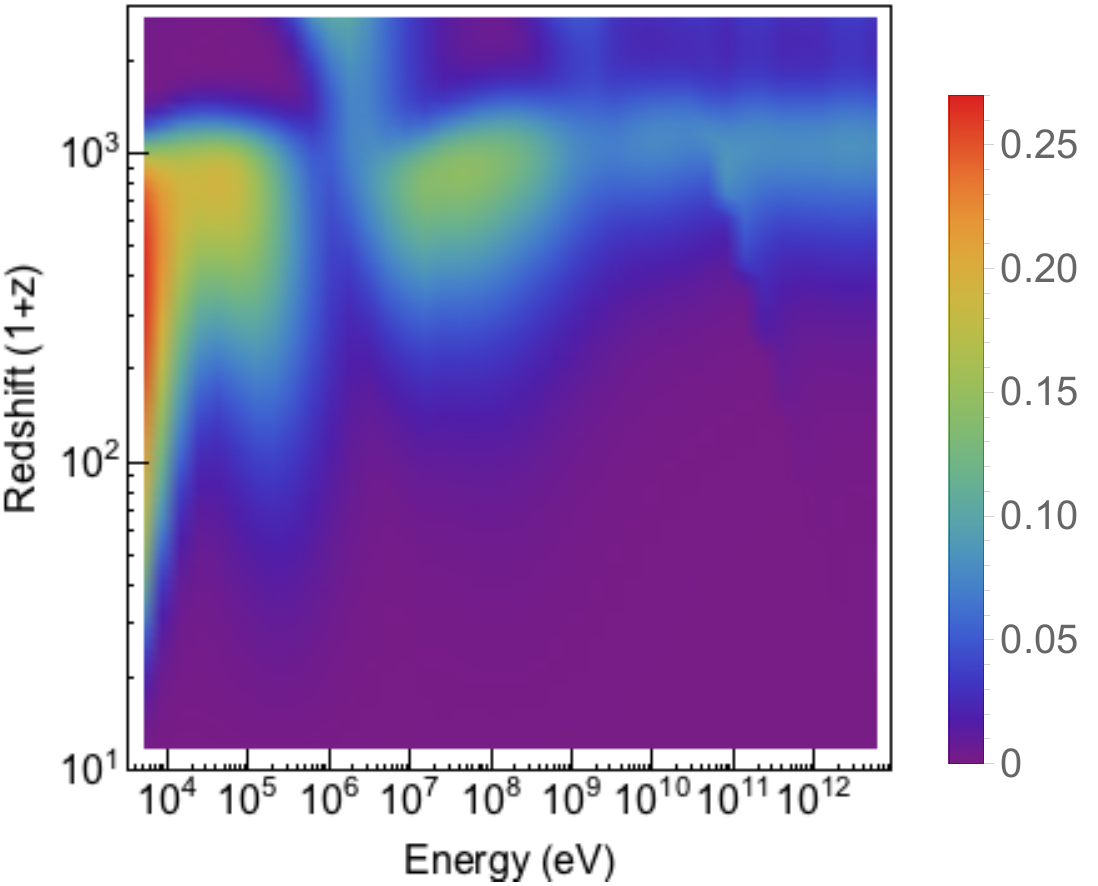} \\
\includegraphics[width=0.3\textwidth]{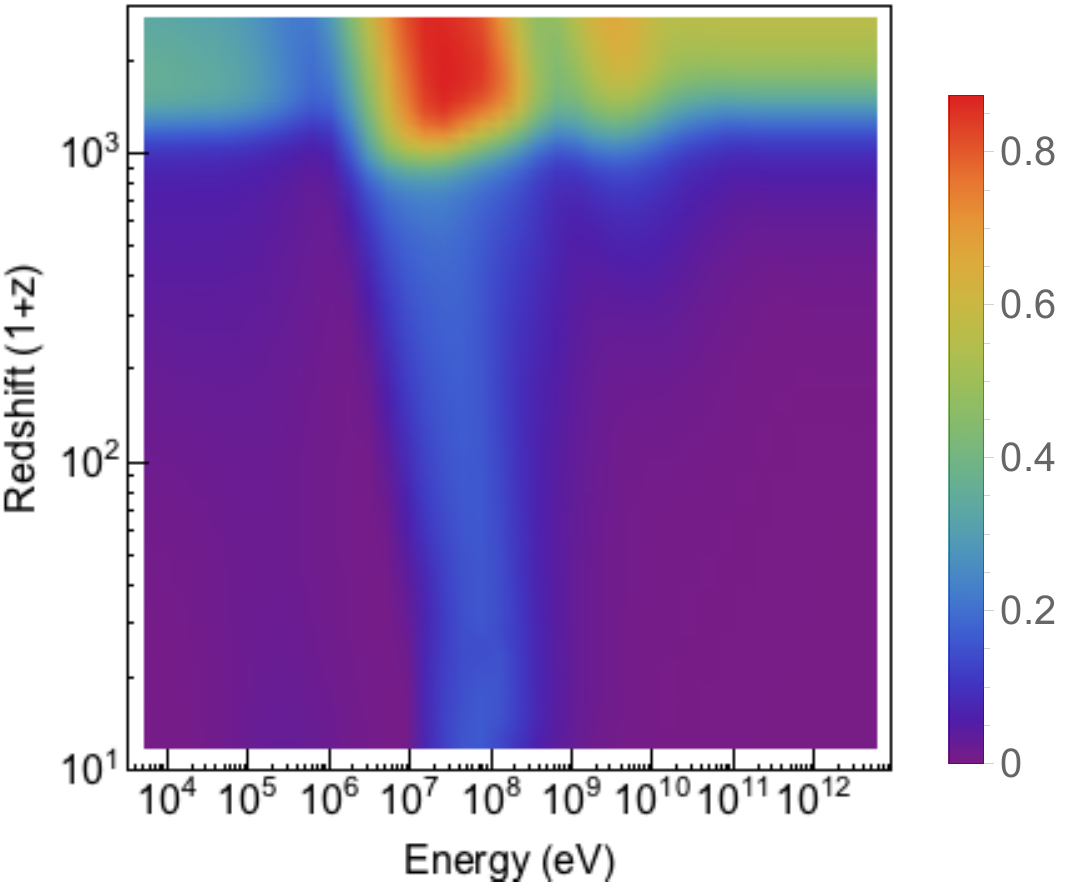}
\includegraphics[width=0.3\textwidth]{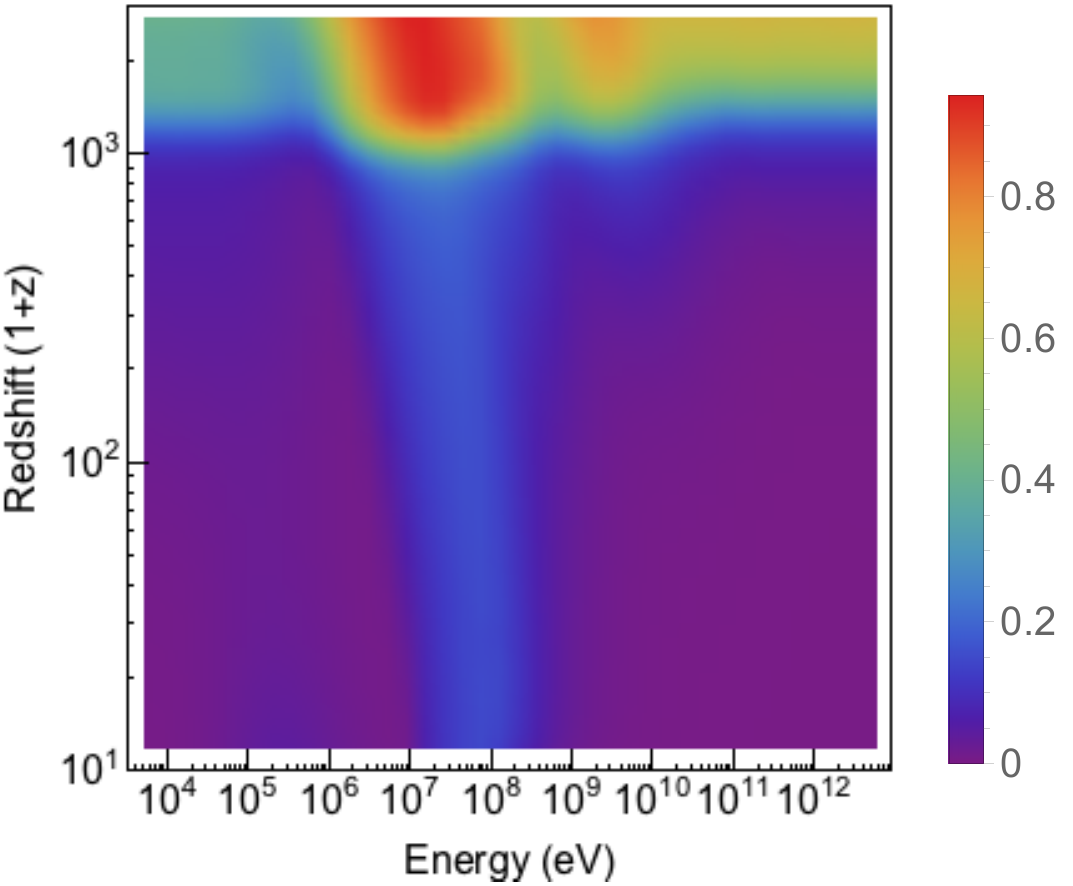}
\includegraphics[width=0.3\textwidth]{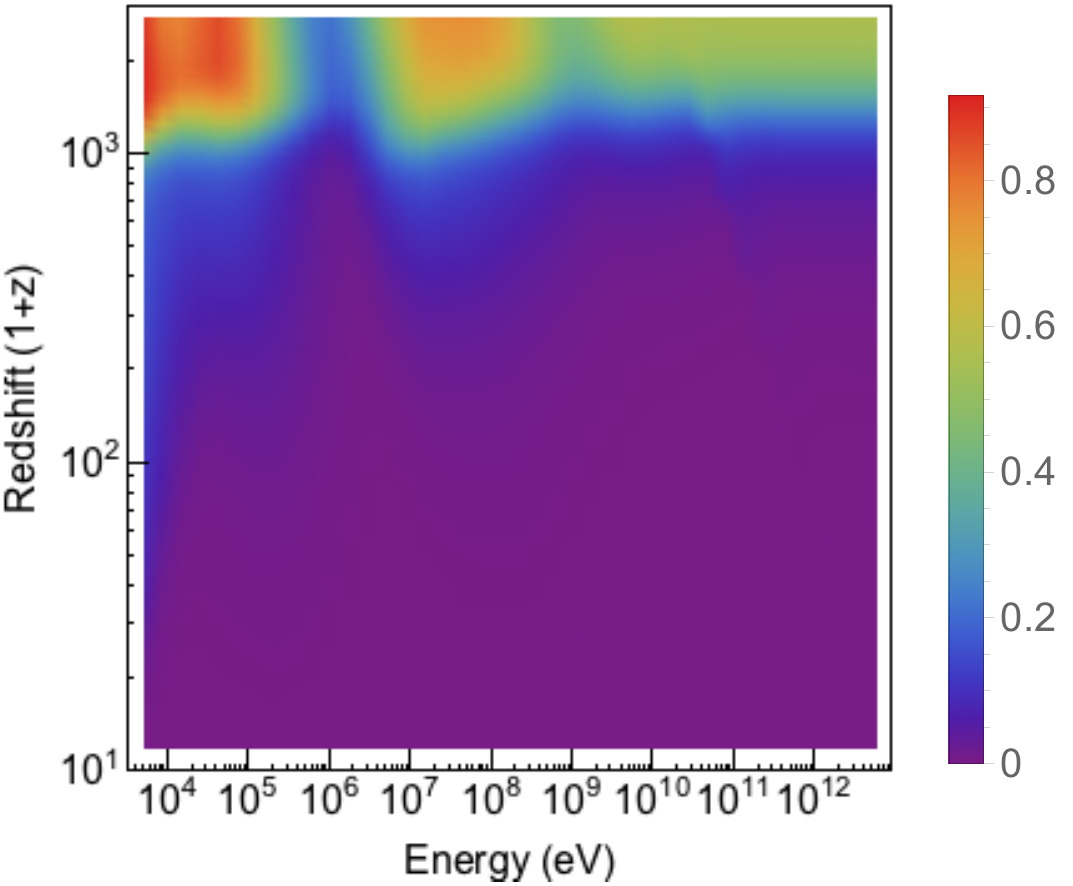} \\
\includegraphics[width=0.3\textwidth]{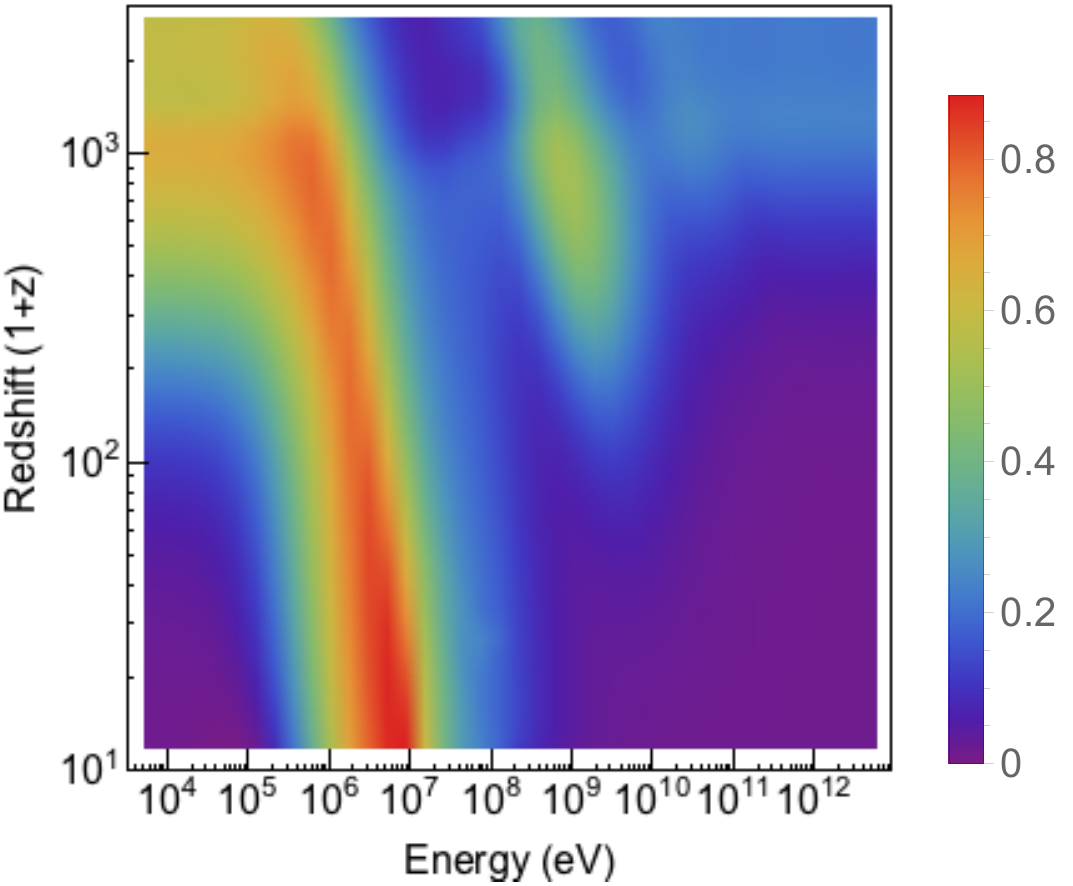}
\includegraphics[width=0.3\textwidth]{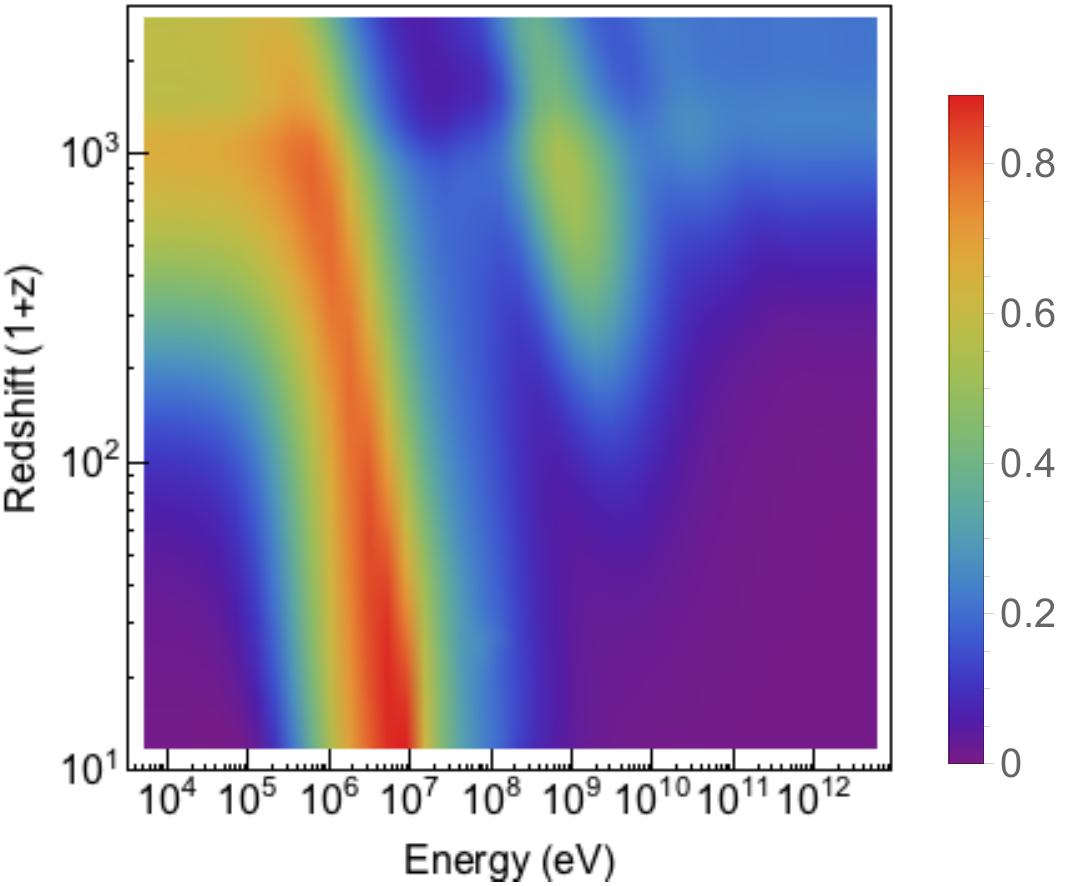}
\includegraphics[width=0.3\textwidth]{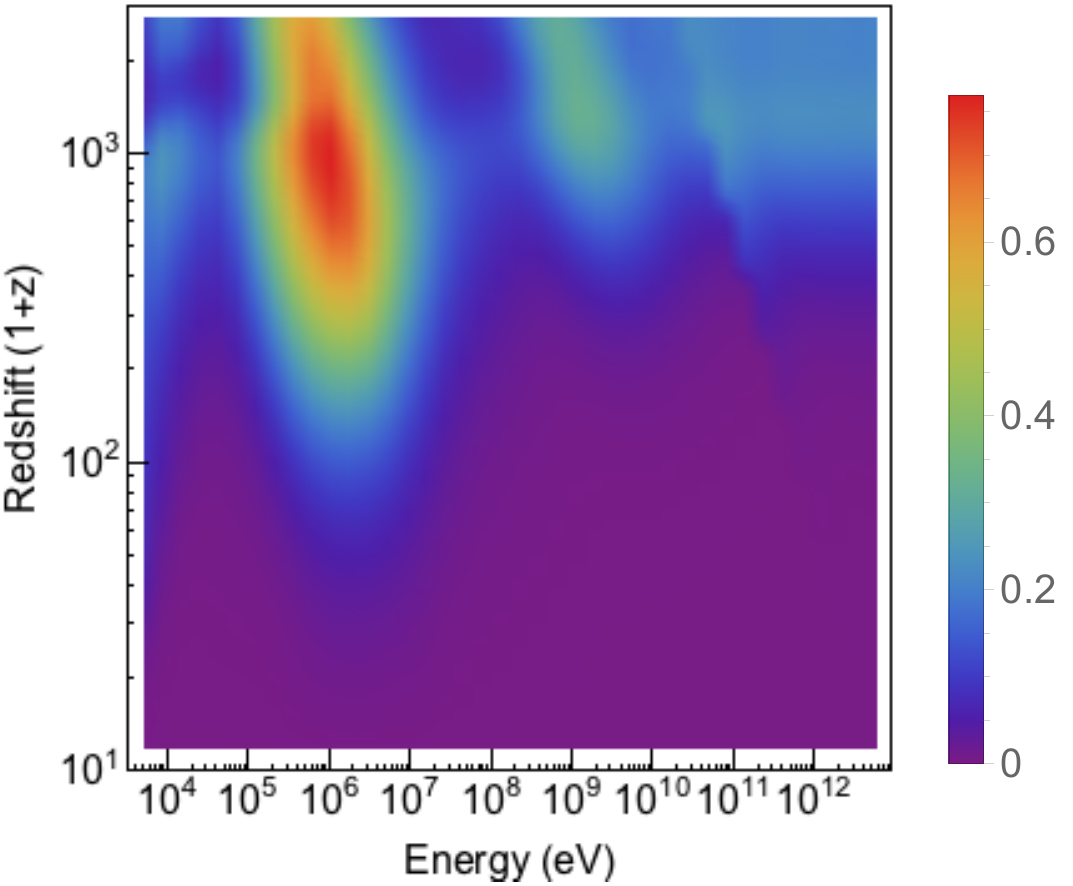}
\caption{\label{fig:totaldep}
Total power deposited (down to $1+z=10$) into each channel, as a function of injection energy and redshift. From top to bottom, the rows correspond to H ionization, He ionization, Lyman-$\alpha$ photons, heating, and sub-10.2 eV continuum photons. The left column describes energy absorption for  $e^+e^-$ pairs (the $x$-axis ``energy'' label here indicates the kinetic energy of a single member of the pair at injection), while the right column describes energy absorption for photons. The center column is an alternate method of estimating the same quantities shown in the left column, as described in Section \ref{subsubsec:approx}, and should be regarded as a cross-check.
}
\end{figure*}

As was previously done for the overall energy deposition \cite{2013PhRvD..87l3513S}, we can ask e.g. what fraction of a particle's energy is eventually deposited to each of these five channels; alternatively, upon specifying a redshift history for the energy injection we can ask how much power is deposited to each channel at a given redshift. Figure \ref{fig:totaldep} shows the former for $e^+e^-$ pairs and photons, for the ``best'' approach. As a cross-check, we also display the results of the ``approx'' approach for $e^+ e^-$ pairs using the ``3 keV'' baseline, re-adding the power lost to continuum photons for channel 5; the photon and ``SSCK'' cases are qualitatively similar.

In general, we see that heating of the gas and distortions of the CMB spectrum dominate the energy losses at high redshifts prior to recombination, while excitation and ionization become more significant after recombination; in general, the fraction of power eventually deposited falls for all channels at lower redshifts of injection (with some limited exceptions), as the universe becomes more transparent. Especially for $e^+ e^-$ pairs, but also for photons, there is a striking structure at injection energies around 10-100 MeV, where the fraction of injected power proceeding into ionization and excitation is quite high (and the fraction of power proceeding into heating is also enhanced); at lower energies, $\sim 1-10$ MeV, these channels are suppressed, and instead the production of low-energy continuum photons is enhanced. This accords with the behavior shown in Figure \ref{fig:contfrac}. As expected, the results of the ``approx'' and ``best'' methods are in general very similar for redshifts $z\lesssim 1000$.

%

\section{Corrected deposition-efficiency functions}
\label{sec:fcurves}

\subsection{Deposition efficiency by channel}

Above we have presented results for the fraction of injected power deposited to the gas over the age of the universe; however, as discussed in Section \ref{sec:review}, the key figure of merit is instead the power deposited at any given redshift. Once an energy injection history is specified (as a function of redshift), one can use the results presented above to integrate over the redshift of injection, and determine the deposited power originating from energy injections at all earlier times.

It is often convenient to normalize the deposited power at a given redshift to the injected power at the same redshift (where both quantities are defined within a given comoving volume, per baryon, etc). This ratio defines an \emph{effective deposition efficiency} curve $f(z)$. Since deposition at a given redshift may have contributions from power injected at much earlier redshifts, $f(z)$ can in some cases be greater than 1, but it is typically $\mathcal{O}(0.1-1)$ \cite{Slatyer:2009yq}.\footnote{In the context of annihilating DM, $f(z)$ curves are usually defined with respect to the injected power from the smooth DM distribution, since this is easy to characterize, even though the onset of structure formation may greatly increase the injected and hence deposited power. In this paper we will not take these effects into account, and so will always define $f(z)$ curves with respect to the total injected power.} 

As discussed above, it is not in general sufficient to derive the deposition-efficiency curve and then multiply the result by a model-independent prescription $\chi_c^\mathrm{base}(z)$ for the fraction of deposited power proceeding into the various deposition channels. Instead, we will define $f_c(z)$ curves corresponding to the individual channels, which give the power deposited at a given redshift \emph{to a specific channel}, normalized to the total injected power at the same redshift. By definition, $f(z) = \sum_c f_c(z)$.

Given an injection of some species with redshift and energy dependence such that the rate of particle injection per unit time per unit volume is given by $\frac{dN}{dE dV dt} = I(z, E)$, the corresponding $f_c(z)$ curves can be approximated by:
\begin{align} f_c(z^i) & \approx \frac{ \sum_j \sum_k E^j I(z^k,E^j) dV(z^k) dt(z^k) T^\mathrm{species}_{c,ijk} dE^j }{\sum_j E^j I(z^i,E^j) dE^j dV(z^i) dt(z^i)} \nonumber \\
& = \frac{ H(z^i) (1+z^i)^3 }{\sum_j E^j I(z^i,E^j) dE^j} \nonumber \\
& \times \sum_k \frac{1}{(1+z^k)^3 H(z^k)} \sum_j E^j I(z^k,E^j)  T^\mathrm{species}_{c,ijk} dE^j. \label{eq:fcurve} \end{align}
Here the indices $i, j, k$ label redshift of deposition, energy of injection and redshift of injection, as above; $dt(z)$ is the time interval corresponding to $d\ln(1+z)$, and we have employed the relations $H(z) = - d\ln(1+z)/dt$ and $dV(z_1)/dV(z_2) = (1+z_2)^3/(1+z_1)^3$.

\begin{figure*}
\includegraphics[width=0.3\textwidth]{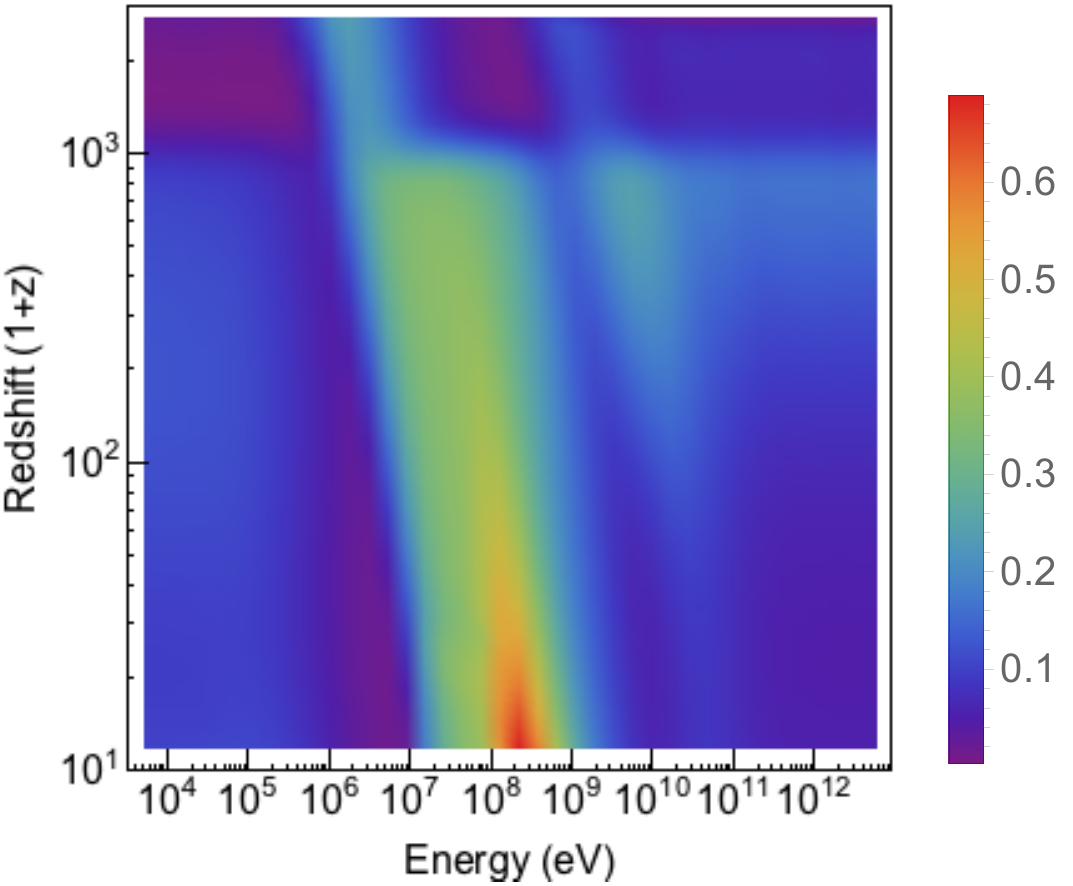}
\includegraphics[width=0.3\textwidth]{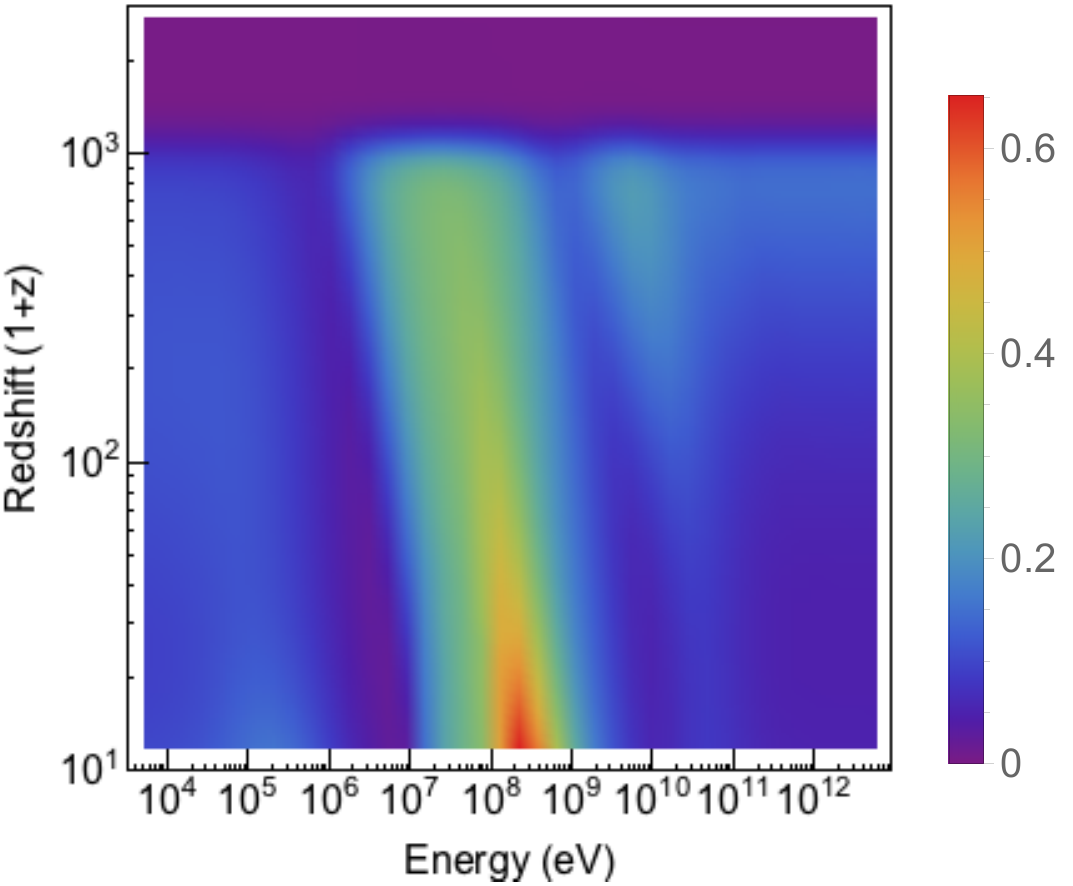}
\includegraphics[width=0.3\textwidth]{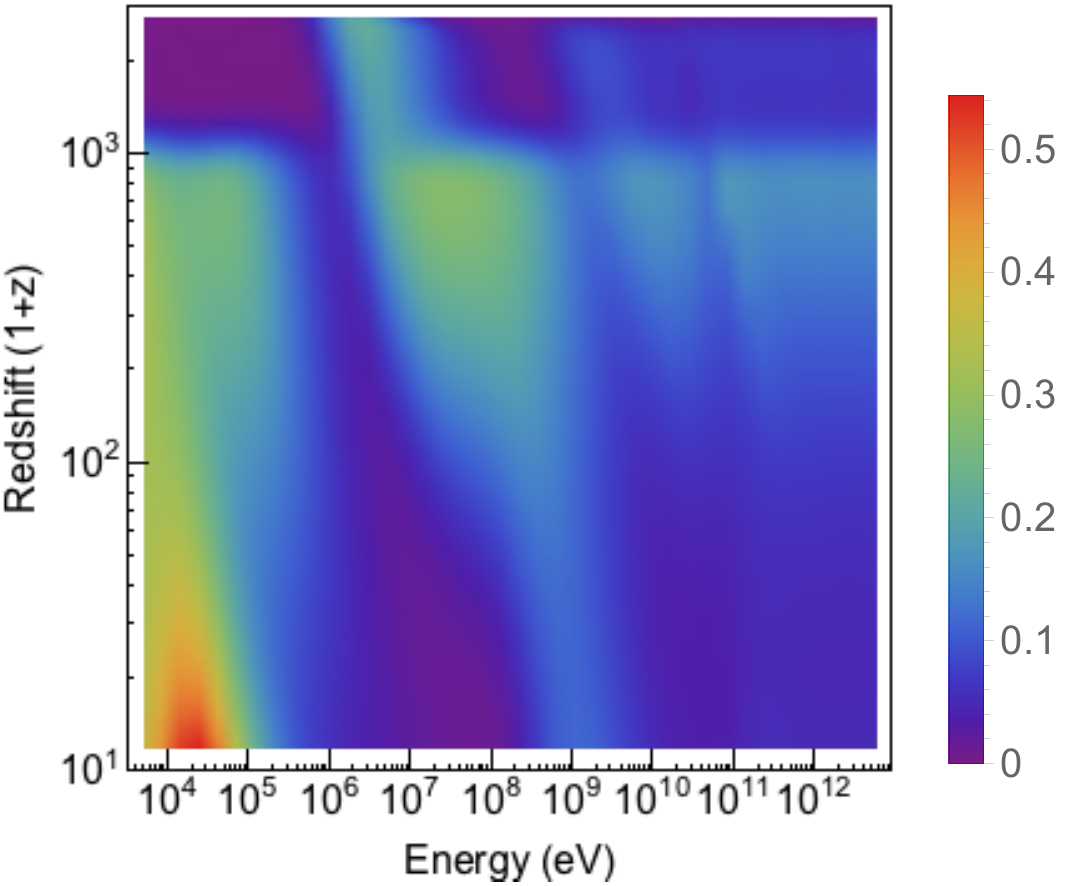} \\
\includegraphics[width=0.3\textwidth]{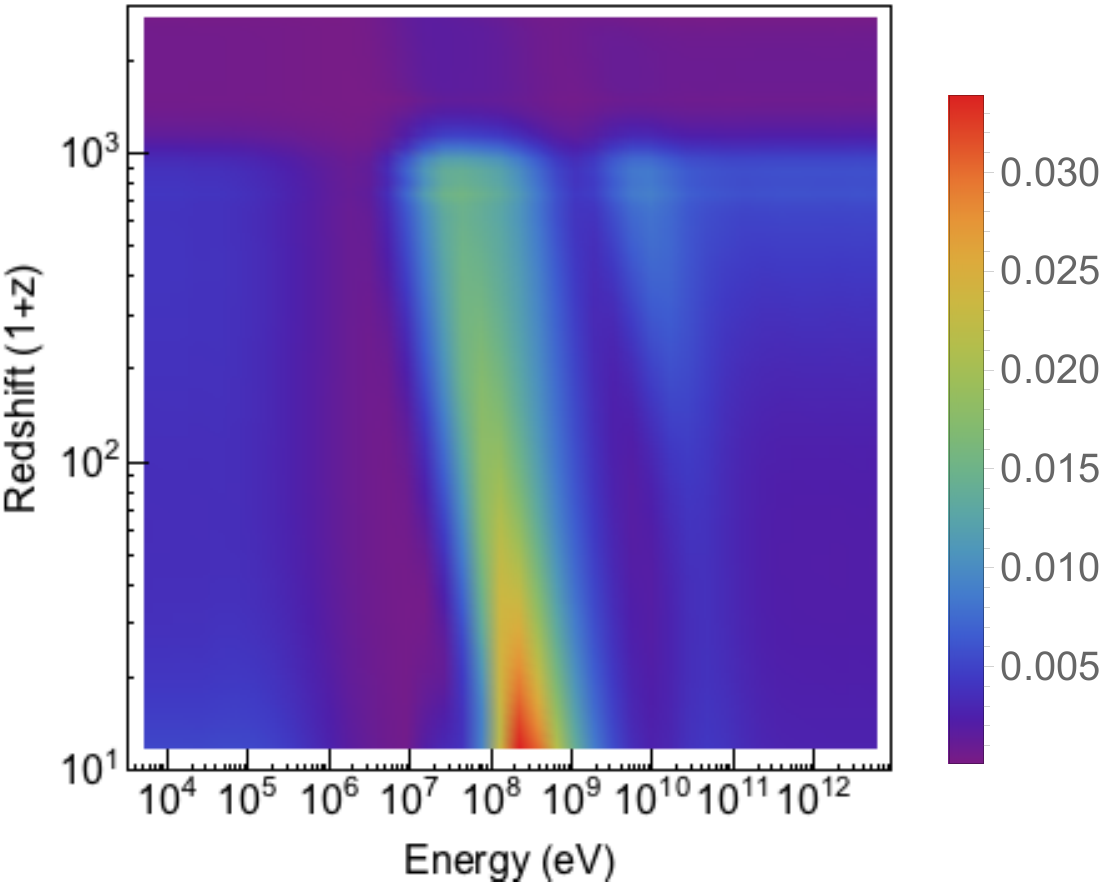}
\includegraphics[width=0.3\textwidth]{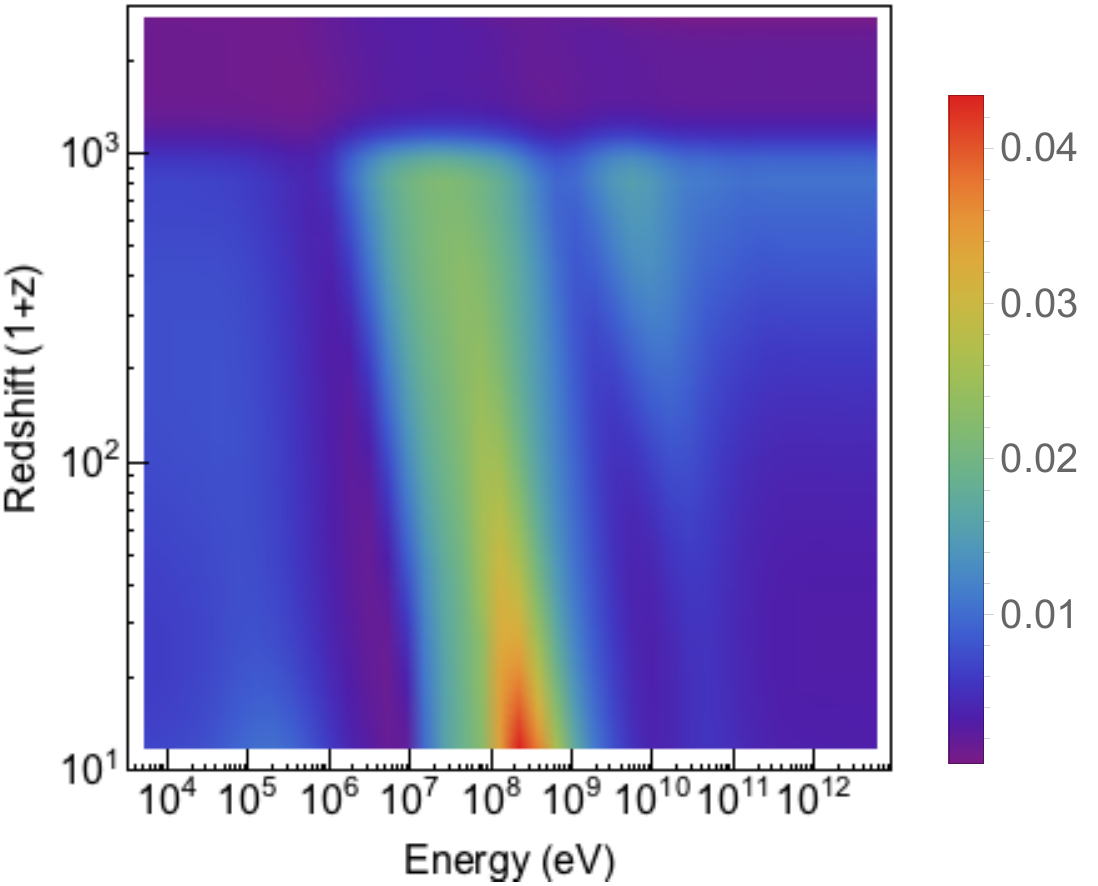}
\includegraphics[width=0.3\textwidth]{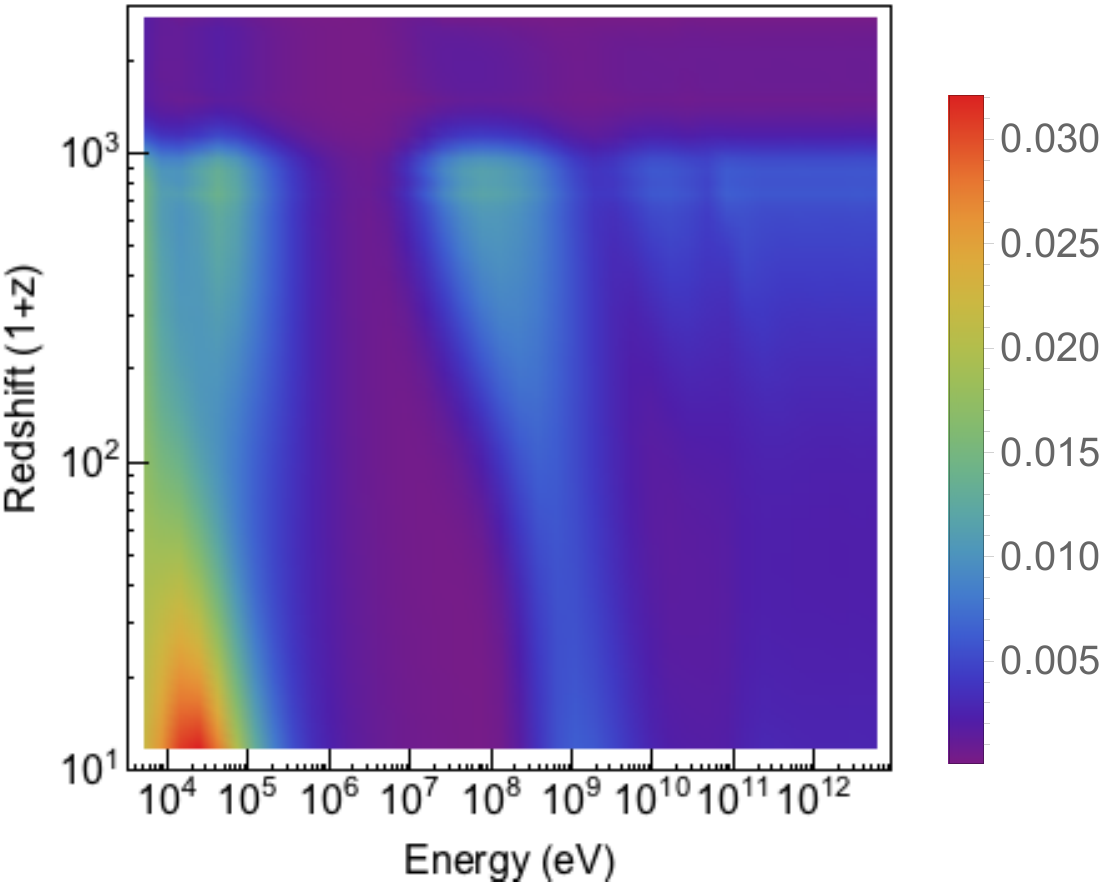} \\
\includegraphics[width=0.3\textwidth]{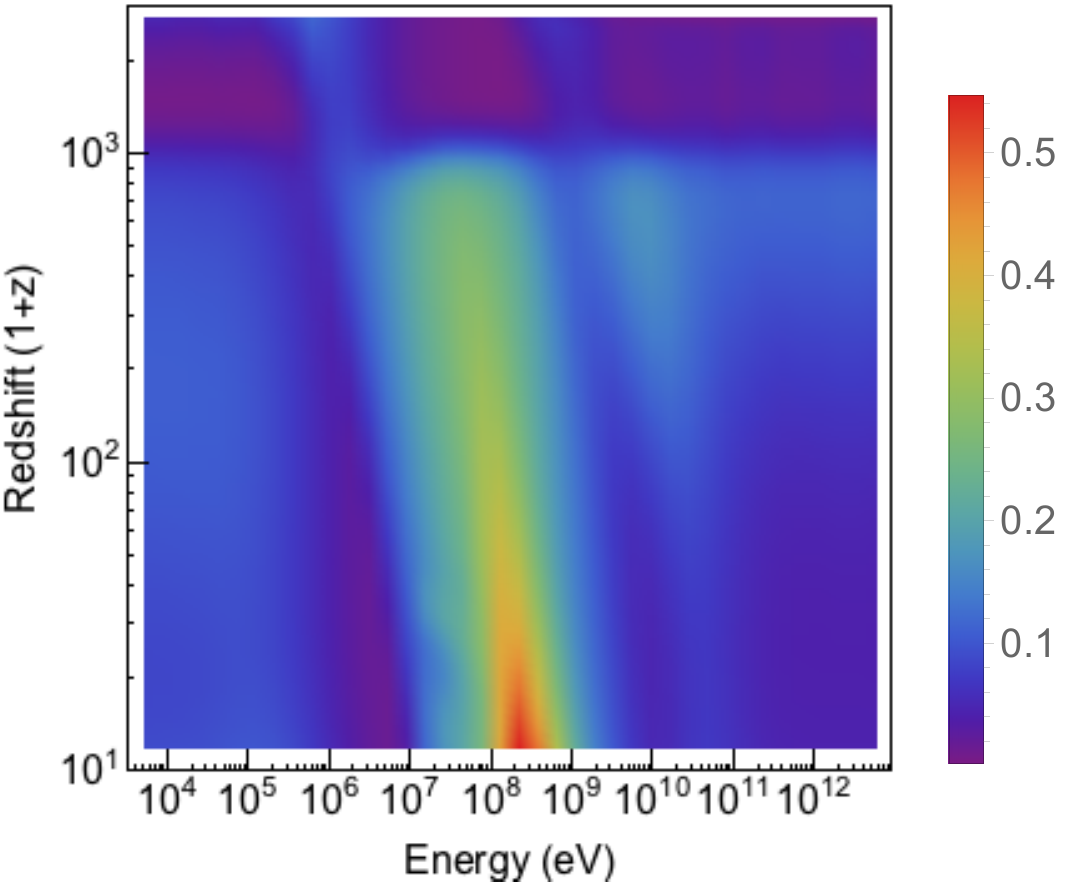}
\includegraphics[width=0.3\textwidth]{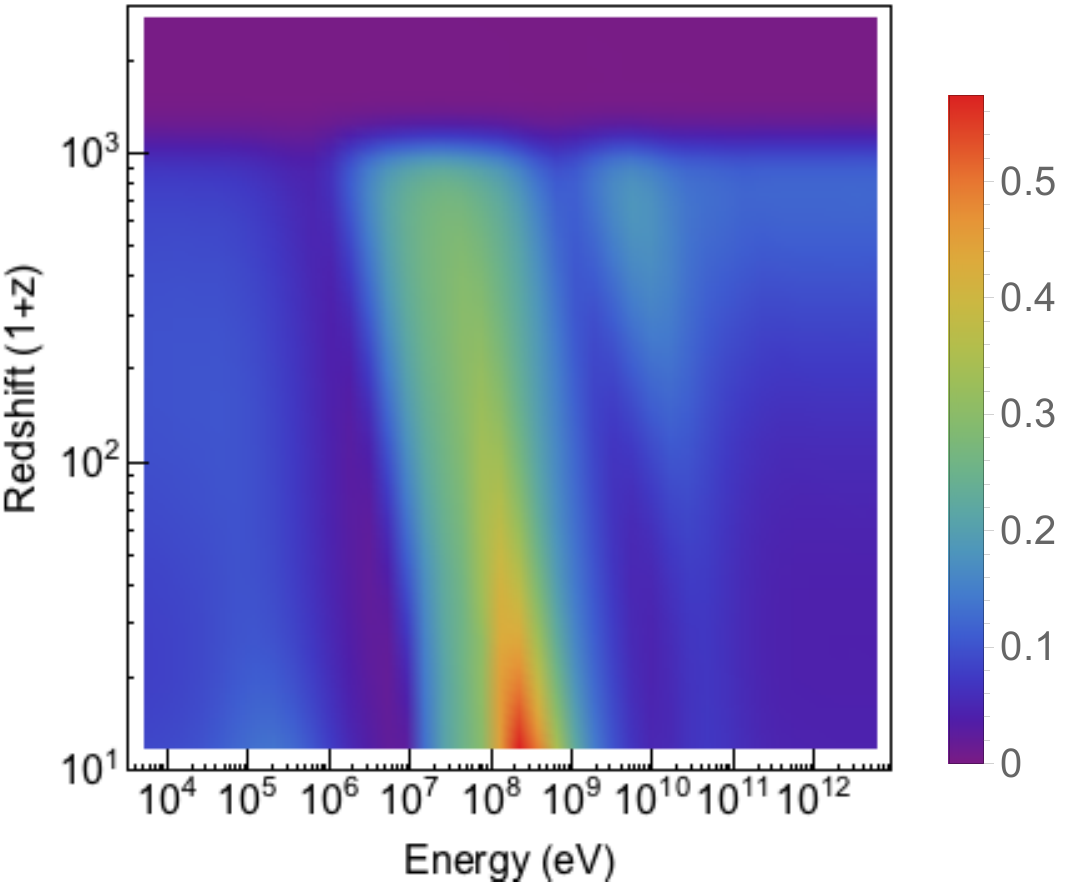}
\includegraphics[width=0.3\textwidth]{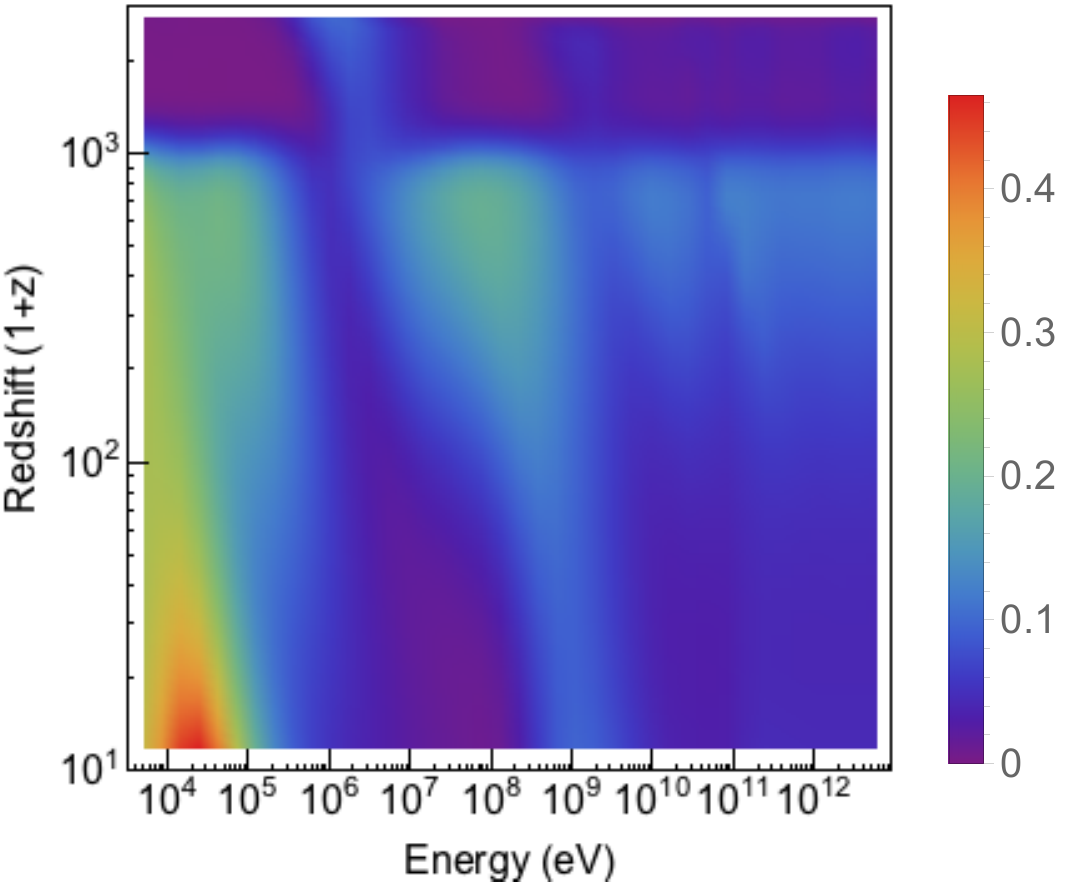} \\
\includegraphics[width=0.3\textwidth]{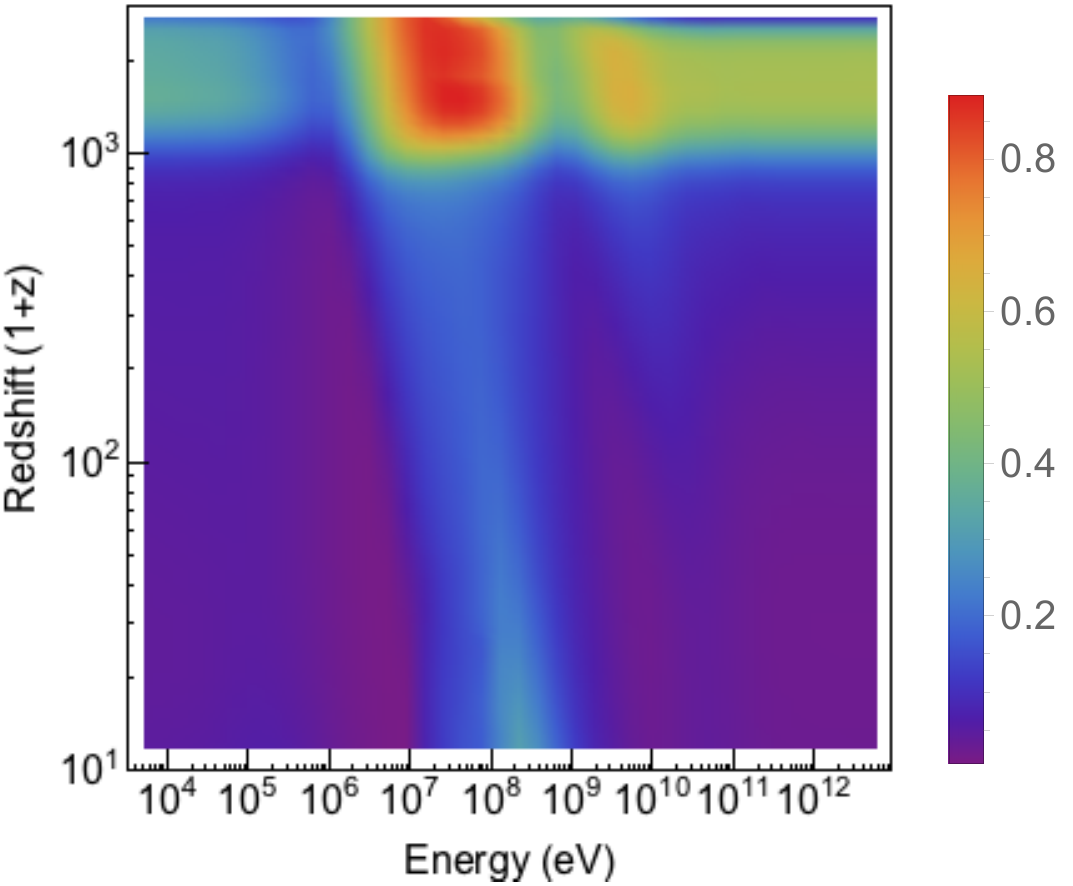}
\includegraphics[width=0.3\textwidth]{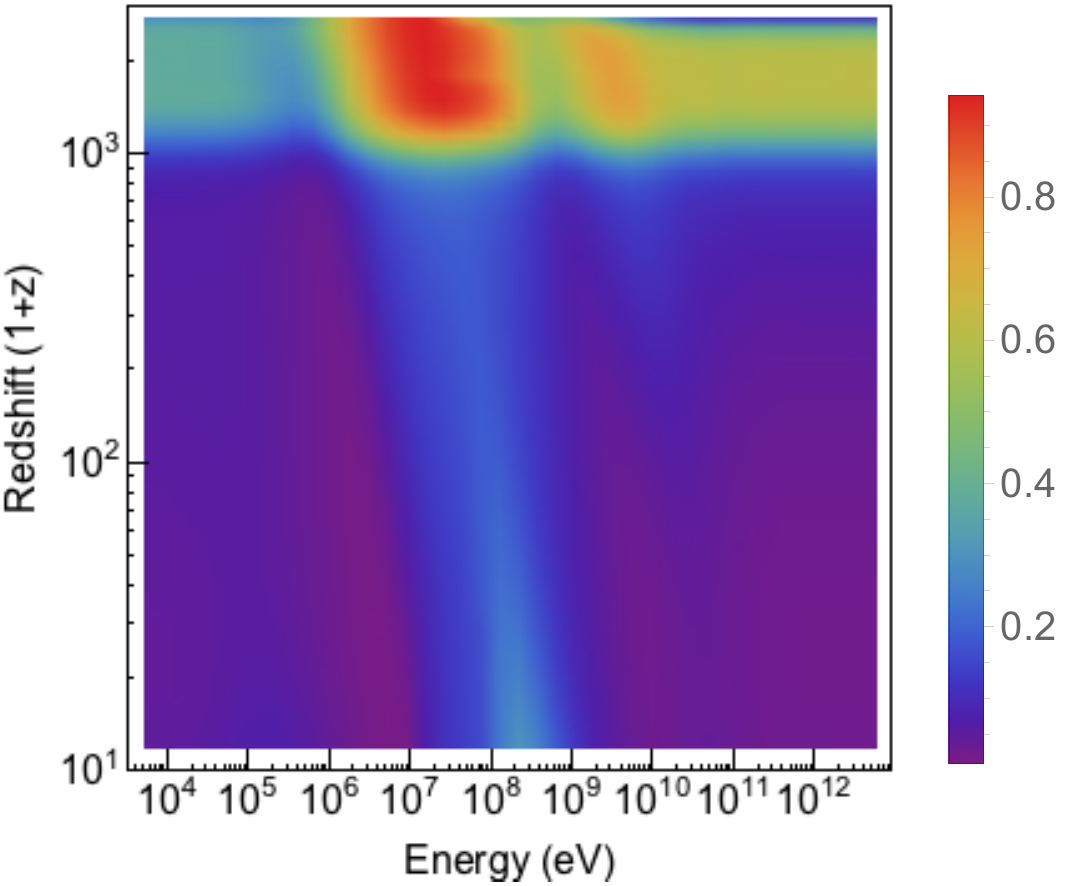}
\includegraphics[width=0.3\textwidth]{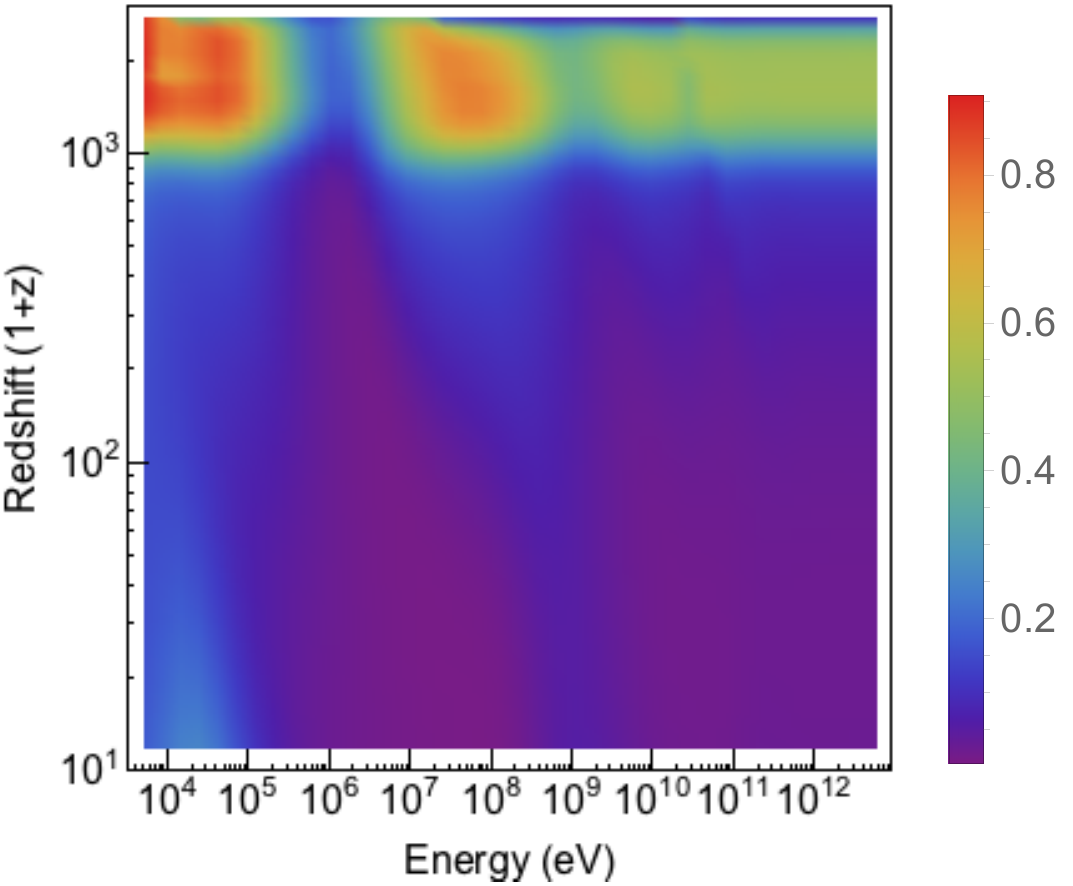} \\
\includegraphics[width=0.3\textwidth]{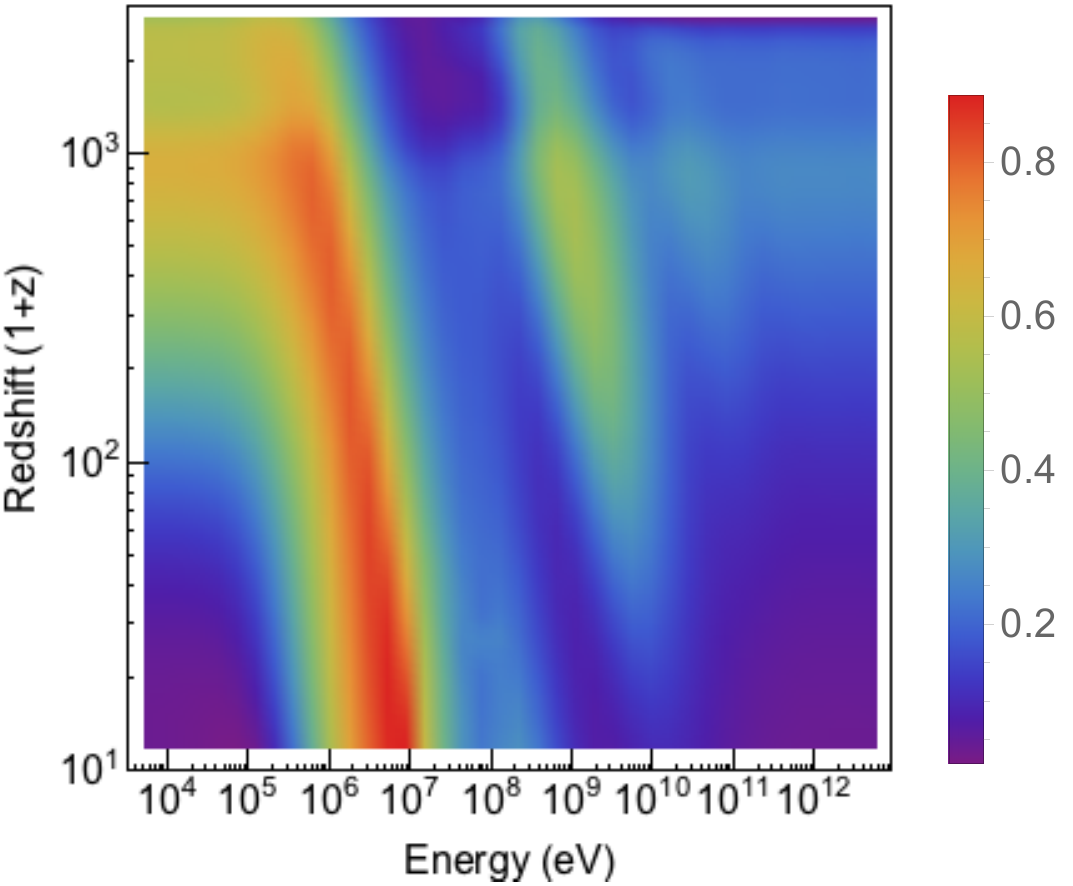}
\includegraphics[width=0.3\textwidth]{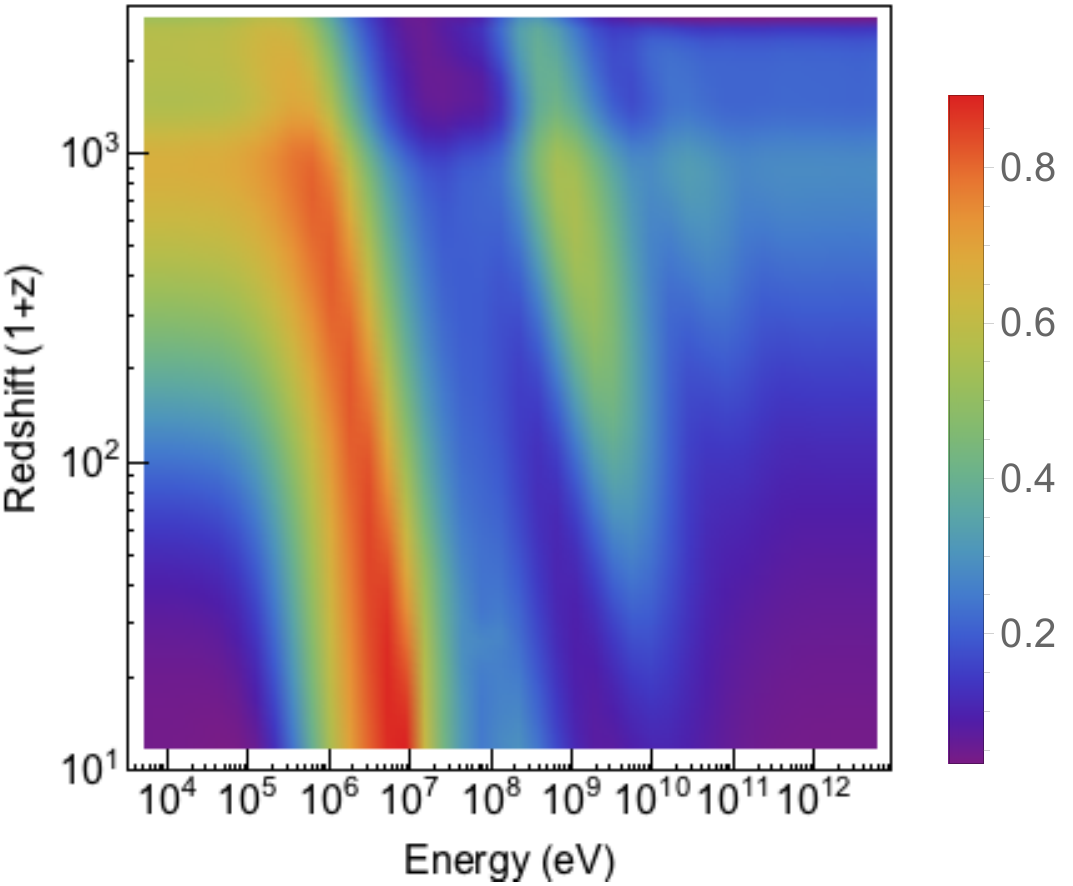}
\includegraphics[width=0.3\textwidth]{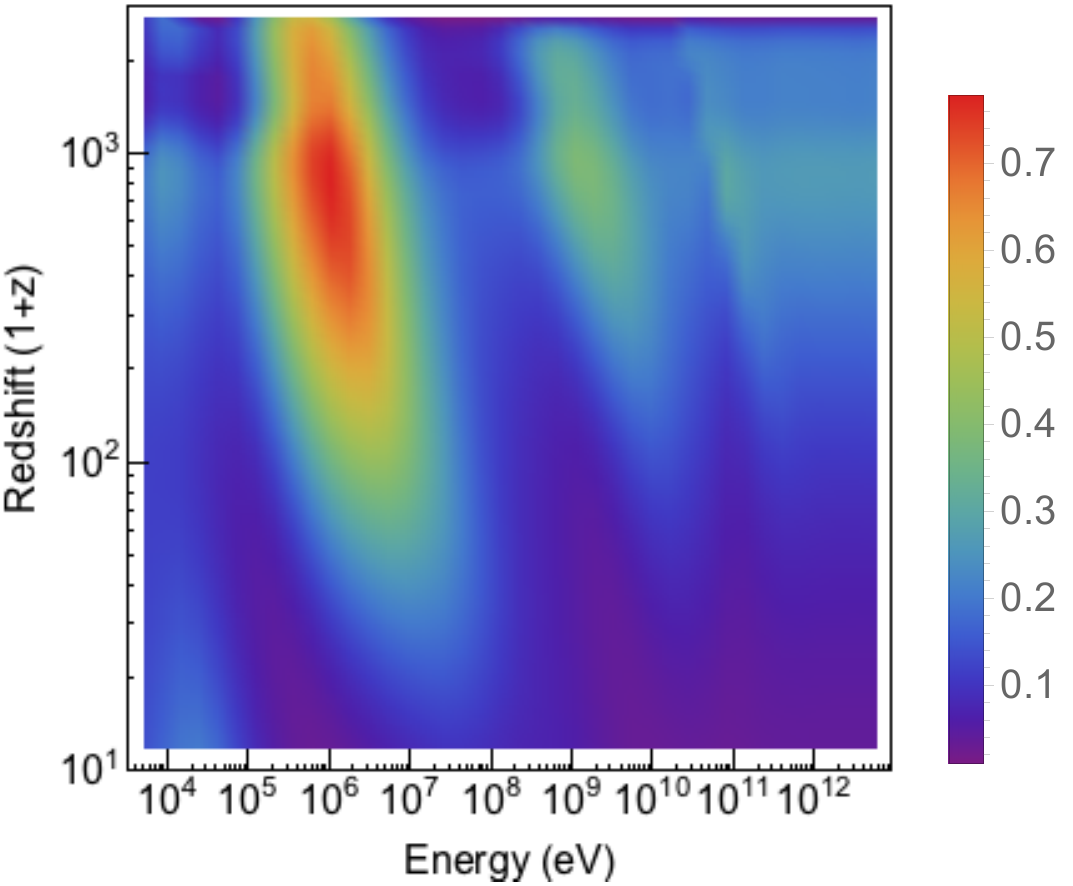}
\caption{\label{fig:fbychannel}
Power absorbed into each channel from particles injected by DM annihilation (or another process scaling as cosmological density squared), as a function of injection energy and redshift of absorption, normalized to the total injected power at the same redshift. From top to bottom, the rows correspond to H ionization, He ionization, Lyman-$\alpha$ photons, heating, and sub-10.2 eV continuum photons. The left column describes energy absorption for  $e^+e^-$ pairs (the $x$-axis ``energy'' label here indicates the kinetic energy of a single member of the pair at injection), while the right column describes energy absorption for photons. The center column is an alternate method of estimating the same quantities shown in the left column, as described in Section \ref{subsubsec:approx}, and should be regarded as a cross-check.
}
\end{figure*}

For example, for conventional DM annihilation, with a rate that scales as the square of the density, $I(z,E) \propto (1+z)^6 n(E)$, with the remaining factors being redshift- and energy-independent. For DM decay with a lifetime much longer than the age of the universe, $I(z,E) \propto (1+z)^3 n(E)$. In both cases $n(E)$ describes the spectrum of injected particles for the species in question. Substituting these expressions into Equation \ref{eq:fcurve} and summing over all deposition channels reproduces the results of \cite{2013PhRvD..87l3513S} (up to numerical error associated with the discretization of the transfer function $T$). If we consider only particles injected at a single energy, so $I(z,E)$ is proportional to a delta function in energy at $E^j$, the $f_c(z)$ curves can be simplified to:
\begin{align} f_c(z^i,E^j) & = \frac{ H(z^i) (1+z^i)^3 }{I(z^i,E^j)} \sum_k \frac{I(z^k,E^j)  T^\mathrm{species}_{c,ijk}}{(1+z^k)^3 H(z^k)} . \label{eq:fcurvebyenergy} \end{align}
More generally, for any energy injection history $I(z,E)$ that is a separable function of $z$ and $E$, i.e. $I(z,E) = I(z) n(E)$, the $f_c(z)$ curve can be written in the form:
\begin{align} f_c(z^i) &= \frac{\sum_j f_c(z^i, E^j) E^j n(E^j) dE^j}{\sum_j E^j n(E^j) dE^j}.\end{align}
Thus characterizing the $f_c(z)$ curves for individual energies is sufficient to describe all separable energy injection histories. We plot these curves for a range of injection energies, for the injection profile corresponding to DM annihilation, in Figure \ref{fig:fbychannel}. (As a default, we present results based on the ``best'' method described earlier, but as a cross-check, we also show results for the ``approx'' method for injected $e^+ e^-$ and the ``3 keV'' baseline prescription.)

By replacing $T^\mathrm{species}_{c,ijk}$ in Equation \ref{eq:fcurve} with $T^\mathrm{species; base}_{c, ijk}$ as defined in Equation \ref{eq:tbase}, we can construct alternative $f^\mathrm{sim}_c(z)$ curves, which correspond to rescaling the overall $f(z)$ by the channel-dependent $\chi_c^\mathrm{base}(z)$ factors and a correction factor to account for the losses into continuum photons. From Equation \ref{eq:tbase}, it follows that,
\begin{equation} f^\mathrm{sim}_c(z) = \left[ f(z) - f_\mathrm{corr}(z) \right] \chi_c^\mathrm{base}(z),\end{equation}
where $f_\mathrm{corr}(z)$ is obtained by replacing $T^\mathrm{species; base}_{c, ijk}$ with $S^\mathrm{species}_{corr, ijk}$ in Equation \ref{eq:fcurve}.

We can now define new, model-dependent fractions $\chi_c(z)$ by $\chi_c(z) = f_c(z)/f(z)$; i.e. the fraction of deposited power proceeding into each of the deposition channels. Likewise we define $\chi_\mathrm{corr}(z) = f_\mathrm{corr}(z)/f(z)$.

\begin{figure*}
\includegraphics[width=0.305\textwidth]{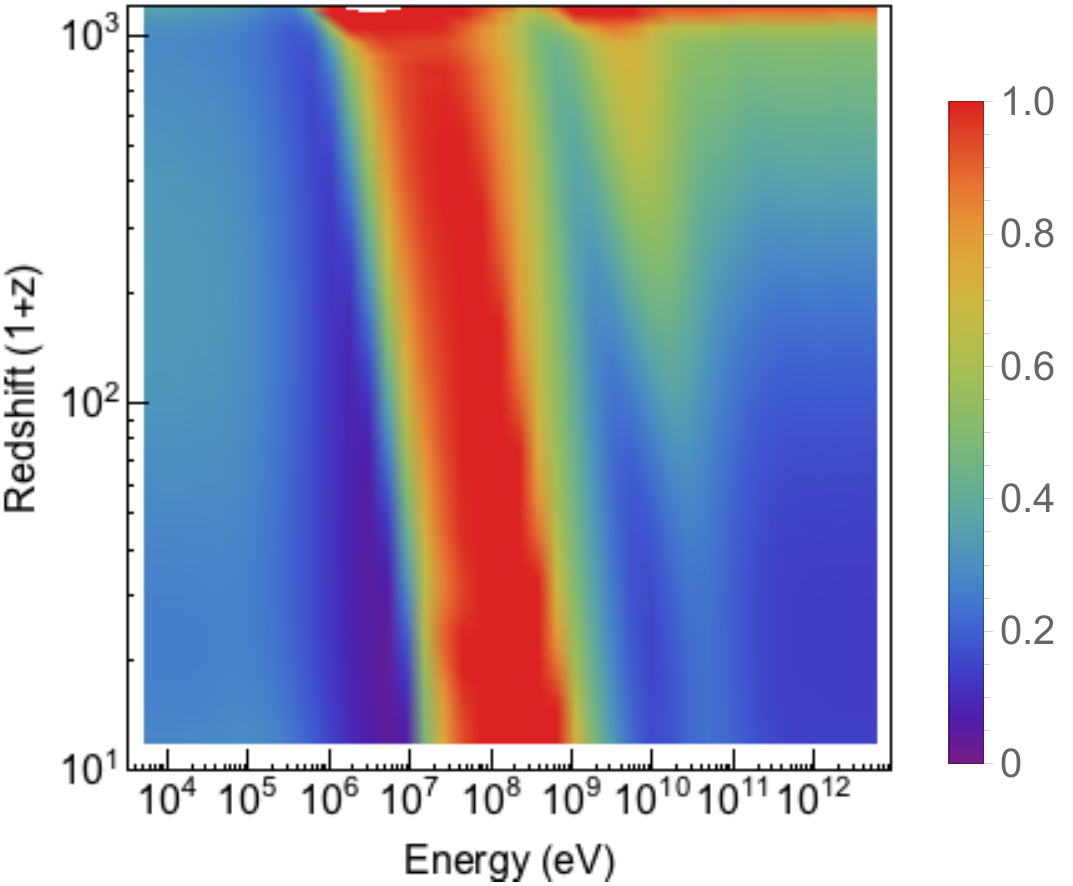}
\includegraphics[width=0.305\textwidth]{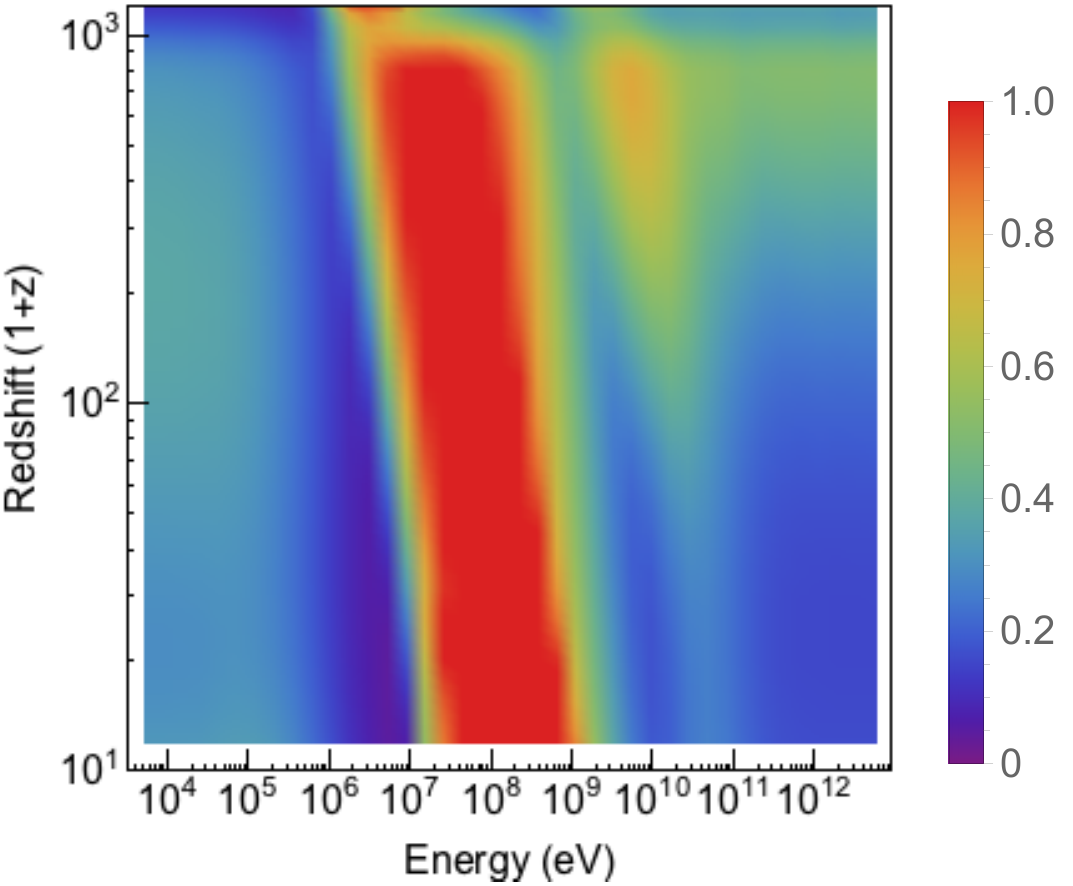}
\includegraphics[width=0.31\textwidth]{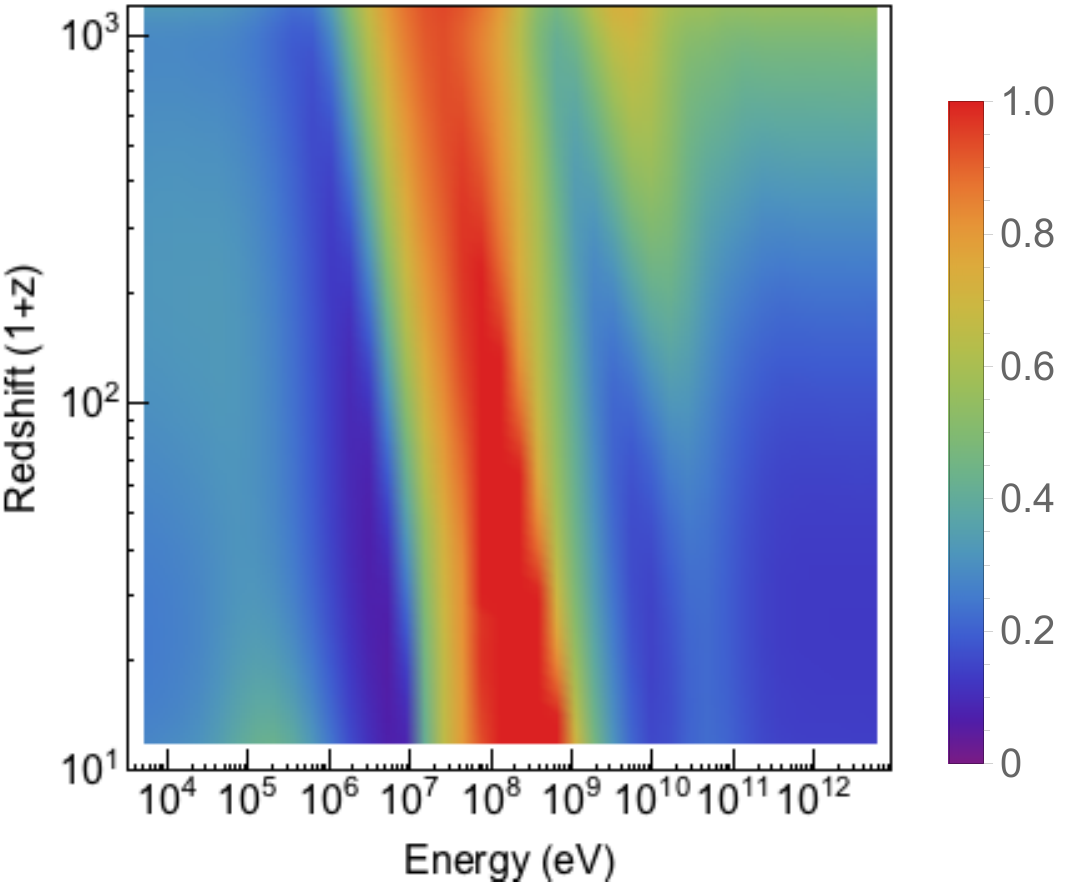} \\
\includegraphics[width=0.305\textwidth]{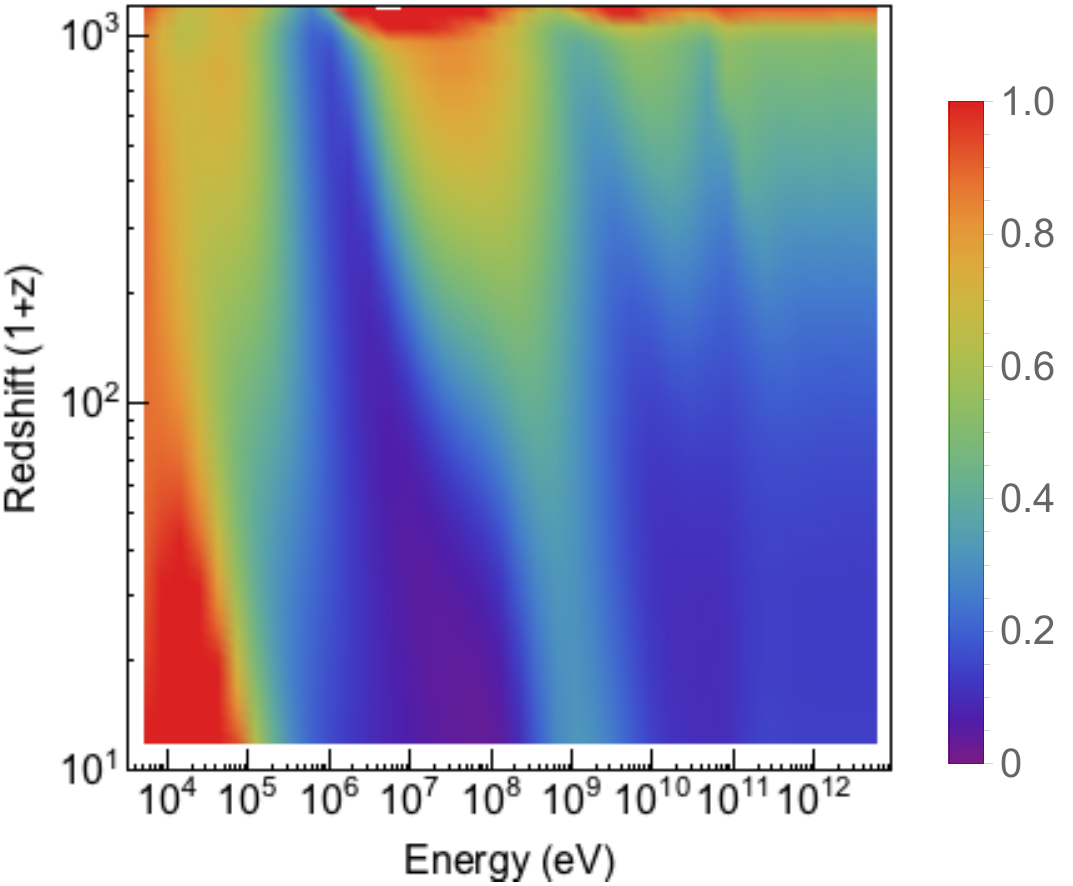}
\includegraphics[width=0.305\textwidth]{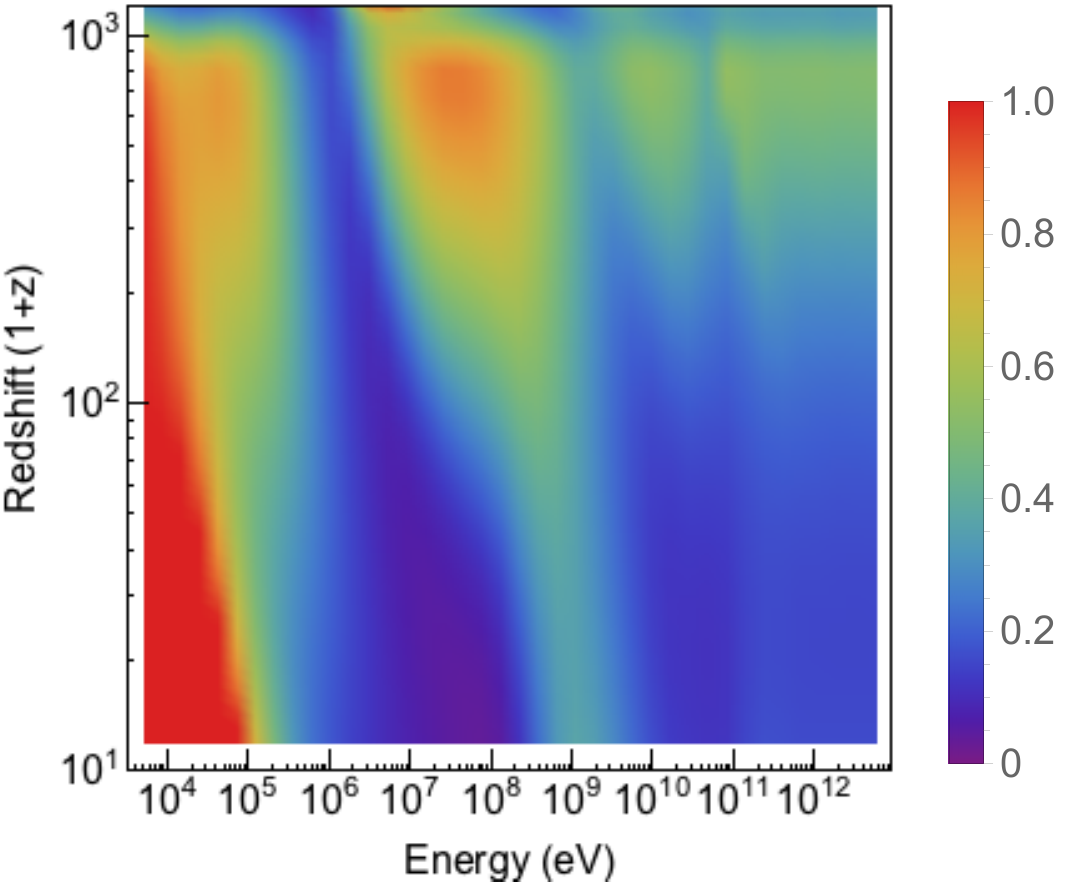}
\includegraphics[width=0.31\textwidth]{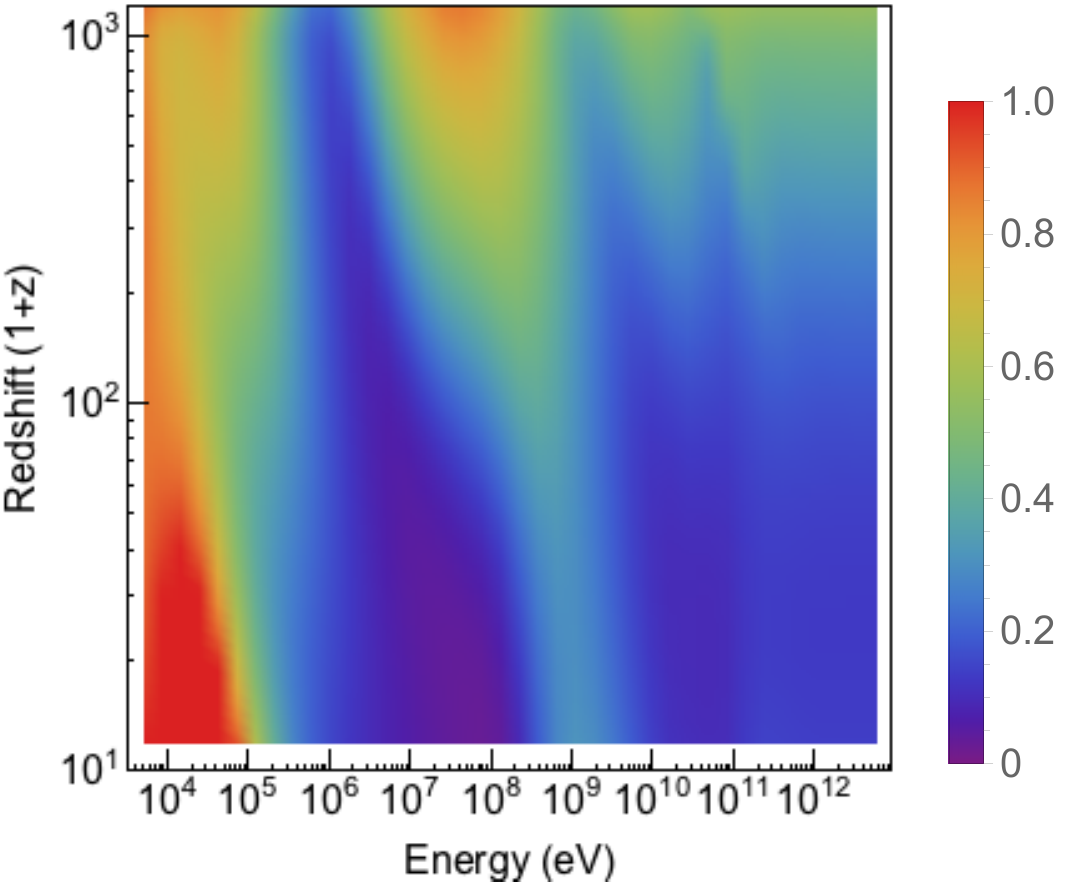}
\caption{\label{fig:fDM}
Corrected $f(z)$ curve for particles injected by DM annihilation, as a function of injection energy and redshift of absorption. In the left panel we use the ``3 keV'' baseline ionization fractions (so these $f(z)$ curves should be used with analyses that employed the same prescription); in the center panel we use the ``SSCK'' baseline. In the right panel we correct for the continuum losses using the results of Figure \ref{fig:fcorr}, and thus derive an alternate channel-independent $f^\mathrm{sim}(z)$ curve. The upper row describes $e^+ e^-$ pairs (the $x$-axis ``energy'' label here indicates the kinetic energy of a single member of the pair at injection), the lower row describes photons.
}
\end{figure*}

\subsection{Correcting the $f(z)$ curves for use with earlier studies}

Since the $f_c(z)$ are the quantities that determine the observable effects of energy injection (ionization, heating, etc), a correct computation of the constraints using the $f(z)$ deposition-efficiency curves would employ the $\chi_c(z)$ fractions derived here. However, many constraints have already been set assuming older, model-independent forms for the $\chi_c^\mathrm{base}(z)$ fractions. Since only the product $f(z) \chi_c(z)$ matters, the use of incorrect $\chi_c(z)$ fractions can be compensated by a correction to the $f(z)$ deposition-efficiency curve (this observation was also made in \cite{Galli:2013dna, Madhavacheril:2013cna}). However, since $f(z)$ is channel-independent, only the power deposited to a single channel $c$ can be completely described in this way; the deposition to other channels will be only approximate. Fortunately, this is not a problem for constraints that depend almost entirely on a single deposition channel. For example, constraints from the CMB anisotropies are primarily determined by ionization; while in principle the H and He ionization contributions should be treated separately (as in e.g. \cite{Padmanabhan:2005es}), in practice the contribution from He ionization is negligible, and so we can simply consider the sum of channels 1 and 2.

When the key figure of merit for a particular constraint is set by $f_c(z)$ for some channel $c$, in order to adapt older studies performed using some ``base'' prescription for the $\chi^\mathrm{base}_c(z)$ fractions, one should define a new corrected deposition-efficiency curve $f^{c,\mathrm{base}}(z)$ by:
\begin{equation}f^{c,\mathrm{base}}(z) \equiv f(z) \chi_c(z) / \chi^\mathrm{base}_c(z), \end{equation}
so that $\chi^\mathrm{base}_c(z) f^{c,\mathrm{base}}(z) = f_c(z)$.

The alternate simplified method described above for computing the energy deposition fractions by channel, i.e. rescaling a simplified ``base'' prescription to account for losses to continuum photons, corresponds to using a channel-independent corrected deposition-efficiency curve $f^\mathrm{sim}(z)$ given by:
\begin{align} f^\mathrm{sim}(z) & \equiv f^\mathrm{sim}_c(z)/\chi^\mathrm{base}_c(z) \nonumber \\
& = f(z) - f_\mathrm{corr}(z) \nonumber \\
& = f(z) \left[1 - \chi_\mathrm{corr}(z) \right]. \label{eq:fsim} \end{align}
Note that in this prescription the correction factor to $f(z)$ is independent of channel (for channels 1-4), and is also independent of the choice of ``base'' prescription to describe the energy deposition by channel. This means that slightly different constraints will be obtained if this simplified prescription is combined with studies using different ``base'' prescriptions; in contrast, if the correct $f^{c,\mathrm{base}}(z)$ curve is employed, the dependence on the ``base'' prescription will cancel out between the corrected $f(z)$ curve and the choice of the $\chi^\mathrm{base}_c(z)$ factors in the original analysis.

\begin{figure*}
\includegraphics[width=0.35\textwidth]{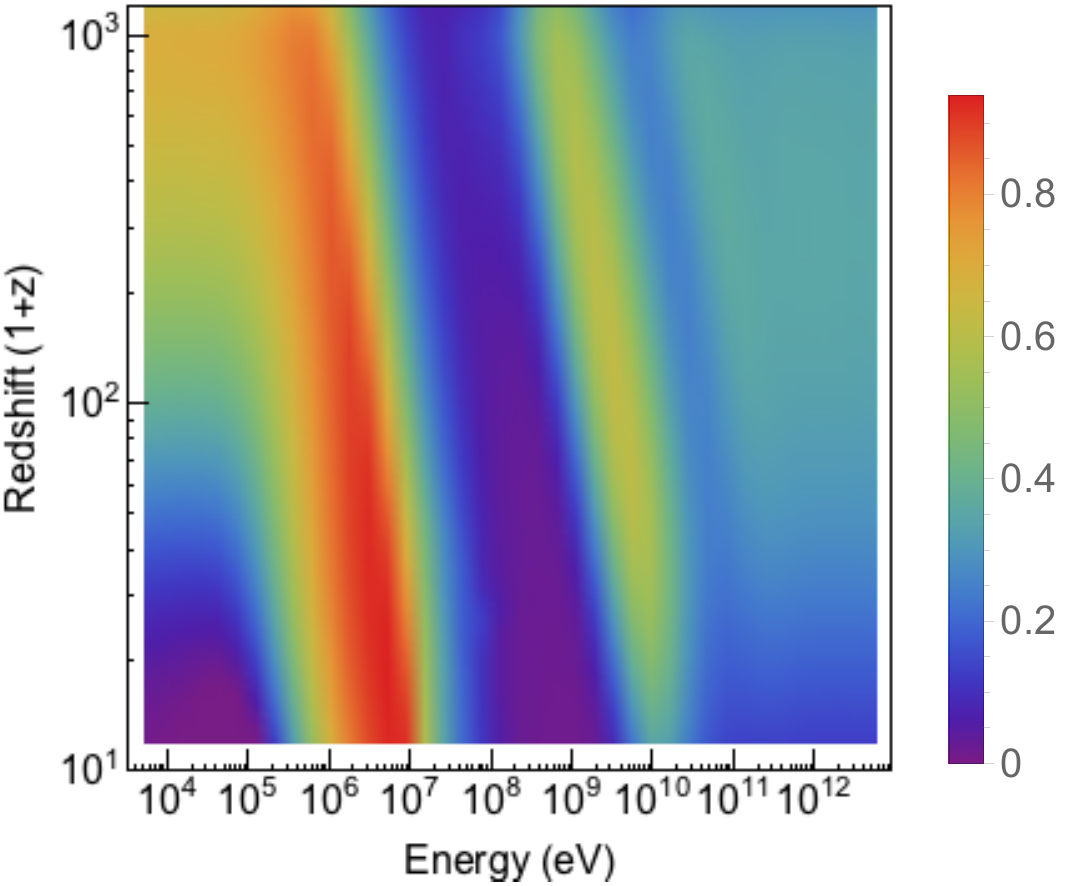}
\includegraphics[width=0.35\textwidth]{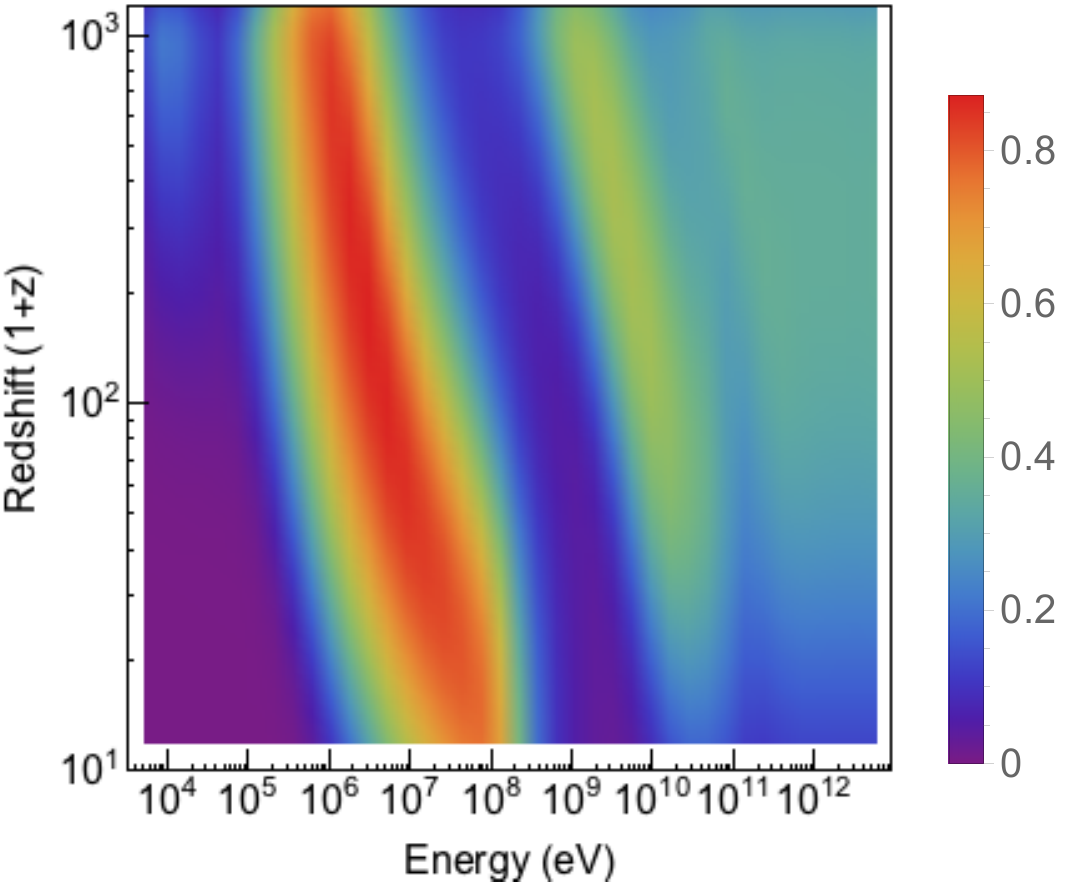}
\caption{\label{fig:fcorr}
Fraction of deposited energy proceeding into previously-unaccounted-for continuum photons, $\chi_\mathrm{corr}(z)$, for a DM-annihilation-like injection history. The upper panel gives results for $e^+ e^-$ pairs (the $x$-axis ``energy'' label here indicates the kinetic energy of a single member of the pair at injection), the lower panel for photons.
}
\end{figure*}

In Figure \ref{fig:fDM} we plot the $f^{\mathrm{ion},\mathrm{base}}(z)$ curves for the ``SSCK'' and ``3 keV'' choices of ``base'' prescription, again for an annihilation-like history: these curves constitute our best estimate of the appropriately corrected deposition-efficiency curves for the purposes of computing CMB constraints on DM annihilation. We also display the $f^\mathrm{sim}(z)$ curves obtained as described in Equation \ref{eq:fsim}. In Figure \ref{fig:fcorr} we plot the approximate correction factor $\chi_\mathrm{corr}(z)$, which should be interpreted as the fraction of deposited energy proceeding into previously unaccounted-for continuum photons, for the energy injection history corresponding to conventional DM annihilation.

From Figure \ref{fig:fcorr} we see that the correction to $f(z)$ due to continuum losses is largest at injection energies around $1-100$ MeV (depending on redshift) for photons, and at slightly lower energies ($\sim 1-10$ MeV) for $e^+ e^-$ pairs. This is consistent with the discussion of Figure \ref{fig:contfrac}; the correction is smaller than one might expect for non-relativistic $e^+ e^-$ pairs (with injection kinetic energies well below 1 MeV) because most of the injected energy is bound up in their mass, and thus the deposition of the kinetic energy is almost irrelevant. Such particles deposit their energy primarily through the annihilation of the positron, producing photons at 511 keV and below.

The correction factor falls abruptly at both lower and higher energies, although it is still appreciable at the highest energies. The correction factor for high-energy injected particles is somewhat complex, as it depends on the discontinuous cooling of the high-energy photons and electrons.  
For example, a 5 GeV electron injected at $z \sim 1000$ will upscatter a $\sim 1$ eV CMB photon to an energy of order $\gamma^2 \sim 10^8$ eV; the resulting photon will dominantly lose its energy by pair production or Compton scattering \cite{Slatyer:2009yq}, partitioning its energy into lower-energy electrons and positrons. The upper end of this spectrum, corresponding to 10-100 MeV electrons, will again upscatter CMB photons, now to energies ranging from hundreds of eV to tens of keV; these photons will efficiently ionize the gas. 
(All numbers in this paragraph are approximate and for illustration only.) 
An order of magnitude reduction in initial electron energy would reduce the typical energy of upscattered photons by two orders of magnitude, to around 1 MeV; the resulting secondary electrons would upscatter CMB photons to energies too low to further interact with the gas, thus losing a large fraction of their energy to the continuum. 
However, another order of magnitude reduction in the initial electron energy would mean that CMB photons would be upscattered to $\mathcal{O}$(10 keV) energies, at which point (at $z=1000$) they would be efficient photoionizers themselves and produce electrons in an energy range where atomic processes dominate the cooling. 
Patterns of this type are the reason for the ``striping'' visible in Figures \ref{fig:fDM}-\ref{fig:fcorr}.

\subsection{Low-energy particle spectra from arbitrary energy injections}
\label{subsec:spectra}

In the same spirit as the $f(z)$ curves, which describe total absorbed power, one can integrate over injection redshift to determine the low-energy electron and photon spectra produced by a specific energy injection history, at any redshift. Likewise, one can produce a $f_\mathrm{loss}(z)$ curve describing depletion of the CMB spectrum by scattering, and $f_{\mathrm{high},c}(z)$ curves that include only the power deposited by the cooling of high-energy electrons, from the initial outputs of the code described in Section \ref{sec:code}. Such results can be converted into the $f_c(z)$ curves, using results from codes that describe the cooling of low-energy electrons, just as described earlier in Section \ref{sec:deposition}; the photon spectrum at very low energies may also constitute an observable in its own right, as a distortion to the CMB blackbody spectrum. For convenience we can normalize the photon and electron spectra to the number of injected pairs at the ``output'' redshift (when the low-energy electrons and photons are produced), thus canceling out model-dependent normalization factors in the energy injection rate; we denote these normalized low-energy spectra by $F^\mathrm{sec}(z,E_\mathrm{sec})$. (Note this choice is not identical to the $f(z)$ curves, where we normalize to injected power rather than number of annihilations.) 

Specifically, we define these curves by:
\begin{align} f_\mathrm{loss}(z^i) & = \frac{ H(z^i) (1+z^i)^3 }{\sum_j E^j I(z^i,E^j) dE^j} \sum_k \frac{1}{(1+z^k)^3 H(z^k)}  \nonumber \\
& \times \sum_j E^j I(z^k,E^j)  S^\mathrm{species}_{\mathrm{loss},ijk} dE^j, \end{align}
\begin{align}f_{\mathrm{high},c}(z^i) & =  \frac{ H(z^i) (1+z^i)^3 }{\sum_j E^j I(z^i,E^j) dE^j}   \sum_k \frac{1}{(1+z^k)^3 H(z^k)}\nonumber \\
& \times \sum_j E^j I(z^k,E^j)  S^\mathrm{species}_{c,ijk} dE^j, \end{align}
\begin{align} F^\mathrm{sec}(z^i,E_\mathrm{sec}^l) & = \frac{ H(z^i) (1+z^i)^3 }{\sum_j I(z^i,E^j) dE^j} \sum_k \frac{1}{(1+z^k)^3 H(z^k)} \nonumber \\
& \times \sum_j I(z^k,E^j)  S^\mathrm{species}_{\mathrm{sec},ijkl} dE^j. \end{align}

For convenience, we include these arrays in our processed results, for the energy injection history corresponding to conventional DM annihilation (i.e. a fixed spectrum of annihilation products and a $I(z,E) \propto (1+z)^6$ redshift dependence).

\section{Conclusions}

We have presented the results of a comprehensive numerical study of the energy losses of keV-TeV photons and $e^+ e^-$ pairs in the cosmic dark ages, from their injection energies down to the $\sim 3$ keV scale where interactions with the gas begin to dominate the energy losses of electrons. All our results are provided in \texttt{.fits} and \texttt{.dat} format to facilitate the matching of these results onto detailed models of the low-energy cooling. We have employed previously published results for the low-energy cooling to estimate the partition of deposited energy between ionization, heating and the production of Lyman-$\alpha$ and continuum photons. We have demonstrated how to use these results to compute corrected deposition-efficiency $f(z)$ curves for use with studies of constraints on energy injection, and similarly made these processed results public.

\section*{Acknowledgements}

The author is grateful to the Mainz Institute for Theoretical Physics (MITP) and the Korea Institute for Advanced Study (KIAS) for their hospitality and support during the completion of this work, and thanks Aaron Vincent, Bhaskar Dutta, Patrick Fox, Silvia Galli, Fabio Iocco, Hongwan Liu and Nicholas Rodd for helpful discussions. This work is supported by the U.S. Department of Energy under grant Contract Numbers DE$-$SC00012567 and DE$-$SC0013999. This research made use of the IDL Astronomy UserÕs Library at Goddard.

\onecolumngrid
\appendix

\section{Summary of output files}
\label{app:files}

We provide three types of output files: \texttt{resultsgrid\_species\_fulloutput.fits} (where ``species'' can be either ``phot'' or ``elec'', corresponding to injection of photons and $e^+ e^-$ pairs respectively), \texttt{species\_results\_derived.fits} (likewise), and \texttt{supplementary\_deposition\_fractions.fits}. The first describes the direct outputs of our code -- energy deposited by electrons above 3 keV, and low-energy electron and photon spectra produced in each timestep -- whereas the second describes the processed results, giving our estimate of the energy deposited into each of the five channels described in Section \ref{sec:code}. The third file provides the reference ionization history we used, the resulting ``SSCK'' fractions, and the table of $\chi_c$ fractions as a function of redshift and electron energy derived from \cite{Galli:2013dna}. The contents of each type of file are as follows:

\vskip 5mm

{\bf  \texttt{species\_data.fits}}
\begin{itemize}
\item \texttt{OUTPUT\_REDSHIFT}: this 63-element array provides the abscissa for output redshift, i.e. the value of $1+z$ at which the energy is deposited.
\item \texttt{ENERGY}: this 40-element array provides the abscissa in energy, with values given by $\log_{10}(\text{energy in eV})$. Note that this is \emph{kinetic} energy of one of the two particles in the case of $e^+ e^-$ pairs; a particle annihilating or decaying to $e^+ e^-$ would need a mass sufficient to provide this energy in addition to the mass energy of the pair.
\item \texttt{INPUT\_REDSHIFT}: this 63-element array provides the abscissa for input redshift, i.e. the value of $1+z$ at which the energy is injected.
\item \texttt{CHANNELS:} this 5-element array lists the five deposition channels: hydrogen ionization, helium ionization, Lyman-$\alpha$ / excitation, heating and continuum photons.
\item \texttt{DEPOSITION\_FRACTIONS\_ORIG}: this $63 \times 40 \times 63$ array provides the table $T^\mathrm{species}_{ijk} = \sum_c T^\mathrm{species}_{c,ijk}$ for the appropriate species: that is, for a particle injected at some input redshift and energy (given by the abscissa arrays), the fraction of its initial energy deposited to \emph{all} channels in the (log-spaced) timestep associated with the output redshift. Up to small numerical differences, this table should match the \texttt{DEPOSITION\_FRACTIONS} table in the files associated with \cite{2013PhRvD..87l3513S}.
\item \texttt{F\_ORIG}: this $63 \times 40$ array describes the original depositon-efficiency $f(z)$-curve for DM annihilation to the species in question, with DM mass given by \texttt{ENERGY}, sampled at the redshift points given by the \texttt{OUTPUT\_REDSHIFT} array. This is based on the total energy deposition and contains no corrections to account for model-dependent $\chi_c(z)$ fractions.
\item \texttt{CMBLOSS\_FRACTIONS}: this $63 \times 40 \times 63$ array, labeled $S^\mathrm{species}_{\mathrm{loss},ijk}$ in the text, describes the power scattered \emph{out} of the CMB, as a fraction of the initial energy of the injected particle, in the timestep corresponding to the deposition redshift. Its index structure is the same as that of \texttt{DEPOSITION\_FRACTIONS\_ORIG}.
\item \texttt{HIGHDEP\_FRACTIONS}: this $63 \times 40 \times 63 \times 5$ array provides the table $S^\mathrm{species}_{c,ijk}$, as defined in Table \ref{tab:definitions}: that is, the power deposited into each of the channels $c=1-5$ by cooling of high-energy electrons (see Section \ref{sec:code}), as a fraction of the initial energy of the injected particle, in the timestep corresponding to the deposition redshift. This array's first three indices are the same as those for \texttt{DEPOSITION\_FRACTIONS\_ORIG}; the last index corresponds to $c$, and lists the 5 different channels. As noted in Section \ref{sec:code}, channel 2 is always empty.
\item \texttt{PHOTENG}: this $40 \times 500$ array provides the abscissa for the low-energy photon spectrum, the energy in eV for the low-energy secondary photons (500 energy bins), at each of the 40 injection energies defined by the \texttt{ENERGY} array. 
\item \texttt{ELECENG}: this $40 \times 500$ array provides the abscissa for the low-energy photon spectrum, the kinetic energy in eV for the low-energy secondary electrons (500 energy bins), at each of the 40 injection energies defined by the \texttt{ENERGY} array. 
\item \texttt{LOWENGPHOT\_SPEC}: this $63 \times 40 \times 63 \times 500$ array provides the table $S^\mathrm{species}_{\gamma,ijkl}$, as defined in Table \ref{tab:definitions}: that is, the spectrum $dN/dE$ of low-energy photons produced per injected pair of particles, in the timestep corresponding to the deposition redshift. This array's first three indices are the same as those for \texttt{DEPOSITION\_FRACTIONS}; the last index $l$ corresponds to the energy of the secondary photons, defined by the $(j, l)$th element of the \texttt{PHOTENG} array.
\item \texttt{LOWENGELEC\_SPEC}: this $63 \times 40 \times 63 \times 500$ array provides the table $S^\mathrm{species}_{e,ijkl}$, as defined in Table \ref{tab:definitions}: that is, the spectrum $dN/dE$ of low-energy electrons produced per injected pair of particles, in the timestep corresponding to the deposition redshift. This array's first three indices are the same as those for \texttt{DEPOSITION\_FRACTIONS}; the last index $l$ corresponds to the kinetic energy of the secondary electrons, defined by the $(j, l)$th element of the \texttt{ELECENG} array.
\end{itemize}

\vskip 5mm

{\bf  \texttt{species\_processed\_results.fits}}
\begin{itemize}
\item \texttt{OUTPUT\_REDSHIFT}: this 63-element array provides the abscissa for output redshift, i.e. the value of $1+z$ at which the energy is deposited.
\item \texttt{ENERGY}: this 40-element array provides the abscissa in energy, with values given by $\log_{10}(\text{energy in eV})$. Note that this is \emph{kinetic} energy of one of the two particles in the case of $e^+ e^-$ pairs; a particle annihilating or decaying to $e^+ e^-$ would need a mass sufficient to provide this energy in addition to the mass energy of the pair.
\item \texttt{INPUT\_REDSHIFT}: this 63-element array provides the abscissa for input redshift, i.e. the value of $1+z$ at which the energy is injected.
\item \texttt{CHANNELS:} this 5-element array lists the five deposition channels: hydrogen ionization, helium ionization, Lyman-$\alpha$, heating and continuum photons.
\item \texttt{DEPOSITION\_FRACTIONS\_ORIG}: as above.
\item \texttt{F\_ORIG}: as above.
\item \texttt{DEPOSITION\_FRACTIONS\_NEW}: this $63 \times 40 \times 63 \times 5$ array provides the $T^\mathrm{species}_{c,ijk}$ array, summing the contributions from high-energy deposition and integrating over the low-energy photon and electron spectra. This array's first three indices are the same as those for \texttt{DEPOSITION\_FRACTIONS\_ORIG} defined above; the last index corresponds to $c$, and lists the 5 different channels. 
\item \texttt{CONT\_CORR\_FRACTIONS}: this $63 \times 40 \times 63$ array provides the $T^\mathrm{species}_{\mathrm{corr},ijk}$ array, the contribution to channel $c=5$ arising from the low-energy photons produced at each timestep by cooling of electrons above 3 keV (divided as usual by the initial injected energy, and having corrected for the original energy of these low-energy photons). This array's indices are the same as those for \texttt{DEPOSITION\_FRACTIONS\_ORIG} defined above.
\item \texttt{F\_ION}: this $63 \times 40$ array describes the deposition-efficiency $f$-curve for DM annihilation to the species in question, corrected to give the true power into ionization combined with the ``3 keV'' prescription. This quantity is denoted $f^\mathrm{ion,3keV}(z)$ in Section \ref{sec:fcurves}. The DM mass is given by the \texttt{ENERGY} array, and the function is sampled at the redshift points given by the \texttt{OUTPUT\_REDSHIFT} array.
\item \texttt{F\_CORR}: this $63 \times 40$ array describes the deposition-efficiency $f$-curve for DM annihilation to the species in question, approximately corrected by rescaling the deposited energy according to the unaccounted losses into continuum photons. This quantity is denoted $f^\mathrm{sim}(z)$ in Section \ref{sec:fcurves}. The DM mass is given by the \texttt{ENERGY} array, and the function is sampled at the redshift points given by the \texttt{OUTPUT\_REDSHIFT} array.
\item \texttt{F\_CMBLOSS}: this $63 \times 40$ array describes the power scattered \emph{out} of the CMB at each redshift, as a fraction of the power injected at that redshift, for DM annihilation to the species in question. This quantity is denoted $f_\mathrm{loss}(z)$ in Section \ref{subsec:spectra}.  The DM mass is given by the \texttt{ENERGY} array, and the function is sampled at the redshift points given by the \texttt{OUTPUT\_REDSHIFT} array.
\item \texttt{F\_HIGHDEP}: this $63 \times 40 \times 5$ array describes the power deposited to each channel by the cooling of high-energy electrons at each redshift, as a fraction of the power injected at that redshift, for DM annihilation to the species in question. This quantity is denoted $f_\mathrm{high,c}(z)$ in Section \ref{subsec:spectra}. The DM mass is given by the \texttt{ENERGY} array, and the function is sampled at the redshift points given by the \texttt{OUTPUT\_REDSHIFT} array.
\item \texttt{PHOTENG}: this $40 \times 500$ array provides the abscissa for the low-energy photon spectrum, the energy in eV for the low-energy secondary photons (500 energy bins), at each of the 40 injection energies defined by the \texttt{ENERGY} array. 
\item \texttt{ELECENG}: this $40 \times 500$ array provides the abscissa for the low-energy photon spectrum, the kinetic energy in eV for the low-energy secondary electrons (500 energy bins), at each of the 40 injection energies defined by the \texttt{ENERGY} array. 
\item \texttt{FSPEC\_PHOT}: this $63 \times 40 \times 500$ array provides the spectrum $dN/dE$ of low-energy photons produced at each redshift, normalized to the number of DM annihilations occurring at that redshift, for DM annihilation to the species in question. This quantity is denoted $F^\gamma(z,E_\gamma)$ in Section \ref{subsec:spectra}. The DM mass is given by the \texttt{ENERGY} array, and the function is sampled at the redshift points given by the \texttt{OUTPUT\_REDSHIFT} array; the energies of the secondary photons are given by the corresponding elements of the \texttt{PHOTENG} array.
\item \texttt{FSPEC\_ELEC}: this $63 \times 40 \times 500$ array provides the spectrum $dN/dE$ of low-energy electrons produced at each redshift, normalized to the number of DM annihilations occurring at that redshift, for DM annihilation to the species in question. This quantity is denoted $F^e(z,E_e)$ in Section \ref{subsec:spectra}. The DM mass is given by the \texttt{ENERGY} array, and the function is sampled at the redshift points given by the \texttt{OUTPUT\_REDSHIFT} array; the kinetic energies of the secondary electrons are given by the corresponding elements of the \texttt{ELECENG} array.
\end{itemize}

\vskip 5mm

{\bf  \texttt{deposition\_fractions\_supplement.fits}}
\begin{itemize} 
\item \texttt{REDSHIFT}: this 63-element array provides the abscissa for deposition redshift.
\item \texttt{XH}: this 63-element array provides the hydrogen gas ionization fraction as a function of redshift, for the baseline ionization history with no energy injection (calculated using RECFAST, as in \cite{Wong:2007ym}).
\item \texttt{CHANNELS:} this 5-element array lists the five deposition channels: hydrogen ionization, helium ionization, Lyman-$\alpha$, heating and continuum photons.
\item \texttt{SSCK:} this $63 \times 5$ array describes the fraction of deposited power proceeding into channels $1-5$ under the simple ``SSCK'' prescription, as a function of redshift (channels 2 and 5 are not populated by this prescription), i.e. $\chi^\mathrm{SSCK}_c(z)$.
\item \texttt{ELECTRON\_ENERGY:} this 6-element array provides the kinetic energy values at which the \texttt{CHANNEL\_FRACTIONS} array is evaluated.
\item \texttt{CHANNEL\_FRACTIONS:} this $6 \times 63 \times 5$ array describes the fraction of deposited power proceeding into channels $1-5$ in a detailed calculation of the low-energy cooling (presented in \cite{Galli:2013dna} and based on \cite{Valdes:2008cr}), for the six electron injection energies listed in \texttt{ELECTRON\_ENERGY}, as a function of redshift. 
\end{itemize}

The files are available online in \texttt{.fits} format, at \texttt{http://nebel.rc.fas.harvard.edu/epsilon}; we supply a Mathematica notebook demonstrating how to read the \texttt{.fits} files and reproduce the calculations in this note. Finally, we also provide the key results as \texttt{.dat} files, as described below:

\vskip 5mm

{\bf Overall best estimate for the energy deposition fractions:}
\begin{itemize}
\item \texttt{species\_channel=c\_deposition\_fractions.dat}: holds $T^\mathrm{species}_{c,ijk}$ for each choice of $c$ and species (``phot'' corresponds to photons, ``elec'' to $e^+ e^-$ pairs). Starting with the third column, the entries in the first row give the injection redshift $1+z$. Starting with the second row, the first column lists the deposition redshift $1+z$, and the second the injection energy (as defined in the \texttt{ENERGY} array in the \texttt{.fits} files); subsequent columns list $T_{c,ijk}$ for the appropriate triple of injection redshift, deposition redshift and injection energy. (Holds the same information as \texttt{DEPOSITION\_FRACTIONS\_NEW} in the \texttt{.fits} files.)
\item \texttt{species\_continuum\_correction\_fractions.dat}: as \texttt{species\_channel=c\_highE\_deposition\_fractions.dat}, except $T^\mathrm{species}_{c,ijk}$ is replaced with $T^\mathrm{species}_{\mathrm{corr},ijk}$. (Holds the same information as \texttt{CONT\_CORR\_FRACTIONS} in the \texttt{.fits} files.)
\end{itemize}

{\bf Raw outputs of the high-energy code:}
\begin{itemize}
\item \texttt{species\_channel=c\_highE\_deposition\_fractions.dat}: as \texttt{species\_channel=c\_highE\_deposition\_fractions.dat}, except $T^\mathrm{species}_{c,ijk}$ is replaced with $S^\mathrm{species}_{c,ijk}$. (Holds the same information as \texttt{HIGHDEP\_FRACTIONS} in the \texttt{.fits} files.)
\item \texttt{species\_cmbloss\_fractions.dat}: as \texttt{species\_channel=c\_highE\_deposition\_fractions.dat}, except $T^\mathrm{species}_{c,ijk}$  is replaced with $S^\mathrm{species}_{\mathrm{loss},ijk}$. (Holds the same information as \texttt{CMBLOSS\_FRACTIONS} in the \texttt{.fits} files.)
\item \texttt{species\_lowEsecspectra.dat}: holds $S^\mathrm{species}_{\mathrm{sec},ijkl}$ for each choice of ``sec'' (``electron'' for secondary electrons and ``photon'' for secondary photons) and each injected species. Starting with the fourth column, the entries in the first row give the injection redshift $1+z$. Starting with the second row, the first column lists the deposition redshift $1+z$, and the second the injection energy (as defined in the \texttt{ENERGY} array in the \texttt{.fits} files), the third the secondary (kinetic) energy, expressed as log$_{10}(E_\mathrm{sec}$/eV); subsequent columns list $S^\mathrm{species}_{\mathrm{sec},ijkl}$ for the appropriate quadruple of injection redshift, deposition redshift, injection energy and secondary energy. (Holds the same information as \texttt{LOWENGELEC\_SPEC} and \texttt{LOWENGPHOT\_SPEC} in the \texttt{.fits} files.)
\end{itemize}

{\bf f(z) curves for DM annihilation:}
\begin{itemize}
\item \texttt{species\_bestf\_DMann\_3keV.dat}: holds $f^\mathrm{ion,3keV}(z)$, our best-estimate corrected $f(z)$ curve for DM-annihilation-like energy-injection histories, for studies which assume the ``3 keV'' baseline prescription to set constraints on energy injection via the ionization channel (e.g. the recent constraints on DM annihilation presented by the \emph{Planck} Collaboration \cite{Planck:2015xua}). Starting with the second column, the entries in the first row give the deposition redshift $1+z$. Starting with the second row, the first column lists the injection energy (as defined in the \texttt{ENERGY} array in the \texttt{.fits} files); subsequent columns list $f(z)$ for the appropriate pair of injection energy and redshift. (Holds the same information as \texttt{F\_ION} in the \texttt{.fits} files.)
\item \texttt{species\_simplef\_DMann.dat}: holds $f^\mathrm{sim}(z)$, our simplified approximate prescription for the corrected $f(z)$ curve for DM-annihilation-like energy-injection histories. Layout is the same as \texttt{species\_bestf\_DMann\_3keV.dat}.  (Holds the same information as \texttt{F\_CORR} in the \texttt{.fits} files.)
\end{itemize}


\bibliography{deposition}

\begin{thebibliography}{36}
\expandafter\ifx\csname natexlab\endcsname\relax\def\natexlab#1{#1}\fi
\expandafter\ifx\csname bibnamefont\endcsname\relax
  \def\bibnamefont#1{#1}\fi
\expandafter\ifx\csname bibfnamefont\endcsname\relax
  \def\bibfnamefont#1{#1}\fi
\expandafter\ifx\csname citenamefont\endcsname\relax
  \def\citenamefont#1{#1}\fi
\expandafter\ifx\csname url\endcsname\relax
  \def\url#1{\texttt{#1}}\fi
\expandafter\ifx\csname urlprefix\endcsname\relax\def\urlprefix{URL }\fi
\providecommand{\bibinfo}[2]{#2}
\providecommand{\eprint}[2][]{\url{#2}}

\bibitem[{\citenamefont{Adams et~al.}(1998)\citenamefont{Adams, Sarkar, and
  Sciama}}]{Adams:1998nr}
\bibinfo{author}{\bibfnamefont{J.~A.} \bibnamefont{Adams}},
  \bibinfo{author}{\bibfnamefont{S.}~\bibnamefont{Sarkar}}, \bibnamefont{and}
  \bibinfo{author}{\bibfnamefont{D.}~\bibnamefont{Sciama}},
  \bibinfo{journal}{Mon.Not.Roy.Astron.Soc.} \textbf{\bibinfo{volume}{301}},
  \bibinfo{pages}{210} (\bibinfo{year}{1998}), \eprint{astro-ph/9805108}.

\bibitem[{\citenamefont{Chen and Kamionkowski}(2004)}]{Chen:2003gz}
\bibinfo{author}{\bibfnamefont{X.-L.} \bibnamefont{Chen}} \bibnamefont{and}
  \bibinfo{author}{\bibfnamefont{M.}~\bibnamefont{Kamionkowski}},
  \bibinfo{journal}{Phys. Rev.} \textbf{\bibinfo{volume}{D70}},
  \bibinfo{pages}{043502} (\bibinfo{year}{2004}), \eprint{astro-ph/0310473}.

\bibitem[{\citenamefont{Padmanabhan and Finkbeiner}(2005)}]{Padmanabhan:2005es}
\bibinfo{author}{\bibfnamefont{N.}~\bibnamefont{Padmanabhan}} \bibnamefont{and}
  \bibinfo{author}{\bibfnamefont{D.~P.} \bibnamefont{Finkbeiner}},
  \bibinfo{journal}{Phys. Rev.} \textbf{\bibinfo{volume}{D72}},
  \bibinfo{pages}{023508} (\bibinfo{year}{2005}), \eprint{astro-ph/0503486}.

\bibitem[{\citenamefont{Furlanetto et~al.}(2006)\citenamefont{Furlanetto, Oh,
  and Pierpaoli}}]{Furlanetto:2006wp}
\bibinfo{author}{\bibfnamefont{S.~R.} \bibnamefont{Furlanetto}},
  \bibinfo{author}{\bibfnamefont{S.~P.} \bibnamefont{Oh}}, \bibnamefont{and}
  \bibinfo{author}{\bibfnamefont{E.}~\bibnamefont{Pierpaoli}},
  \bibinfo{journal}{Phys. Rev.} \textbf{\bibinfo{volume}{D74}},
  \bibinfo{pages}{103502} (\bibinfo{year}{2006}), \eprint{astro-ph/0608385}.

\bibitem[{\citenamefont{Valdes et~al.}(2007)\citenamefont{Valdes, Ferrara,
  Mapelli, and Ripamonti}}]{Valdes:2007cu}
\bibinfo{author}{\bibfnamefont{M.}~\bibnamefont{Valdes}},
  \bibinfo{author}{\bibfnamefont{A.}~\bibnamefont{Ferrara}},
  \bibinfo{author}{\bibfnamefont{M.}~\bibnamefont{Mapelli}}, \bibnamefont{and}
  \bibinfo{author}{\bibfnamefont{E.}~\bibnamefont{Ripamonti}},
  \bibinfo{journal}{Mon. Not. Roy. Astron. Soc.}
  \textbf{\bibinfo{volume}{377}}, \bibinfo{pages}{245} (\bibinfo{year}{2007}),
  \eprint{astro-ph/0701301}.

\bibitem[{\citenamefont{Natarajan and Schwarz}(2009)}]{Natarajan:2009bm}
\bibinfo{author}{\bibfnamefont{A.}~\bibnamefont{Natarajan}} \bibnamefont{and}
  \bibinfo{author}{\bibfnamefont{D.~J.} \bibnamefont{Schwarz}},
  \bibinfo{journal}{Phys.Rev.} \textbf{\bibinfo{volume}{D80}},
  \bibinfo{pages}{043529} (\bibinfo{year}{2009}), \eprint{0903.4485}.

\bibitem[{\citenamefont{Valdes et~al.}(2012)\citenamefont{Valdes, Evoli,
  Mesinger, Ferrara, and Yoshida}}]{Valdes:2012zv}
\bibinfo{author}{\bibfnamefont{M.}~\bibnamefont{Valdes}},
  \bibinfo{author}{\bibfnamefont{C.}~\bibnamefont{Evoli}},
  \bibinfo{author}{\bibfnamefont{A.}~\bibnamefont{Mesinger}},
  \bibinfo{author}{\bibfnamefont{A.}~\bibnamefont{Ferrara}}, \bibnamefont{and}
  \bibinfo{author}{\bibfnamefont{N.}~\bibnamefont{Yoshida}}
  (\bibinfo{year}{2012}), \eprint{1209.2120}.

\bibitem[{\citenamefont{Zavala et~al.}(2010)\citenamefont{Zavala, Vogelsberger,
  and White}}]{Zavala:2009mi}
\bibinfo{author}{\bibfnamefont{J.}~\bibnamefont{Zavala}},
  \bibinfo{author}{\bibfnamefont{M.}~\bibnamefont{Vogelsberger}},
  \bibnamefont{and} \bibinfo{author}{\bibfnamefont{S.~D.~M.}
  \bibnamefont{White}}, \bibinfo{journal}{Phys. Rev.}
  \textbf{\bibinfo{volume}{D81}}, \bibinfo{pages}{083502}
  (\bibinfo{year}{2010}), \eprint{0910.5221}.

\bibitem[{\citenamefont{Hannestad and Tram}(2011)}]{Hannestad:2010zt}
\bibinfo{author}{\bibfnamefont{S.}~\bibnamefont{Hannestad}} \bibnamefont{and}
  \bibinfo{author}{\bibfnamefont{T.}~\bibnamefont{Tram}},
  \bibinfo{journal}{JCAP} \textbf{\bibinfo{volume}{1101}}, \bibinfo{pages}{016}
  (\bibinfo{year}{2011}), \eprint{1008.1511}.

\bibitem[{\citenamefont{{Chluba} and {Sunyaev}}(2012)}]{2012MNRAS.419.1294C}
\bibinfo{author}{\bibfnamefont{J.}~\bibnamefont{{Chluba}}} \bibnamefont{and}
  \bibinfo{author}{\bibfnamefont{R.~A.} \bibnamefont{{Sunyaev}}},
  \bibinfo{journal}{\mnras} \textbf{\bibinfo{volume}{419}},
  \bibinfo{pages}{1294} (\bibinfo{year}{2012}), \eprint{1109.6552}.

\bibitem[{\citenamefont{Chluba}(2013)}]{Chluba:2013wsa}
\bibinfo{author}{\bibfnamefont{J.}~\bibnamefont{Chluba}},
  \bibinfo{journal}{Mon.Not.Roy.Astron.Soc.} \textbf{\bibinfo{volume}{436}},
  \bibinfo{pages}{2232} (\bibinfo{year}{2013}), \eprint{1304.6121}.

\bibitem[{\citenamefont{Slatyer et~al.}(2009)\citenamefont{Slatyer,
  Padmanabhan, and Finkbeiner}}]{Slatyer:2009yq}
\bibinfo{author}{\bibfnamefont{T.~R.} \bibnamefont{Slatyer}},
  \bibinfo{author}{\bibfnamefont{N.}~\bibnamefont{Padmanabhan}},
  \bibnamefont{and} \bibinfo{author}{\bibfnamefont{D.~P.}
  \bibnamefont{Finkbeiner}}, \bibinfo{journal}{Phys. Rev.}
  \textbf{\bibinfo{volume}{D80}}, \bibinfo{pages}{043526}
  (\bibinfo{year}{2009}), \eprint{0906.1197}.

\bibitem[{\citenamefont{{Slatyer}}(2013)}]{2013PhRvD..87l3513S}
\bibinfo{author}{\bibfnamefont{T.~R.} \bibnamefont{{Slatyer}}},
  \bibinfo{journal}{\prd} \textbf{\bibinfo{volume}{87}}, \bibinfo{eid}{123513}
  (\bibinfo{year}{2013}), \eprint{1211.0283}.

\bibitem[{\citenamefont{Galli et~al.}(2013)\citenamefont{Galli, Slatyer,
  Valdes, and Iocco}}]{Galli:2013dna}
\bibinfo{author}{\bibfnamefont{S.}~\bibnamefont{Galli}},
  \bibinfo{author}{\bibfnamefont{T.~R.} \bibnamefont{Slatyer}},
  \bibinfo{author}{\bibfnamefont{M.}~\bibnamefont{Valdes}}, \bibnamefont{and}
  \bibinfo{author}{\bibfnamefont{F.}~\bibnamefont{Iocco}},
  \bibinfo{journal}{Phys.Rev.} \textbf{\bibinfo{volume}{D88}},
  \bibinfo{pages}{063502} (\bibinfo{year}{2013}), \eprint{1306.0563}.

\bibitem[{\citenamefont{{Shull} and {van
  Steenberg}}(1985)}]{1985ApJ...298..268S}
\bibinfo{author}{\bibfnamefont{J.~M.} \bibnamefont{{Shull}}} \bibnamefont{and}
  \bibinfo{author}{\bibfnamefont{M.~E.} \bibnamefont{{van Steenberg}}},
  \bibinfo{journal}{\apj} \textbf{\bibinfo{volume}{298}}, \bibinfo{pages}{268}
  (\bibinfo{year}{1985}).

\bibitem[{\citenamefont{{Furlanetto} and
  {Stoever}}(2010)}]{2010MNRAS.404.1869F}
\bibinfo{author}{\bibfnamefont{S.~R.} \bibnamefont{{Furlanetto}}}
  \bibnamefont{and} \bibinfo{author}{\bibfnamefont{S.~J.}
  \bibnamefont{{Stoever}}}, \bibinfo{journal}{\mnras}
  \textbf{\bibinfo{volume}{404}}, \bibinfo{pages}{1869} (\bibinfo{year}{2010}),
  \eprint{0910.4410}.

\bibitem[{\citenamefont{Valdes and Ferrara}(2008)}]{Valdes:2008cr}
\bibinfo{author}{\bibfnamefont{M.}~\bibnamefont{Valdes}} \bibnamefont{and}
  \bibinfo{author}{\bibfnamefont{A.}~\bibnamefont{Ferrara}},
  \bibinfo{journal}{Mon.Not.Roy.Astron.Soc.} \textbf{\bibinfo{volume}{387}},
  \bibinfo{pages}{8} (\bibinfo{year}{2008}), \eprint{0803.0370}.

\bibitem[{\citenamefont{Valdes et~al.}(2010)\citenamefont{Valdes, Evoli, and
  Ferrara}}]{Valdes:2009cq}
\bibinfo{author}{\bibfnamefont{M.}~\bibnamefont{Valdes}},
  \bibinfo{author}{\bibfnamefont{C.}~\bibnamefont{Evoli}}, \bibnamefont{and}
  \bibinfo{author}{\bibfnamefont{A.}~\bibnamefont{Ferrara}},
  \bibinfo{journal}{Mon.Not.Roy.Astron.Soc.} \textbf{\bibinfo{volume}{404}},
  \bibinfo{pages}{1569} (\bibinfo{year}{2010}), \eprint{0911.1125}.

\bibitem[{\citenamefont{Evoli et~al.}(2012)\citenamefont{Evoli, Valdes,
  Ferrara, and Yoshida}}]{MNR:MNR20624}
\bibinfo{author}{\bibfnamefont{C.}~\bibnamefont{Evoli}},
  \bibinfo{author}{\bibfnamefont{M.}~\bibnamefont{Valdes}},
  \bibinfo{author}{\bibfnamefont{A.}~\bibnamefont{Ferrara}}, \bibnamefont{and}
  \bibinfo{author}{\bibfnamefont{N.}~\bibnamefont{Yoshida}},
  \bibinfo{journal}{Monthly Notices of the Royal Astronomical Society}
  \textbf{\bibinfo{volume}{422}}, \bibinfo{pages}{420} (\bibinfo{year}{2012}),
  ISSN \bibinfo{issn}{1365-2966},
  \urlprefix\url{http://dx.doi.org/10.1111/j.1365-2966.2012.20624.x}.

\bibitem[{\citenamefont{Madhavacheril et~al.}(2014)\citenamefont{Madhavacheril,
  Sehgal, and Slatyer}}]{Madhavacheril:2013cna}
\bibinfo{author}{\bibfnamefont{M.~S.} \bibnamefont{Madhavacheril}},
  \bibinfo{author}{\bibfnamefont{N.}~\bibnamefont{Sehgal}}, \bibnamefont{and}
  \bibinfo{author}{\bibfnamefont{T.~R.} \bibnamefont{Slatyer}},
  \bibinfo{journal}{Phys.Rev.} \textbf{\bibinfo{volume}{D89}},
  \bibinfo{pages}{103508} (\bibinfo{year}{2014}), \eprint{1310.3815}.

\bibitem[{\citenamefont{Weniger et~al.}(2013)\citenamefont{Weniger, Serpico,
  Iocco, and Bertone}}]{Weniger:2013hja}
\bibinfo{author}{\bibfnamefont{C.}~\bibnamefont{Weniger}},
  \bibinfo{author}{\bibfnamefont{P.~D.} \bibnamefont{Serpico}},
  \bibinfo{author}{\bibfnamefont{F.}~\bibnamefont{Iocco}}, \bibnamefont{and}
  \bibinfo{author}{\bibfnamefont{G.}~\bibnamefont{Bertone}},
  \bibinfo{journal}{Phys.Rev.} \textbf{\bibinfo{volume}{D87}},
  \bibinfo{pages}{123008} (\bibinfo{year}{2013}), \eprint{1303.0942}.

\bibitem[{\citenamefont{Hutsi et~al.}(2011)\citenamefont{Hutsi, Chluba, Hektor,
  and Raidal}}]{Hutsi:2011vx}
\bibinfo{author}{\bibfnamefont{G.}~\bibnamefont{Hutsi}},
  \bibinfo{author}{\bibfnamefont{J.}~\bibnamefont{Chluba}},
  \bibinfo{author}{\bibfnamefont{A.}~\bibnamefont{Hektor}}, \bibnamefont{and}
  \bibinfo{author}{\bibfnamefont{M.}~\bibnamefont{Raidal}},
  \bibinfo{journal}{\aap} \textbf{\bibinfo{volume}{535}}, \bibinfo{eid}{A26}
  (\bibinfo{year}{2011}), \eprint{1103.2766}.

\bibitem[{\citenamefont{{Seager} et~al.}(1999)\citenamefont{{Seager},
  {Sasselov}, and {Scott}}}]{Seager:1999}
\bibinfo{author}{\bibfnamefont{S.}~\bibnamefont{{Seager}}},
  \bibinfo{author}{\bibfnamefont{D.~D.} \bibnamefont{{Sasselov}}},
  \bibnamefont{and} \bibinfo{author}{\bibfnamefont{D.}~\bibnamefont{{Scott}}},
  \bibinfo{journal}{\apjl} \textbf{\bibinfo{volume}{523}}, \bibinfo{pages}{L1}
  (\bibinfo{year}{1999}), \eprint{arXiv:astro-ph/9909275}.

\bibitem[{\citenamefont{{Chluba} and {Thomas}}(2011)}]{Chluba:2010ca}
\bibinfo{author}{\bibfnamefont{J.}~\bibnamefont{{Chluba}}} \bibnamefont{and}
  \bibinfo{author}{\bibfnamefont{R.~M.} \bibnamefont{{Thomas}}},
  \bibinfo{journal}{\mnras} \textbf{\bibinfo{volume}{412}},
  \bibinfo{pages}{748} (\bibinfo{year}{2011}), \eprint{1010.3631}.

\bibitem[{\citenamefont{Ali-Haimoud and Hirata}(2011)}]{AliHaimoud:2010dx}
\bibinfo{author}{\bibfnamefont{Y.}~\bibnamefont{Ali-Haimoud}} \bibnamefont{and}
  \bibinfo{author}{\bibfnamefont{C.~M.} \bibnamefont{Hirata}},
  \bibinfo{journal}{Phys. Rev.} \textbf{\bibinfo{volume}{D83}},
  \bibinfo{pages}{043513} (\bibinfo{year}{2011}), \eprint{1011.3758}.

\bibitem[{\citenamefont{Zhang et~al.}(2007)\citenamefont{Zhang, Chen,
  Kamionkowski, Si, and Zheng}}]{Zhang:2007zzh}
\bibinfo{author}{\bibfnamefont{L.}~\bibnamefont{Zhang}},
  \bibinfo{author}{\bibfnamefont{X.}~\bibnamefont{Chen}},
  \bibinfo{author}{\bibfnamefont{M.}~\bibnamefont{Kamionkowski}},
  \bibinfo{author}{\bibfnamefont{Z.-g.} \bibnamefont{Si}}, \bibnamefont{and}
  \bibinfo{author}{\bibfnamefont{Z.}~\bibnamefont{Zheng}},
  \bibinfo{journal}{Phys. Rev.} \textbf{\bibinfo{volume}{D76}},
  \bibinfo{pages}{061301} (\bibinfo{year}{2007}), \eprint{0704.2444}.

\bibitem[{\citenamefont{Galli et~al.}(2009)\citenamefont{Galli, Iocco, Bertone,
  and Melchiorri}}]{Galli:2009zc}
\bibinfo{author}{\bibfnamefont{S.}~\bibnamefont{Galli}},
  \bibinfo{author}{\bibfnamefont{F.}~\bibnamefont{Iocco}},
  \bibinfo{author}{\bibfnamefont{G.}~\bibnamefont{Bertone}}, \bibnamefont{and}
  \bibinfo{author}{\bibfnamefont{A.}~\bibnamefont{Melchiorri}},
  \bibinfo{journal}{Phys. Rev.} \textbf{\bibinfo{volume}{D80}},
  \bibinfo{pages}{023505} (\bibinfo{year}{2009}), \eprint{0905.0003}.

\bibitem[{\citenamefont{Kanzaki et~al.}(2010)\citenamefont{Kanzaki, Kawasaki,
  and Nakayama}}]{Kanzaki:2009hf}
\bibinfo{author}{\bibfnamefont{T.}~\bibnamefont{Kanzaki}},
  \bibinfo{author}{\bibfnamefont{M.}~\bibnamefont{Kawasaki}}, \bibnamefont{and}
  \bibinfo{author}{\bibfnamefont{K.}~\bibnamefont{Nakayama}},
  \bibinfo{journal}{Prog.Theor.Phys.} \textbf{\bibinfo{volume}{123}},
  \bibinfo{pages}{853} (\bibinfo{year}{2010}), \eprint{0907.3985}.

\bibitem[{\citenamefont{Hisano et~al.}(2011)\citenamefont{Hisano, Kawasaki,
  Kohri, Moroi, Nakayama et~al.}}]{Hisano:2011dc}
\bibinfo{author}{\bibfnamefont{J.}~\bibnamefont{Hisano}},
  \bibinfo{author}{\bibfnamefont{M.}~\bibnamefont{Kawasaki}},
  \bibinfo{author}{\bibfnamefont{K.}~\bibnamefont{Kohri}},
  \bibinfo{author}{\bibfnamefont{T.}~\bibnamefont{Moroi}},
  \bibinfo{author}{\bibfnamefont{K.}~\bibnamefont{Nakayama}},
  \bibnamefont{et~al.}, \bibinfo{journal}{Phys.Rev.}
  \textbf{\bibinfo{volume}{D83}}, \bibinfo{pages}{123511}
  (\bibinfo{year}{2011}), \eprint{1102.4658}.

\bibitem[{\citenamefont{{Galli} et~al.}(2011)\citenamefont{{Galli}, {Iocco},
  {Bertone}, and {Melchiorri}}}]{Galli:2011rz}
\bibinfo{author}{\bibfnamefont{S.}~\bibnamefont{{Galli}}},
  \bibinfo{author}{\bibfnamefont{F.}~\bibnamefont{{Iocco}}},
  \bibinfo{author}{\bibfnamefont{G.}~\bibnamefont{{Bertone}}},
  \bibnamefont{and}
  \bibinfo{author}{\bibfnamefont{A.}~\bibnamefont{{Melchiorri}}},
  \bibinfo{journal}{\prd} \textbf{\bibinfo{volume}{84}},
  \bibinfo{pages}{027302} (\bibinfo{year}{2011}), \eprint{1106.1528}.

\bibitem[{\citenamefont{Lopez-Honorez et~al.}(2013)\citenamefont{Lopez-Honorez,
  Mena, Palomares-Ruiz, and Vincent}}]{Lopez-Honorez:2013cua}
\bibinfo{author}{\bibfnamefont{L.}~\bibnamefont{Lopez-Honorez}},
  \bibinfo{author}{\bibfnamefont{O.}~\bibnamefont{Mena}},
  \bibinfo{author}{\bibfnamefont{S.}~\bibnamefont{Palomares-Ruiz}},
  \bibnamefont{and} \bibinfo{author}{\bibfnamefont{A.~C.}
  \bibnamefont{Vincent}}, \bibinfo{journal}{JCAP}
  \textbf{\bibinfo{volume}{1307}}, \bibinfo{pages}{046} (\bibinfo{year}{2013}),
  \eprint{1303.5094}.

\bibitem[{\citenamefont{Diamanti et~al.}(2014)\citenamefont{Diamanti,
  Lopez-Honorez, Mena, Palomares-Ruiz, and Vincent}}]{Diamanti:2013bia}
\bibinfo{author}{\bibfnamefont{R.}~\bibnamefont{Diamanti}},
  \bibinfo{author}{\bibfnamefont{L.}~\bibnamefont{Lopez-Honorez}},
  \bibinfo{author}{\bibfnamefont{O.}~\bibnamefont{Mena}},
  \bibinfo{author}{\bibfnamefont{S.}~\bibnamefont{Palomares-Ruiz}},
  \bibnamefont{and} \bibinfo{author}{\bibfnamefont{A.~C.}
  \bibnamefont{Vincent}}, \bibinfo{journal}{JCAP}
  \textbf{\bibinfo{volume}{1402}}, \bibinfo{pages}{017} (\bibinfo{year}{2014}),
  \eprint{1308.2578}.

\bibitem[{\citenamefont{Ade et~al.}(2015)}]{Planck:2015xua}
\bibinfo{author}{\bibfnamefont{P.}~\bibnamefont{Ade}} \bibnamefont{et~al.}
  (\bibinfo{collaboration}{Planck}) (\bibinfo{year}{2015}),
  \eprint{1502.01589}.

\bibitem[{\citenamefont{Wong et~al.}(2008)\citenamefont{Wong, Moss, and
  Scott}}]{Wong:2007ym}
\bibinfo{author}{\bibfnamefont{W.~Y.} \bibnamefont{Wong}},
  \bibinfo{author}{\bibfnamefont{A.}~\bibnamefont{Moss}}, \bibnamefont{and}
  \bibinfo{author}{\bibfnamefont{D.}~\bibnamefont{Scott}},
  \bibinfo{journal}{Mon.Not.Roy.Astron.Soc.} \textbf{\bibinfo{volume}{386}},
  \bibinfo{pages}{1023} (\bibinfo{year}{2008}), \eprint{0711.1357}.

\bibitem[{\citenamefont{{Evoli} et~al.}(2013)\citenamefont{{Evoli}, {Pandolfi},
  and {Ferrara}}}]{Evoli:2012qh}
\bibinfo{author}{\bibfnamefont{C.}~\bibnamefont{{Evoli}}},
  \bibinfo{author}{\bibfnamefont{S.}~\bibnamefont{{Pandolfi}}},
  \bibnamefont{and}
  \bibinfo{author}{\bibfnamefont{A.}~\bibnamefont{{Ferrara}}},
  \bibinfo{journal}{\mnras} \textbf{\bibinfo{volume}{433}},
  \bibinfo{pages}{1736} (\bibinfo{year}{2013}), \eprint{1210.6845}.

\bibitem[{\citenamefont{Finkbeiner et~al.}(2012)\citenamefont{Finkbeiner,
  Galli, Lin, and Slatyer}}]{Finkbeiner:2011dx}
\bibinfo{author}{\bibfnamefont{D.~P.} \bibnamefont{Finkbeiner}},
  \bibinfo{author}{\bibfnamefont{S.}~\bibnamefont{Galli}},
  \bibinfo{author}{\bibfnamefont{T.}~\bibnamefont{Lin}}, \bibnamefont{and}
  \bibinfo{author}{\bibfnamefont{T.~R.} \bibnamefont{Slatyer}},
  \bibinfo{journal}{Phys.Rev.} \textbf{\bibinfo{volume}{D85}},
  \bibinfo{pages}{043522} (\bibinfo{year}{2012}), \eprint{1109.6322}.

\end{thebibliography}

\end{document}